\documentclass{article}
\usepackage{geometry}
\usepackage{authblk}
\usepackage{amsmath,amsfonts,amssymb}
\usepackage{bm}
\usepackage{graphicx}
\usepackage{color}
\usepackage{subcaption}
\usepackage{hyperref}
\usepackage[square,numbers,compress]{natbib}
\usepackage{fltpage}
\usepackage{xr}
\usepackage{sectsty}
\usepackage{setspace}
\hypersetup{
	colorlinks = true,
	citecolor= cyan,
	linkcolor= black
	}

\date{\today}

\begin{document}
	
\title{\textbf{Controlling collective dynamical states of mesoscale brain networks with local perturbations}}

\author[1]{Lia Papadopoulos}
\author[2,3]{Demian Battaglia}
\author[4,5,6,7,8,9]{Dani S. Bassett}
\affil[1]{Institute of Neuroscience, University of Oregon, Eugene, Oregon}
\affil[2]{Institute for Systems Neuroscience, Aix-Marseille University, Marseille, France}
\affil[3]{University of Strasbourg Institute for Advanced Studies, Strasbourg, France}
\affil[4]{Department of Bioengineering, University of Pennsylvania, Philadelphia, PA}
\affil[5]{Department of Electrical \& Systems Engineering, University of Pennsylvania, Philadelphia, PA}
\affil[6]{Department of Physics \& Astronomy, University of Pennsylvania, Philadelphia, PA}
\affil[7]{Department of Neurology, University of Pennsylvania, Philadelphia, PA}
\affil[8]{Department of Psychiatry, University of Pennsylvania, Philadelphia, PA}
\affil[9]{Santa Fe Institute, Santa Fe, NM}

\maketitle

\section*{}
\begin{center}
\large{\textit{\textbf{Note:} This report has been adapted from Chapter 3 of the dissertation of L. Papadopoulos, which can be found in Ref. \cite{Papadopoulos_PhDThesis}.}}
\end{center}
	
\section*{Abstract}

Oscillatory synchrony is hypothesized to support the flow of information between brain regions, with different phase-locked configurations enabling activation of different effective interactions. Along these lines, past work has proposed multistable phase-locking as a means for hardwired brain networks to flexibly support multiple functional patterns, without having to reconfigure their anatomical connections. Given the potential link between interareal communication and phase-locked states, it is thus important to understand how those states might be controlled to achieve rapid alteration of functional connectivity in interareal circuits. Here, we study functional state control in small networks of coupled neural masses that display collective multistability under determinstic conditions, and that display more biologically-realistic irregular oscillations and transient phase-locking when conditions are stochastic. In particular, we investigate the global responses of these mesoscale circuits to external signals that target only a single subunit. Focusing mainly on the more realistic scenario wherein network dynamics are stochastic, we identify conditions under which local inputs \textit{(i)} can trigger fast transitions to topologically distinct functional connectivity motifs that are temporarily stable (“state switching”), \textit{(ii)} can smoothly adjust the spatial pattern of phase-relations for a particular set of lead-lag relationships (“state morphing”), and \textit{(iii)} fail to regulate global phase-locking states. In total, our results add to a growing literature highlighting that the modulation of multistable, interareal coherence patterns could provide a basis for flexible brain network operation.

\section{Introduction}
\label{s:intro}

Brain function depends not only upon the behaviors and capabilities of single areas in isolation, but also on the ability of many distributed regions to coordinate and exchange information with one another. Nonetheless, understanding how interareal communication is established and reconfigured based on the dynamics and anatomical coupling of multiregional brain circuits remains an open question. One popular hypothesis for how information may be routed between different neuronal populations is that of \textit{communication-through-coherence} (CTC) \cite{Fries:2005a,Fries2015:Rhythms,Maris2016:DiversePhase,Varela2001:TheBrainweb,Bastos2015:Communication,Engel:2001a,Engel2013:IntrinsicCoupling,Cannon2014:Neurosystems}. This mechanism proposes that communication between coupled neuronal areas can be established when the areas' population activities oscillate and become phase-locked (coherent) with a certain phase relation that allows output from one area to arrive at a receiving area during times when the receiving site is excitable. The hypothesized link between coordinated oscillations and communication thus suggests that the pattern of phase relations emergent in an interareal network defines a ``functional connectivity" state of the system. This collective state then determines which areas can effectively exchange information with which others \cite{Battaglia2020:Functional,Battaglia2014:FunctionFollows}. Indeed, by utilizing information-theoretic measures, prior studies have demonstrated a link between the lead-lag relationships among collective neuronal rhythms and the directionality of information flow among the corresponding areas in an anatomical circuit \cite{Battaglia2012:DynamicEffective,Palmigiano2017:FlexibleInformation,Kirst2016:Dynamic}. 
In addition to establishing a mechanism for interareal communication, it is also desirable that networks of anatomically-linked brain regions are able to perform multiple functions. For example, how can these circuits quickly modulate the direction or extent of information flow at different times or depending on context? This question is especially interesting given that long-range structural couplings that determine direct routes for signal propagation cannot be drastically rewired on rapid time scales. This fact precludes structural changes from underlying the capability of large-scale brain networks to achieve many stable and rapidly adjustable functional states. In thinking about potential mechanisms by which an interareal network could generate a number of different dynamical configurations despite an inflexible anatomical scaffold, an exciting collection of past studies has suggested multistable phase-locking \cite{Battaglia2012:DynamicEffective,Palmigiano2017:FlexibleInformation,Kirst2016:Dynamic,Witt2013:Controlling,Battaglia2007:TemporalDecorrelation}. There has also been some experimental work suggesting that multiple phase relations could exist within the same anatomical circuit, lending some initial support to this theoretical idea \cite{Dotson2014:FrontoParietal}. But, given that distinct phase-locked configurations in interareal brain networks may enable distinct computations or functional outcomes, critical questions arise. For example, how can specific collective states from a larger set be tuned or rapidly selected for -- without rewiring anatomical connectivity -- in order to achieve a desired pattern of functional interactions? And, is it possible to effectively control interareal coherence in biologically-realistic scenarios where oscillations are noisy and irregular, and where consistent phase relations may be relatively short-lived?

Building upon previous work \cite{Battaglia2012:DynamicEffective,Palmigiano2017:FlexibleInformation,Kirst2016:Dynamic,Witt2013:Controlling,Lisitsyn2019:Causally}, here we examine this question of functional state control by considering reduced computational models of small, multiarea brain circuits composed of oscillatory subunits.  We focus in particular on parameter regimes that yield multistable phase-locking when the network dynamics are deterministic and in the absence of any state control signal. Though studying the deterministic version of the circuit model is useful to acquire a basic understanding of its dynamical behaviors, our primary interest is the more realistic and complicated scenario in which the network operates in the presence of a stochastically fluctuating background environment. This setting gives rise to more biologically-plausible dynamics: the oscillations of individual units become noisy and less precisely timed, and phase-locking in a given configuration lasts only for a finite amount of time due to spontaneous (noise-induced) switches between its assortment of collective states. 

After characterizing the behavior of the model networks under baseline conditions, we proceed to examine how their collective dynamical state is affected by different types of external signals that target only a single region. These local perturbations could represent, for example, sensory inputs, artificial stimulation, or targeted, modulatory signals coming from other brain areas in a larger network. We focus specifically on how these \textit{locally-applied} external inputs could be used to precisely control the collective dynamical states of the network as a \textit{whole}, without altering its structural connectivity. 

We study the consequences of two kinds of focal perturbations. To begin, we analyze the effects of brief, excitatory pulses injected into one region of the circuit. Given that the pulse is applied at a specific phase of the receiving area's ongoing oscillation, these perturbations can induce a global reconfiguration of the activity pattern that manifests as a transition to a different dynamical attractor of the entire system. Here we find that even when network dynamics are stochastic, carefully timed pulse inputs can still be used to trigger state transitions with efficacies that exceed chance levels and that occur on much faster timescales than would happen spontaneously. However, we also describe, for the stochastic networks, a parameter regime in which this state control can break down and become ineffective. In addition to pulse inputs, we also consider the effects of sustained rhythmic stimulation applied to a single area. Depending upon its amplitude and frequency, we find that this type of external drive can lead to a shifting of the spatial configuration of the phase differences in the network. By inducing this morphing of the collective state, rhythmic input thus enables a different type of tuning of the temporal relations between regional activities. When background noise leads to spontaneous state-switching, steady sinusoidal input can still cause symmetry-breaking that shifts the phase relations of the collective states in which the network tends to dwell. Taken together, our results provide additional support to a collection of studies that have considered the control of dynamical multistability as a path towards flexible, coordinated behavior in networks of neural populations that exhibit oscillatory coherence.

\section{Model and methods}

\begin{figure}[h!]
	\centering
	\includegraphics[width=\columnwidth]{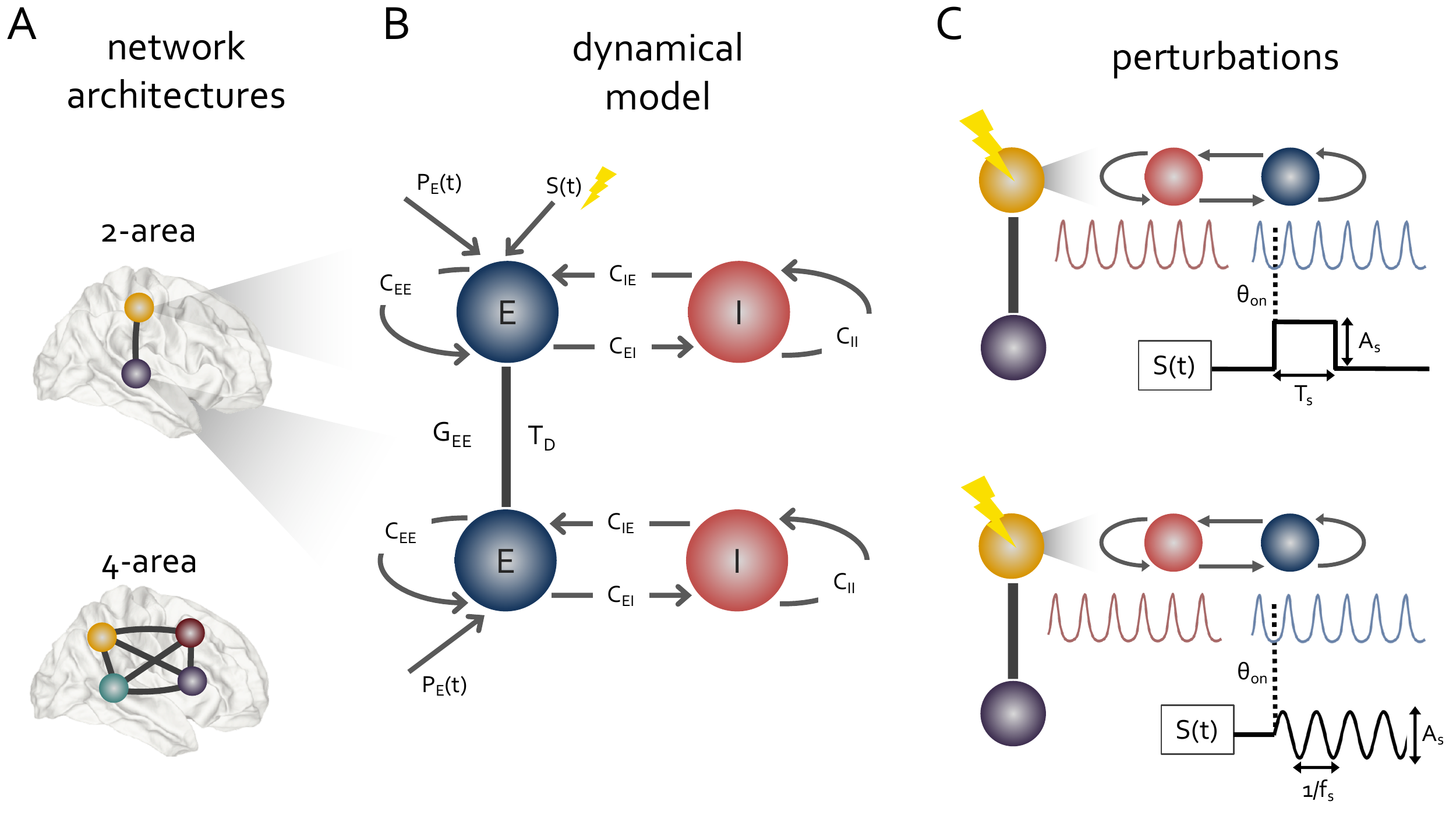}
	\caption[Schematics of the model setup.]{\textbf{Schematics of the model setup.} \textbf{(A)} The two different network architectures examined in this study: a bidrectional 2-area network (top) and an all-to-all 4-area network (bottom). Each node in each network represents a large neuronal population or brain area, and edges represent anatomical connections. \textbf{(B)} A schematic of the computational model of network dynamics for the 2-area system. Brain areas are modeled as Wilson-Cowan neural mass units, each of which are composed of coupled subpopulations of excitatory (E) and inhibitory (I) neural populations. Anatomical coupling between distinct regions is introduced by linking the excitatory pools of the corresponding areas. This interareal connectivity is characterized by a coupling strength $G_{EE}$ and a time-delay $T_{D}$. In addition to the generic background drive $P_{E}(t)$ that enters each region's excitatory pool, the yellow population also receives an external perturbation signal $S(t)$ (indicated by the yellow lightning bolt). \textbf{(C)} Two different types of perturbations $S(t)$ are considered. In the top panel, the excitatory subgroup of the yellow brain area receives a brief square-wave pulse input. This perturbation is characterized by its duration $T_{s}$, its amplitude $A_{s}$, and the phase $\theta_{\mathrm{on}}$ of the region's ongoing oscillation (blue time-series) at which the pulse begins. In the bottom panel, the excitatory subgroup of the yellow brain area instead receives sustained rhythmic stimulation. This signal is characterized by its frequency $f_{s}$, its amplitude $A_{s}$, and the phase $\theta_{\mathrm{on}}$ of the region's ongoing oscillation (blue time-series) at which the stimulation is initiated.}
	\label{f:model_setup}
\end{figure}

\subsection{Biophysical model of neural population activity}

In this study, we numerically simulate neural population dynamics using a biophysically-motivated but coarse-grained model. In particular, each area in a model anatomical brain circuit (Fig.~\ref{f:model_setup}A) is represented as a Wilson-Cowan (WC) neural mass, where each WC unit describes the average activity of large, interacting excitatory and inhibitory neuronal subgroups \cite{Wilson1972:ExcitatoryandInhibitory} (Fig.~\ref{f:model_setup}B). To study the collective activity of multiarea brain circuits, we couple individual WC units according to two simple network architectures: a 2-unit network with homogeneous, bidirectional connectivity and a 4-unit undirected network with homogeneous, all-to-all connectivity (Fig.~\ref{f:model_setup}A). Each of the $N$ units or nodes in these circuits should be thought of as representing a localized brain region or large neuronal assembly that would be composed of tens of thousands of individual neurons. Moreover, edges in these networks correspond to long-range anatomical couplings between the different neural areas. 

This type of framework, in which the activity of a distributed brain network is modeled as a system of interacting neural masses, has been utilized in several previous investigations \cite{Breakspear2017:DynamicModels}. Importantly, there are different ways that one could incorporate long-range coupling between the neural populations. For simplicity (and in line with a number of past studies \cite{Muldoon2016:StimulationBased,Abeysuriya2018:ABiophysicalModel,Deco2009:KeyRoleofCoupling,Gollo2015:DwellingQuietly,Honey2009:Predicting,Hlinka2012:UsingComputationalModels,Roberts2019:Metastable}), here we will consider anatomical connections to link only the excitatory pools of distinct areas (Fig.~\ref{f:model_setup}B). We also assume that distinct regions or populations are distributed in space, such that it may take some finite amount of time for activity to propagate from one area to another. Hence, every anatomical connection is described by both a coupling strength $G_{EE}$ and time-delay $T_{D}$, which we take to be uniform across the network.

Adapting the formulation of the WC model used in Ref. \cite{Heitmann2017:Optogenetic}, the activity of the $j^{th}$ region evolves according to the following set of differential equations:

\begin{subequations}
	\begin{equation}
	\tau_{E} \frac{dE_{j}(t)}{dt} = -E_{j}(t) + \mathcal{S_{E}}[c_{EE} E_{j}(t) - c_{IE} I_{j}(t) - b_{E} + P_{E,j}(t) + \delta_{j1}S(t) + G_{EE} \sum_{i=1}^{N} E_{i}(t - T_{D})]
	\label{eq:WC_E_multistable}
	\end{equation}
	\begin{equation}
	\tau_{I} \frac{dI_{j}(t)}{dt} = -I_{j}(t) + \mathcal{S_{I}}[c_{EI} E_{j}(t) - c_{II} I_{j}(t) - b_{I} + P_{I,j}(t) ],
	\label{eq:WC_I_multistable}
	\end{equation}
	\label{eq:WC_multistable}
\end{subequations}
~\\
where
~\\
\begin{subequations}
	\begin{equation}
	\mathcal{S}_{E}(x) = \frac{1}{1+e^{-x}}
	\label{eq:WC_sigE_multistable}
	\end{equation}
	\begin{equation}
	\mathcal{S}_{I}(x) = \frac{1}{1+e^{-x}}.
	\label{eq:WC_sigI_multistable}
	\end{equation}
\end{subequations}
~\\
Here, $E_{j}(t)$ and $I_{j}(t)$ represent the mean firing rates of the excitatory and inhibitory subgroups of the $j^{\mathrm{th}}$ unit. The internal parameters of a single WC unit are: the excitatory time constant $\tau_{E}$, the inhibitory time constant $\tau_{I}$, the local $E \rightarrow E$ coupling $c_{EE}$, the local $E \rightarrow I$ coupling $c_{EI}$, the local $I \rightarrow E$ coupling $c_{IE}$, a quantity that controls the firing threshold of the excitatory pool $b_{E}$, and a quantity that controls the firing threshold of the inhibitory pool $b_{I}$. In general, the $E$ and $I$ populations also receive generic background drives, represented by the variables $P_{E,j}(t)$ and $P_{I,j}(t)$. Additionally, due to long-range coupling between different neural populations, each area $j$ also receives input (which is potentially delayed) from other areas that connect to it. This network-based influence is captured in the final term of Eq.~\ref{eq:WC_E_multistable}. Specifically, $G_{EE}$ is a global coupling that scales the weight of all interareal links, and the term $E_{i}(t-T_{D})$ represents delayed excitatory input from area $i$ to area $j$, where the time-delay is given by $T_{D}$. 

The final ingredient of the model is the local perturbation signal, which we denote as $S(t)$. Here, a ``local perturbation" refers to an additional external input that targets only a single area in the larger network. Because we consider fully symmetric networks, all units are topologically identical, but for clarity, we always assign area 1 to be the perturbed site (enforced by the $\delta_{j1}S(t)$ term in Eq.~\ref{eq:WC_E_multistable}). For simplicity, note that we also consider the case that this external stimulation targets only the excitatory subpopulation of the perturbed brain region. With this setup, ``baseline" conditions correspond to the scenario that the excitatory pool of all units $j \in \{1,...,N\}$ receives only the background drive $P_{E,j}(t)$ (i.e., $S(t) = 0$). We provide further details on the perturbation signals in Sec.~\ref{s:modeling_perturbations}.

\subsubsection{Background inputs: deterministic and stochastic versions}
\label{s:stochastic_backgroundDrive}

Throughout this study we consider two varieties of the network model presented above: one in which the neural population dynamics evolve deterministically and one in which the network activity is driven instead by stochastic background inputs. In the former version of the model, the background inputs to the $E$ and $I$ populations are deterministic and taken to be constant in time, such that $P_{E,j}(t) := P_{E,j}$ $\forall j \in \{1,...,N\}$ and $P_{I,j}(t) := P_{I}$ $\forall j \in \{1,...,N\}$. Though useful for gaining intuition about the model's behavior, real brain activity is inherently noisy \cite{Faisal2008:Noise}. In order to incorporate this fact into the model and make our system slightly more realistic, we will primarily consider a scenario in which the background inputs to each unit, $P_{E,j}(t)$ and $P_{I,j}(t)$, are stochastic. In the presence of such noisy drive, the regional activities $E_{j}(t)$ and $I_{j}(t)$ will also exhibit stochastic fluctuations.

To operationalize the formulation of the stochastic model, we consider a situation in which all background inputs to the $E$ and $I$ populations are modeled as independent Ornstein-Uhlenbeck (O.U.) processes. In particular, we have

\begin{subequations}
	\begin{equation}
	\frac{dP_{E,j}(t)}{dt} = \alpha[\overline{P_{E,j}} - P_{E,j}(t)] + \sigma \eta_{j}(t) \quad \forall j \in \{1,...,N\}
	\label{eq:OUprocess_E}
	\end{equation}
	\begin{equation}
	\frac{dP_{I,j}(t)}{dt} = \alpha[\overline{P_{I,j}} - P_{I,j}(t)] + \sigma \xi_{j}(t) \quad \forall j \in \{1,...,N\},
	\label{eq:OUprocess_I}
	\end{equation}
\end{subequations}
~\\
\sloppy{ where $\eta_{j}(t)$ and $\xi_{j}(t)$ are Gaussian white noises that satisfy $\langle \eta_{j}(t) \rangle = 0$, $\langle \xi_{j}(t) \rangle = 0$, $\langle \eta_{i}(t) \eta_{j}(t') \rangle = \delta_{ij}\delta(t-t')$, $\langle \xi_{i}(t) \xi_{j}(t') \rangle = \delta_{ij}\delta(t-t')$, and $\langle \eta_{i}(t)\xi_{j}(t') \rangle = 0$.  In the above formulation, $\overline{P_{E,j}}$ and $\overline{P_{I,j}}$ correspond to the long-term means of the input to the $E$ and $I$ population, respectively, $\sigma$ sets the strength of the noise, and $\alpha$ sets the rate of mean reversion. 

\subsubsection{Default model parameters}

Because we are interested in phase-locking between collective rhythms in mesoscale brain circuits, we wish to consider parameter regimes of the model in which individual units exhibit oscillatory dynamics \cite{Wilson1972:ExcitatoryandInhibitory}. To acheive this, we use the parameters for a single WC unit indicated in Ref. \cite{Heitmann2017:Optogenetic}. With those parameter values, it is known that a Hopf bifurcation occurs in an isolated unit as the strength of a constant input $P_{E}$ to its excitatory population is varied \cite{Heitmann2017:Optogenetic}.  In Table~\ref{t:parameters_multistable}, we present all model parameters, and for those that we hold fixed throughout the study, we specify their value. With the chosen time constants, the model tends to oscillate in the gamma frequency range ($\sim$30-80 Hz). 

\begin{table}[h!]
	\begin{center}
		\begin{tabular}{c c c} 
			\hline
			Parameter 				& Description									& Value\\
			\hline
			\hline
			$\tau_{E}$				& time constant	of E pool						&		2 ms \\
			$\tau_{I}$				& time constant	of I pool						&		4 ms \\
			$c_{EE}$				& local E-to-E coupling							&		15 \\
			$c_{IE}$				& local I-to-E coupling							&		15 \\
			$c_{EI}$				& local E-to-I coupling							&		15 \\
			$c_{II}$				& local I-to-I coupling							&		7 \\
			$b_{E}$					& firing threshold of E pool				    &		4 \\
			$b_{I}$					& firing threshold of I pool					&		4 \\
			$\overline{P_{E}}$		& mean input to E pool							&	    varied     \\
			$\overline{P_{I}}$		& mean input to I pool							&		0 		\\
			$S(t)$ 				  & external perturbation								& 		varied \\
			$\alpha$				& inverse correlation time of background drive	&		varied \\
			$\sigma$				& noise strength of background drive			&		varied \\
			$G_{EE}$				& interareal E-to-E coupling strength			&		varied \\
			$T_{d}$					& interareal time delay							&		varied \\
			\hline
			\hline
		\end{tabular}
	\end{center}
	\caption[Model parameters.]{\textbf{Model parameters.} Parameters with a corresponding numerical value are held constant for all simulations and analyses, while parameters labeled as ``varied" indicate that we study model behavior as a function of or for different values of those parameters.}
	\label{t:parameters_multistable}
\end{table}

\subsection{Incorporating the effects of local perturbations into the model}
\label{s:modeling_perturbations}

In this section we detail the two types of local perturbations considered in this study.

\subsubsection{Brief pulse inputs}
\label{s:pulse_input_description}

The first type of perturbation we consider is a brief pulse input injected into the excitatory pool of a single network node. For these perturbations, the external signal $S(t)$ is a square wave pulse characterized by a duration $T_{s} > 0$ and an amplitude $A_{s} > 0$ (see Fig.~\ref{f:model_setup}C, Top). More specifically, we have that

\begin{equation}
S(t) = \left\{\begin{array}{ll}
0 \text{ for } t < t_{s} \text{ and } t > t_{s} + T_{s} \\
A_{s} \text{ for } t_{s} \leq t \leq t_{s} + T_{s} ,
\end{array} \right. \
\label{eq:pulse_stim}
\end{equation}
~\\
where $t_{s}$ indicates the time at which the pulse is turned on. The pulses we consider are brief in the sense that their durations $T_{s}$ are always shorter than the average oscillation cycle. 

For a dynamical unit that exhibits rhythmic activity, it is also critical to consider the timing of incoming stimuli relative to the phase of the unit's ongoing oscillation. In particular, as detailed in the literature on phase-response curves and phase-resetting \cite{Canavier2015:PhaseResetting,Smeal2010:PhaseResponse}, an external input pulse can have differing effects (e.g. induce a significant phase advance, induce a significant phase delay, or have little influence) depending on the oscillation phase at which it is received. We therefore also study network responses to pulse stimuli as a function of the onset phase $\theta_{\mathrm{on}}$ (i.e., the phase of the directly perturbed area's activity when the external input is first turned on).

\subsubsection{Sustained rhythmic inputs}
\label{s:rhythmic_input_description}

In addition to brief stimulation pulses, we also examine the effects of sustained rhythmic inputs that target the excitatory pool of a single region in the multiarea networks. In particular, we consider the external perturbation signal $S(t)$ to be a sine wave characterized by a fixed amplitude $A_{s}$ and a fixed frequency $f_{s}$ (Fig.~\ref{f:model_setup}C, Bottom). More specifically,

\begin{equation}
S(t) = \left\{\begin{array}{ll}
0 \text{ for } t < t_{s}, \\
A_{s} \sin[2 \pi f_{s} (t - t_{s})] \text{ for } t \geq t_{s} ,
\end{array} \right. \
\label{eq:sine_stim}
\end{equation}
~\\
~\\
where $t_{s}$ is again the time at which the external stimulation begins.

\subsection{Numerical methods}

All numerical simulations and subsequent analyses of the computational model were performed using MATLAB version 2020a. Simulations of the model without noise and without perturbations (e.g. baseline parameter sweeps) made use of built-in ordinary differential equation (ode45) and delay differential equation (dde23) solvers. For the stochastic version of the model and/or when perturbations were incorporated, we instead used custom scripts for numerical simulations. In particular, for the deterministic scenario with local external inputs, we used Euler's method for the numerical integration with a time step $dt = 1\times 10^{-5}$ s. For the scenario with noisy background drive, we used the Euler-Maruyama method for stochastic differential equations with a time step of $dt = 1\times 10^{-5}$ s. In all cases, we assumed a constant history for the initial conditions, but chose a random value in the range [0,1] for the constant history of each unit and subpopulation.

\subsection{Defining phase variables from the population activity}
\label{s:defining_phases}

In this study, we are specifically interested in rhythmic population dynamics and phase-locked states in networks of anatomically connected neural populations. In order to study phase-locking, we first need a means of defining instantaneous phases from the real-valued activity time-series generated from the WC-based network model. That is, given a real-valued oscillatory signal $X(t)$, we need to extract a corresponding phase variable $\theta(t)$. In order to do this, we carry out the following three steps. \textit{(1)} Find the set of times $\{t_{n}^{\mathrm{max}}\}$ for $ n \in \{1,...,N_{\mathrm{max}}\}$ that correspond to each of the $N_{\mathrm{max}}$ local maxima of the time-series $X(t)$. Note that we only consider a peak in the activity time series to be a local maxima if the peak in question has the highest activity value within a time window centered on the peak location and spanning half of an average oscillation cycle in either direction \cite{Palmigiano2017:FlexibleInformation}. The average oscillation period is approximated as the inverse of the frequency corresponding to the maximum of the power spectra, where the spectra is estimated using Welch's method. \textit{(2)} For each maxima $ n \in \{1,...,N_{\mathrm{maxima}}-1\}$, let $\theta(t_{n}^{\mathrm{max}}) = 0$ and let $\theta(t_{n+1}^{\mathrm{max}}) = 2\pi$. Then perform a linear interpolation (using the function `interp1' in MATLAB) between the points $[t_{n}^{\mathrm{max}}, \theta(t_{n}^{\mathrm{max}})]$ and $[t_{n+1}^{\mathrm{max}}, \theta(t_{n+1}^{\mathrm{max}})]$. \textit{(3)} From the linear interpolation, extract the value of the phase $\theta(t)$ at each time point $t$ in the range $[t_{n}^{\mathrm{max}}, t_{n+1}^{\mathrm{max}}]$ (in particular, we sample the phases at the same resolution of the original time-series). Throughout the study, we carry out the above procedure on the excitatory signal $E_{i}(t)$ of each area $i \in \{1,...,N\}$ in order to extract the corresponding phase $\theta_{i}(t)$ that tracks the oscillations.

\subsection{Quantifying interareal phase-locking}
\label{s:PLV}

Once instantaneous phases have been computed for each unit in the network, the extent of phase-coherence between each pair of units is computed using the phase-locking value \cite{Lachaux1999:MeasuringPhase}. In particular, the phase-locking value between units $i$ and $j$ -- which we denote as $\rho_{ij}$ -- is defined as

\begin{equation}
\rho_{ij} = \Bigg | \frac{1}{t_{f} - t_{o}} \sum_{t=t_{o}}^{t_{f}} e^{i[\theta_{i}(t)-\theta_{j}(t)]} \Bigg |,
\label{eq:plv}
\end{equation}
~\\
where $\theta_{i}(t)$ and $\theta_{j}(t)$ are the instantaneous phases of unit $i$ and unit $j$, and where $t_{o}$ and $t_{f}$ correspond to the times at the beginning and end of the time-window over which the phase-locking is evaluated. Intuitively, the phase-locking value measures the consistency of the phase difference $\Delta \theta_{ij} = \theta_{i} - \theta_{j}$ across a continuous segment of time. Moreover, $\rho_{ij}$ $\in [0,1]$, such that if the phase relation between the two units is fixed at a constant value across the window, then $\rho_{ij}$ will be equal to 1; if the phase difference is spread uniformly across the window, then $\rho_{ij}$ will be equal to zero; and if there is partial locking between the two units, then $\rho_{ij}$ will acquire a value between these two extremes.

\section{Results}

\subsection{Characterizing the model's baseline dynamical regimes}

To understand how local perturbations affect collective activity patterns in the model brain circuits, we first characterize the baseline dynamical regimes of the model circuits. For both the 2-area and 4-area networks, we separately consider the case for which the background drive is constant and deterministic versus the case for which the background drive is stochastic.

\subsubsection{Behavior of 2-area circuits: Deterministic scenario}
\label{s:2area_deterministic}

\begin{figure}
	\centering
	\includegraphics[width=\textwidth]{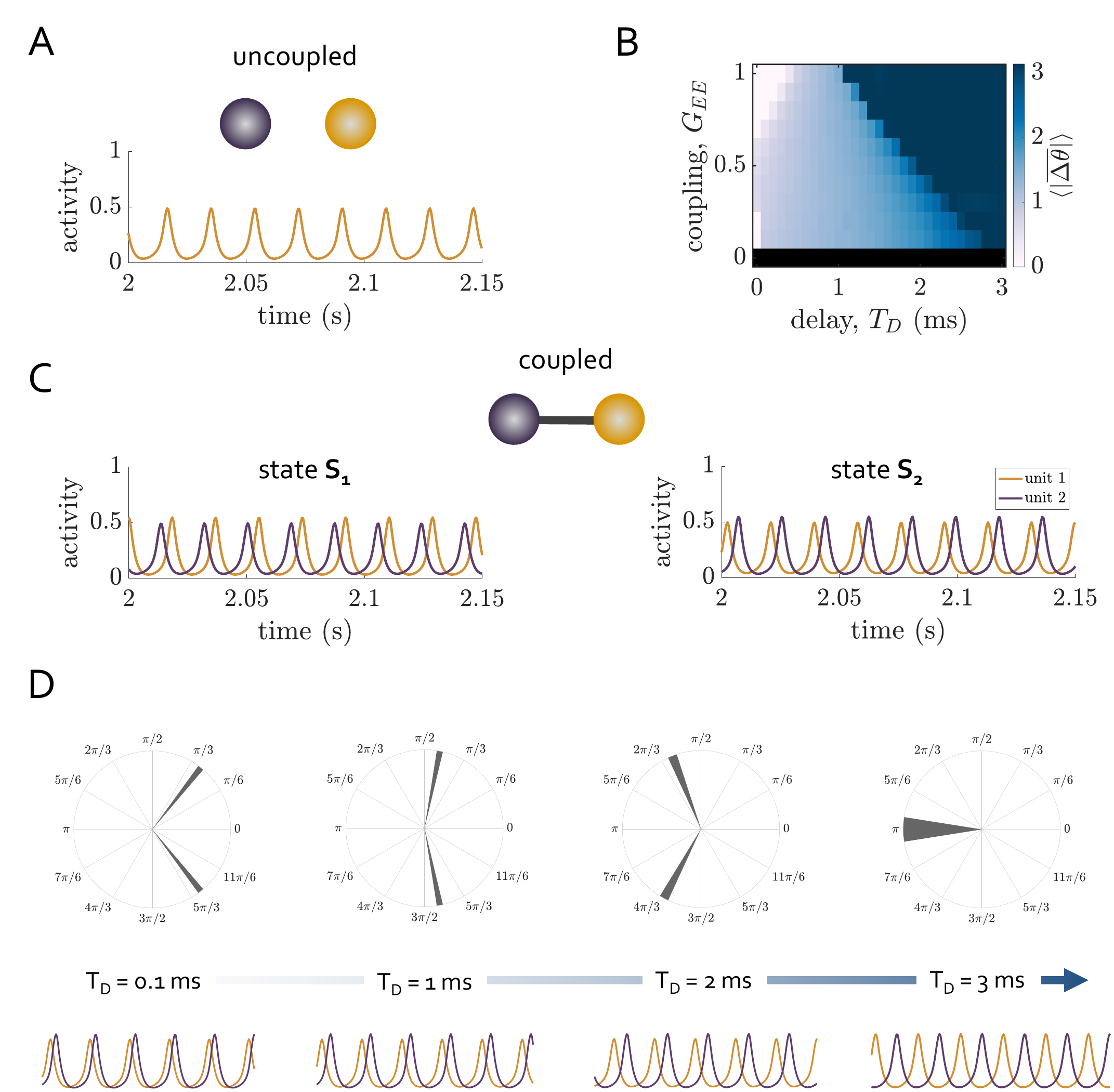}
	\caption[Baseline phase-locking behavior of a 2-area network with deterministic background input.]{\textbf{Baseline phase-locking behavior of a 2-area network with deterministic background input.} \textbf{(A)} We study a 2-area neural mass network, where each unit receives constant, deterministic background drive. The time-series in this first panel shows the oscillatory activity exhibited by an isolated WC unit with $P_{E} = 1.5$. \textbf{(B)} When the two areas are coupled with strength $G_{EE} = 0.2$ and time-delay $T_{D} = 1.5$ ms, different initial conditions give rise to two distinct states $\mathbf{S_{1}}$ and $\mathbf{S_{2}}$. These two asymmetric phase-locking states correspond to a different region taking on the role of the phase-leader. \textbf{(C)} The mean (over initial conditions) of the absolute time-averaged phase difference $\langle | \overline{\Delta \theta} | \rangle $ as a function of the time delay $T_{D}$ and the coupling strength $G_{EE}$.  \textbf{(D)} \textit{Top}: Polar plots of the mean (over initial conditions) of the set of positive and negative time-averaged phase differences for four different values of the delay $T_{D}$ at a fixed $G_{EE} = 0.2$. \textit{Bottom}: Time-series of the two units for the four different values of the delay. For the first three delays, the time-series depict the system in one of its two asymmetric states ($\mathbf{S_{1}}$). For the last delay $T_{D} = 3$ ms, the system is in an anti-phase state where there is no global leader or lagger.}
	\label{f:2node_baseline}
\end{figure}

In this section, we detail the baseline dynamical behaviors of 2-area networks with deterministic background drive. For this analysis, we set $P_{E,i} = 1.5$ for $i \in \{1,2\}$, which corresponds to a situation where each unit would exhibit rhythmic dynamics on its own, even if isolated from the other. An example time-series of an isolated unit is shown in Fig.~\ref{f:2node_baseline}A.

Recall next that, for a network of interacting areas, there are two key parameters that characterize the network's organization. The first is the interareal coupling strength $G_{EE}$, and the second is the signal propagation delay $T_{D}$ between the regions. We first show an example of the network's activity for intermediate $G_{EE} = 0.2$ and $T_{D} = 1.5$ ms in Fig.~\ref{f:2node_baseline}B where the two different panels correspond to initializing the dynamics with two different initial conditions. Upon coupling the two areas, the network locks into one of two out-of-phase configurations, where the lead-lag relationship between the two areas depends on the preparation of the system. Crucially, these asymmetric phase differences arise from an entirely symmetric network. As reported and discussed in several previous studies \cite{Battaglia2007:TemporalDecorrelation,Battaglia2012:DynamicEffective,Kirst2016:Dynamic,Dumont2019:Macroscopic,Witt2013:Controlling,Woodman2011:Effects}, this behavior is an example of multistability arising in the network's collective dynamical state.

We next examine how the network activity varies as a function of the two key network parameters: the coupling strength $G_{EE}$ and the time delay $T_{D}$. Across a range of values for these two quantities, we run 25 different 6-second simulations with random initial conditions, and compute \textit{(1)} the phase-locking value between the two areas over the final 1 second of the simulations, and \textit{(2)} the time-averaged phase difference $\overline{\Delta \theta}$ between the regions over the last 1 second of the simulations. Fig.~\ref{f:2node_baseline}C shows the absolute value of the phase difference $\langle | \overline{\Delta \theta} | \rangle$ between the two units (averaged across initial conditions) as a function of the coupling $G_{EE}$ and delay $T_{D}$. Black areas indicate parameter combinations where the PLV -- averaged across initial conditions -- does not exceed a high threshold of 0.95. 

For fixed coupling strength, increasing the delay has the effect of increasing the absolute phase difference between the two areas' activities until the symmetric anti-phase configuration $\Delta \theta = \pi$ is reached. This effect is not surprising; indeed, other studies have examined the effects of time delays in coupled oscillator systems and have reported similar findings \cite{Dumont2019:Macroscopic}. Additionally, increasing the coupling can also affect the phase difference and push the system into the antiphase state more rapidly as a function of the delay. Fig.~\ref{f:2node_baseline}D shows more explicitly the evolution of the collective dynamics as a function of $T_{D}$ for a fixed $G_{EE} = 0.2$. In sum, similar to other models \cite{Battaglia2007:TemporalDecorrelation,Battaglia2012:DynamicEffective,Palmigiano2017:FlexibleInformation,Witt2013:Controlling,Dumont2019:Macroscopic}, we find here that in a 2-area WC network with delays, there is a range of parameters where the system exhibits bistable phase-locking patterns.

\subsubsection{Behavior of 2-area circuits: Stochastic scenario}
\label{s:2area_baseline_stochastic}

\begin{figure}[h!]
	\centering
	\includegraphics[width=\textwidth]{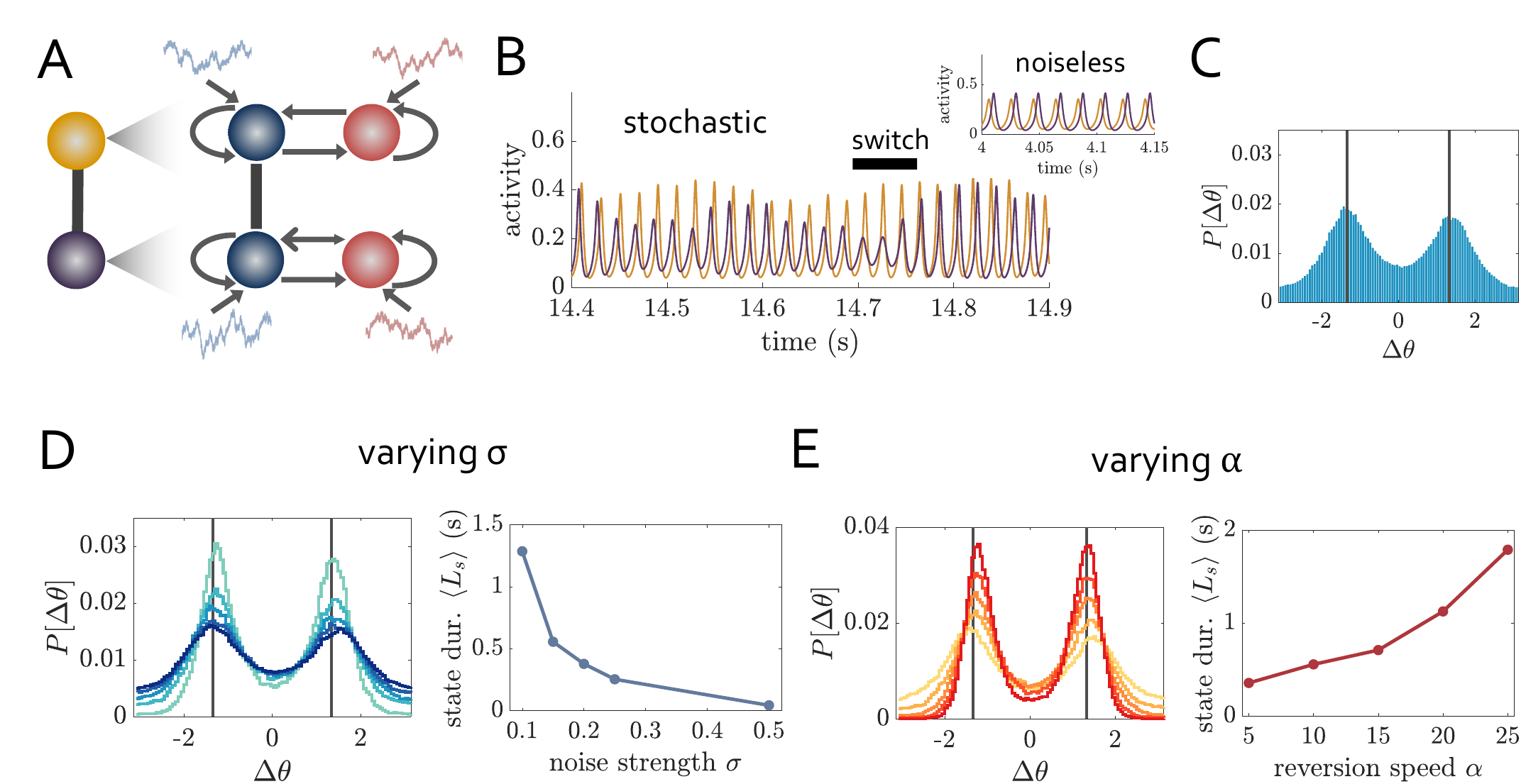}
	\caption[Baseline phase-locking behavior of a 2-area network with stochastic background input.]{\textbf{Baseline phase-locking behavior of a 2-area network with stochastic background input.} \textbf{(A)} We study a 2-area neural mass network where each unit receives an independent stochastic background drive. All panels in this figure correspond to fixed $\overline{P_{E}} = 1.35$, $\overline{P_{I}} = 0$, $G_{EE} = 0.2$, $T_{D} = 1.5$ ms. \textbf{(B)} A segment of the network activity for O.U. parameters $\alpha = 10$ and $\sigma = 0.2$. Note the switch in the lead-lag relationship at $t \approx 14.75$s. \textit{Inset:} The network activity in the deterministic case. \textbf{(B)} For $\alpha = 10$ and $\sigma = 0.2$, the distribution of phase differences between the two coupled areas across a long simulation \textbf{(D)} \textit{Left:} For $\alpha=10$, the distribution of phase differences between the two coupled areas for varying $\sigma$ ($\sigma = \{0.1,0.15,0.2,0.25,0.5\}$, increasing as the curves go from light to dark). \textit{Right:} The mean state duration $\langle L_{s} \rangle$ as a function of the noise strength $\sigma$. \textbf{(E)} \textit{Left:} For $\sigma=0.15$, the distribution of phase differences between the two coupled areas for varying $\alpha$ ($\alpha =\{5,10,15,20,25\}$, increasing as the curves go from light to dark). \textit{Right:} The mean state duration $\langle L_{s} \rangle$ as a function of the rate of mean reversion $\alpha$. \textbf{Note:} In all of the $\Delta \theta$ distributions, the gray vertical lines correspond to the stable phase differences that would exist in the absence of noise.}
	\label{f:2node_baseline_stochastic}
\end{figure}

We turn next to the stochastic 2-area networks, wherein each unit receives fluctuating background drives generated from independent OU processes with rate of mean reversion $\alpha$ and noise strength $\sigma$ (see Eqs.~\ref{eq:OUprocess_E} and ~\ref{eq:OUprocess_I}). A schematic of the model setup for this scenario is shown in Fig.~\ref{f:2node_baseline_stochastic}A. In order to obtain enough statistics to properly characterize system behavior, the results presented in the following exposition are based on very long simulations of 22 minutes in length.

To begin, we consider parameters such that an isolated area with deterministic dynamics intrinsically oscillates, and such that when two areas are coupled, the circuit exhibits out-of-phase locking (see Fig.~\ref{f:2node_baseline_stochastic}B, inset). In this case, stochastic background inputs to the network induce fluctuations in the oscillation amplitude of each unit, and also cause some temporal irregularity of the oscillation period (Fig.~\ref{f:2node_baseline_stochastic}B). However, despite the temporal variation in each areas' activity, periods of approximate phase-locking can still be maintained due to the interareal coupling. In particular, the network exhibits episodes in which one of the two lead-lag relations is maintained for some finite duration, and these epochs are then separated by periods of asynchrony after which the system may re-enter the same lead-lag configuration or spontaneously transition into the other state. 

The effects on phase-locking are most easily appreciated by examining the distribution $P(\Delta \theta)$ of the instantaneous phase difference $\Delta\theta$ between the two units across a very long simulation. As shown in Fig.~\ref{f:2node_baseline_stochastic}C),  $P(\Delta \theta)$ has two clear out-of-phase modes of approximately equal height, with some spread around each peak. The modes are the favored configurations of the emergent dynamics, and are signatures of the bistable collective states that exist in the noiseless system. Before continuing, it is critical to point out that the behaviors found here are not surprising, and have been reported in previous studies that used detailed, biophysical spiking neuron models \cite{Palmigiano2017:FlexibleInformation,Witt2013:Controlling}. These past investigations significantly inspired and laid out key insights for the present work.

The previous observations were based on fixing the two key parameters of the stochastic inputs -- the noise strengh $\sigma$ and the rate of mean reversion $\alpha$ -- to intermediate values. However, it is also important to consider how these parameters affect the phase-difference distribution and the length of time that the network tends to dwell in ones of its two states (i.e., $\Delta \theta > 0$ or $\Delta \theta < 0$) before undergoing a transition. To start, Fig.~\ref{f:2node_baseline_stochastic}D shows the effect of varying $\sigma$ on $P(\Delta \theta)$ (left) and the average state duration $\langle L_{s} \rangle$ (right). In agreement with intuition, increasing $\sigma$ broadens $P(\Delta \theta)$ and decreases the peak heights, and also decreases the average state duration. The effects of varying $\alpha$ are also straightforward. As the rate of mean reversion increases, the phase difference distribution $P(\Delta \theta)$ becomes less broad and more sharply peaked (Fig.~\ref{f:2node_baseline_stochastic}E, left), and the average state duration increases (Fig.~\ref{f:2node_baseline_stochastic}E, right). The intuition for these behaviors is that, the faster the reversion, the less likely it is for the background drive to drift significantly away from its average value and cause the phase relation to shift.

\begin{FPfigure}
	\centering
	\includegraphics[width=\textwidth]{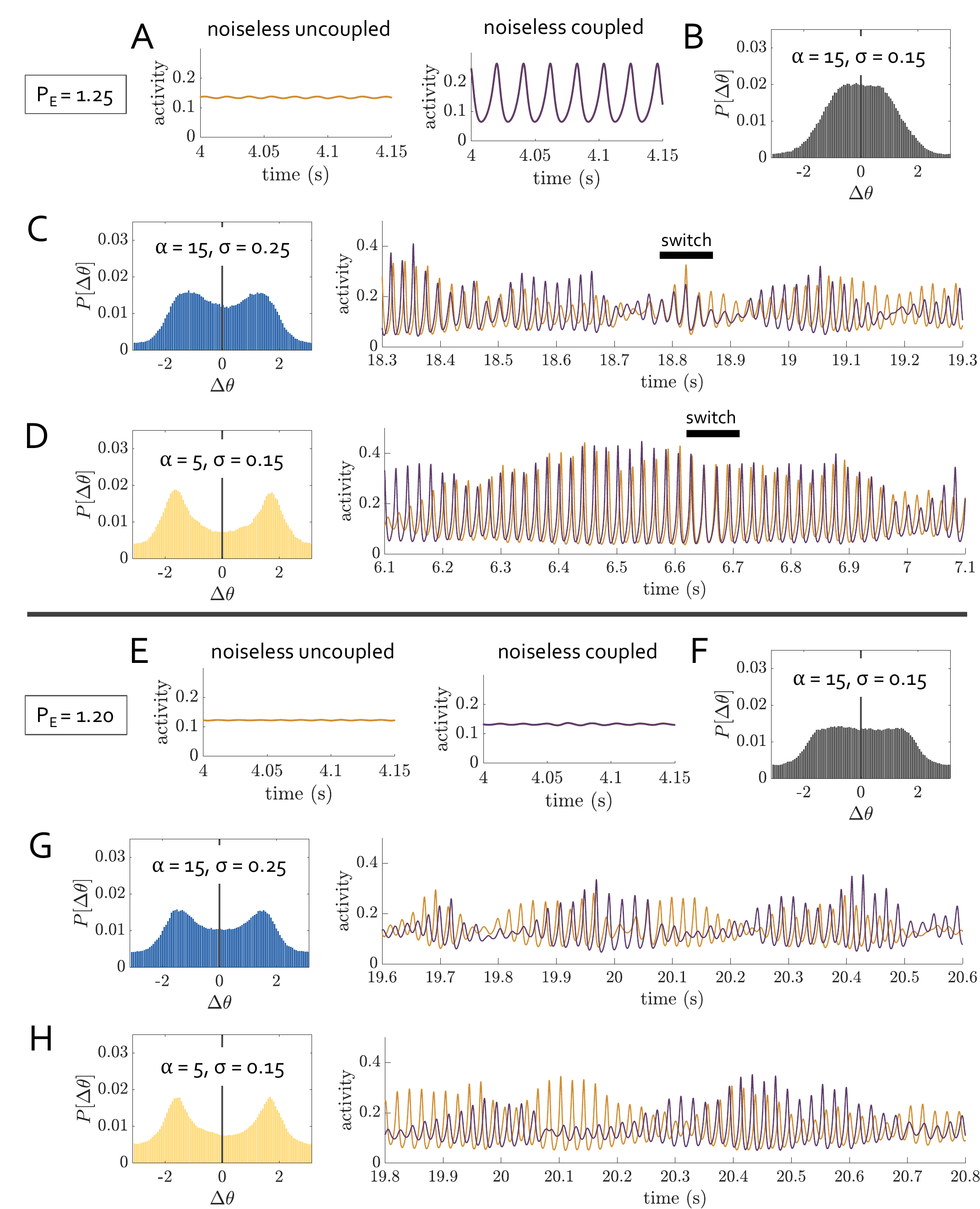}
	\caption[Baseline phase-locking behavior of a 2-area network with stochastic background input in the low drive regime.]{\textbf{Baseline phase-locking behavior of a 2-area network with stochastic background input in the low drive regime.} \textbf{(A)} Here we study a 2-area neural mass network where each unit receives an independent stochastic background drive with $\overline{P_{E}} = 1.25$ and $\overline{P_{I}} = 0$, and where $G_{EE} = 0.3$ and $T_{D} = 1.5$ ms. For context, we first show the time-series of an uncoupled, deterministic WC unit (left) and the time-series of two coupled, deterministic WC units (right). \textbf{(B)} The distribution of phase differences between the two coupled areas across a long simulation for $\alpha = 15$ and $\sigma = 0.15$. \textbf{(C)} \textit{Left:} Activity time-courses of the two areas. \textit{Right:} The distribution of phase differences between the two coupled areas for $\alpha = 15$ and $\sigma = 0.25$.  \textbf{(D)} \textit{Left:} Activity time-courses of the two areas. \textit{Right:} The distribution of phase differences between the two coupled areas for $\alpha = 5$ and $\sigma = 0.15$. \textbf{(E)} Here we study a 2-area neural mass network where each unit receives an independent stochastic background drive with $\overline{P_{E}} = 1.20$ and $\overline{P_{I}} = 0$, and where $G_{EE} = 0.3$ and $T_{D} = 1.5$ ms. For context, we first show the time-series of an uncoupled, deterministic WC unit (left) and the time-series of two coupled, deterministic WC units (right). \textbf{(F)} The distribution of phase differences between the two coupled areas across a long simulation for $\alpha = 15$ and $\sigma = 0.15$. \textbf{(G)} \textit{Left:} Activity time-courses of the two areas. \textit{Right:} The distribution of phase differences between the two coupled areas for $\alpha = 15$ and $\sigma = 0.25$. \textbf{(H)} \textit{Left:} Activity time-courses of the two areas. \textit{Right:} The distribution of phase differences between the two coupled areas for $\alpha = 5$ and $\sigma = 0.15$. \textbf{Note:} In all of the $\Delta \theta$ distributions, the gray vertical lines correspond to the stable phase differences for the selected parameters, but in the absence of noise.}
	\label{f:2node_baseline_stochastic_lowDrive}
\end{FPfigure}

A second import consideraton is how noisy background inputs affect the network dynamics for quantitatively different baseline dynamical regimes of the model. For example, of interest is what happens for lower values of the drive $\overline{P_{E}}$ that place the system at working points close to the onset of oscillatory activity in a noiseless WC unit. For example, in contrast to the previous working point, if $\overline{P_{E,j}} = 1.25$ for $j\in \{1,2\}$ a deterministic, isolated WC exhibits oscillations of almost vanishing amplitude (Fig.~\ref{f:2node_baseline_stochastic_lowDrive}A, Left). When two such areas are then coupled with $G_{EE} = 0.3$ and $T_{D} = 1.5$, the oscillation amplitude in each area increases significantly, but rather than out-of-phase locking, the two regions lock in-phase (Fig.~\ref{f:2node_baseline_stochastic_lowDrive}A, Right). Consequently, there is no collective multistability in the network; only the single in-phase state occurs. Moreover, when stochastic background drive with relatively low $\sigma = 0.15$ and high $\alpha = 15$ is incorporated into the network, the distribution of phase differences exhibits a broad peak centered around $\Delta \theta = 0$ (Fig.~\ref{f:2node_baseline_stochastic_lowDrive}B).

Importantly, though, the general form of the phase difference distribution in this case depends on the noise parameters. For example, when $\sigma$ is increased (Fig.~\ref{f:2node_baseline_stochastic_lowDrive}C) or when $\alpha$ is decreased (Fig.~\ref{f:2node_baseline_stochastic_lowDrive}D) the symmetry around $\Delta \theta = 0$ is broken, and two clear peaks emerge in the distributions of $\Delta \theta$ that correspond to the emergence of favored out-of-phase configurations. However, the out-of-phase states present at this working point arise for different reasons than their appearance at the $\overline{P_{E}} = 1.35$ working point. At the higher drive working point, the network exhibited out-of-phase locking in the absence of noise, and the peaks in $P_{\Delta \theta}$ were signatures of those underlying states. In contrast, the peaks for the lower-drive working point considered here are not simply "blurred" versions of the states that exist in the deterministic scenario. Rather, the out-of-phase configurations are themselves induced by the stochastic fluctuations in the level of the input. 

To conclude this section, we consider a third working point where the background drive is even slightly smaller and set to $\overline{P_{E,j}} = 1.20$ for $j\in \{1,2\}$ (Fig.~\ref{f:2node_baseline_stochastic_lowDrive}E, Left). In contrast to the previous working point, coupling two areas with $G_{EE} = 0.3$ and $T_{D} = 1.5$ yields only very small amplitude oscillations when the background input is deterministic (Fig.~\ref{f:2node_baseline_stochastic_lowDrive}E, Right). Thus, in this regime, the oscillations themselves will be mainly noise-driven. Starting again with relatively low $\sigma = 0.15$ and relatively high $\alpha = 15$ for the background noise, we observe that the phase difference distribution is very flat and unstructured (Fig.~\ref{f:2node_baseline_stochastic_lowDrive}F). But, similar to before, increasing $\sigma$ (Fig.~\ref{f:2node_baseline_stochastic_lowDrive}G) or decreasing $\alpha$ (Fig.~\ref{f:2node_baseline_stochastic_lowDrive}H) brings about the two peaks in $P(\Delta \theta)$. 

One reason for the aforementioned behaviors could be that the input fluctuations transiently push the system into a dynamical regime where -- in a noiseless network subject to the same instantaneous levels of background drive -- multistable out-of-phase locking would arise as it does for the higher drive working point. Alternatively, it could also be the case that transient differences in the levels of instantaneous background drive to each area translate into dynamical asymmetries where one area briefly becomes the leader or lagger. Note that in both of these scenarios, the input needs to fluctuate significantly away from its mean value in order to induce the observed behavior, which provides some intuition for why $\sigma$ ($\alpha$) needs to be high (low) enough to induce the effects. In Sec.~\ref{s:2area_stochastic_pulse}, we will examine how the ability to control phase-locking with brief perturbations depends on the baseline working point and thus the nature of the observed phase-locking.

\subsubsection{Behavior of 4-area circuits: Deterministic scenario}
\label{s:4area_deterministic}

\begin{figure}[h!]
	\centering
	\includegraphics[width=\textwidth]{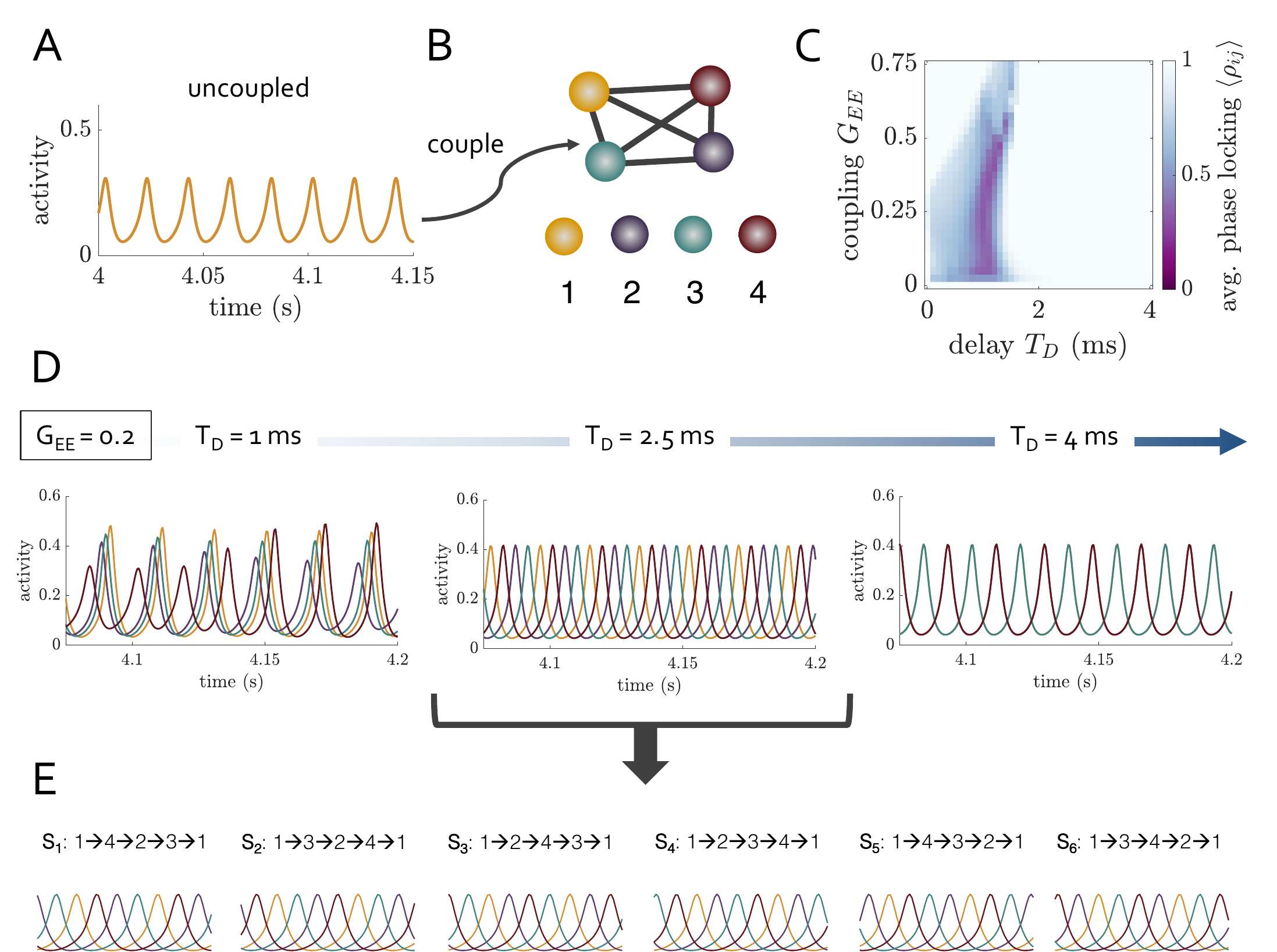}
	\caption[Baseline phase-locking behavior of a 4-area network with deterministic background input.]{\textbf{Baseline phase-locking behavior of a 4-area network with deterministic background input.} \textbf{(A)} Oscillatory activity exhibited by an isolated WC unit subject to deterministic background input with $P_{E} = 1.325$. \textbf{(B)} We study a 4-area neural mass network composed of four units identical to that shown in panel \textbf{(A)} and coupled in an all-to-all fashion. For convenience, we label each of the four areas (depicted in the different colors) ``1", ``2", ``3", and ``4". \textbf{(C)} The average phase-locking value $\langle \rho_{ij} \rangle$ of the network as a function of the time-delay $T_{D}$ and the coupling strength $G_{EE}$. The average was taken across all pairs of units and across 100 simulations with different initial conditions. \textbf{(D)} We consider system behavior for a fixed coupling $G_{EE} = 0.2$ and for three different values of the delay $T_{D}$. For each delay, a segment of the regions' activity time-series from one instantiation of the dynamics is shown. \textbf{(E)} For a delay $T_{D} = 2.5$ ms, there exist six different multistable collective states $\{\mathbf{S_{1}},...,\mathbf{S_{6}}\}$, each characterized by a distinct temporal ordering of the activity peaks (or phases) of different units. We write the phase-ordering and show time-series segments for each state.}
	\label{f:4node_baseline}
\end{figure}

In this section, we consider a slightly more complicated anatomical network composed of four interacting neural populations coupled in an all-to-all topology (Fig.~\ref{f:4node_baseline}B). As for the 2-node system, we first examine the behavior of the 4-area circuit in the deterministic scenario where each excitatory subpopulation receives a constant background drive of strength $P_{E}$. In particular, we consider the case $P_{E,i} = 1.325$ for all $i\in\{1,...4\}$, for which an isolated WC unit with the default parameters exhibits rhythmic population activity (Fig.~\ref{f:4node_baseline}A). 

We begin by investigating whether or not the network phase-locks as a function of the two main network properties: the coupling strength $G_{EE}$ and the time-delay $T_{D}$. For each parameter combination, we run 100, 7-second simulations (where each simulation uses a different set of random initial conditions). For each simulation and each pair of regions, we then compute the PLV $\rho_{ij}$ (see Sec.~\ref{s:defining_phases},~\ref{s:PLV}) using the last 2 seconds of the activity time-series. Finally, we calculate an average phase-locking value $\langle \rho_{ij} \rangle$ by computing the mean across all region pairs and across the 100 simulations. We show $\langle \rho_{ij} \rangle$ as a function of $G_{EE}$ and $T_{D}$ in Fig.~\ref{f:4node_baseline}C, which indicates that different portions of parameter space give rise to different behaviors in terms of the extent of phase-locking. In particular, we find that $\langle \rho_{ij} \rangle \approx 1$ for low delay and high coupling and also for intermediate-to-high delays at all couplings considered. Interestingly, though, there is a regime in between where -- for a range of low-to-intermediate delays -- the average phase-locking clearly dips well below 1. This decrease in $\langle \rho_{ij} \rangle $ indicates that for the corresponding portion of parameter space, there is no fixed phase relation between the different areas' activity and the network is not phase-locked.

To further unpack the nature of the system's collective dynamics, we focus on a fixed coupling strength $G_{EE} = 0.2$ and study network activity for different values of the delay $T_{D}$ (Fig.~\ref{f:4node_baseline}D). As indicated by a phase-locking value that is less than one, when the delay is relatively low (e.g. 1 ms), the activity time-series reveal phase differences that shift across time (Fig.~\ref{f:4node_baseline}D, Left). If we instead consider the opposite end of the spectrum and examine a large delay (e.g. $T_{D} = 4$ ms), the collective dynamics of the network change (Fig.~\ref{f:4node_baseline}D, Right). Here, we observe the emergence of phase-locking between the regional oscillations in the form of a 2-cluster anti-phase state where a given pair of nodes is in-phase with one another and anti-phase with the other pair. The bottom panel of Fig.~\ref{f:4node_baseline}D, Right shows an example where unit 1 and unit 3 are in-phase, and collectively out-of-phase with areas 2 and 4 (which are themselves in-phase). Importantly, though, this is a multistable regime in that a different set of initial conditions could lead to different pairs of units being in-phase with one another. This type of collective state has been observed in other types of coupled neural oscillator systems as well, e.g. \cite{Achuthan2009:PhaseResetting}. At an intermediate delay (e.g. $T_{D} = 2.5$ ms), another dynamical regime is observed where the network is phase-locked, but with a different pattern of phase relationships (Fig.~\ref{f:4node_baseline}D, Middle). In particular, here the system settles into a state where the regional activities peak one after another sequentially in time, and such that the spacing is equal between consecutive pairs. An example activity time-course is shown for one such collective state characterized by the peak-ordering $1 \rightarrow 4 \rightarrow 2 \rightarrow 3 \rightarrow 1$. For this configuration, $\Delta \theta_{1,2} = \pi$, $\Delta \theta_{1,3} = -\pi/2$ and $\Delta \theta_{1,4} = \pi/2$. Crucially, though, the aforementioned pattern of phases again corresponds to one of only a larger set of collective states that are all multistable (Fig.~\ref{f:4node_baseline}E).

Although the model brain circuit exhibits collective dynamical multistability for both the working points in Fig.~\ref{f:4node_baseline}D, Middle and in Fig.~\ref{f:4node_baseline}D, Right, we focus our subsequent analyses on the former scenario, for which the out-of-phase relations may interestingly support directed functional interactions from a nonetheless undirected anatomical coupling structure \cite{Battaglia2012:DynamicEffective,Palmigiano2017:FlexibleInformation,Kirst2016:Dynamic}). For the case of intermediate delay $T_{D} = 2.5$,  we can define six unique stable states of the network $\{\mathbf{S_{1}},...,\mathbf{S_{6}}\}$, where each state corresponds to a distinct temporal ordering of the regions' phases (Fig.~\ref{f:4node_baseline}E). For example, in state $\mathbf{S_{1}}$, the phases arrange in the repeating pattern $1 \rightarrow 4 \rightarrow 2 \rightarrow 3 \rightarrow 1$, whereas in state $\mathbf{S_{2}}$ the configuration is instead  $1 \rightarrow 3 \rightarrow 2 \rightarrow 4 \rightarrow 1$. Working under the assumptions that the phase-locking of neural rhythms establishes a substrate for interareal communication, and that different lead-lag relationships enable different communication channels, each of the six emergent multistable states here could potentially allow for entirely different global information routing capabilities \cite{Battaglia2012:DynamicEffective,Palmigiano2017:FlexibleInformation,Kirst2016:Dynamic}). It is this hypothesis that motivates our examination of how such multistability can be controlled via simple, targeted dynamical perturbations to the network.

\subsubsection{Behavior of 4-area circuits: Stochastic scenario}
\label{s:4area_stochastic_baseline}

\begin{FPfigure}
	\centering
	\includegraphics[width=\textwidth]{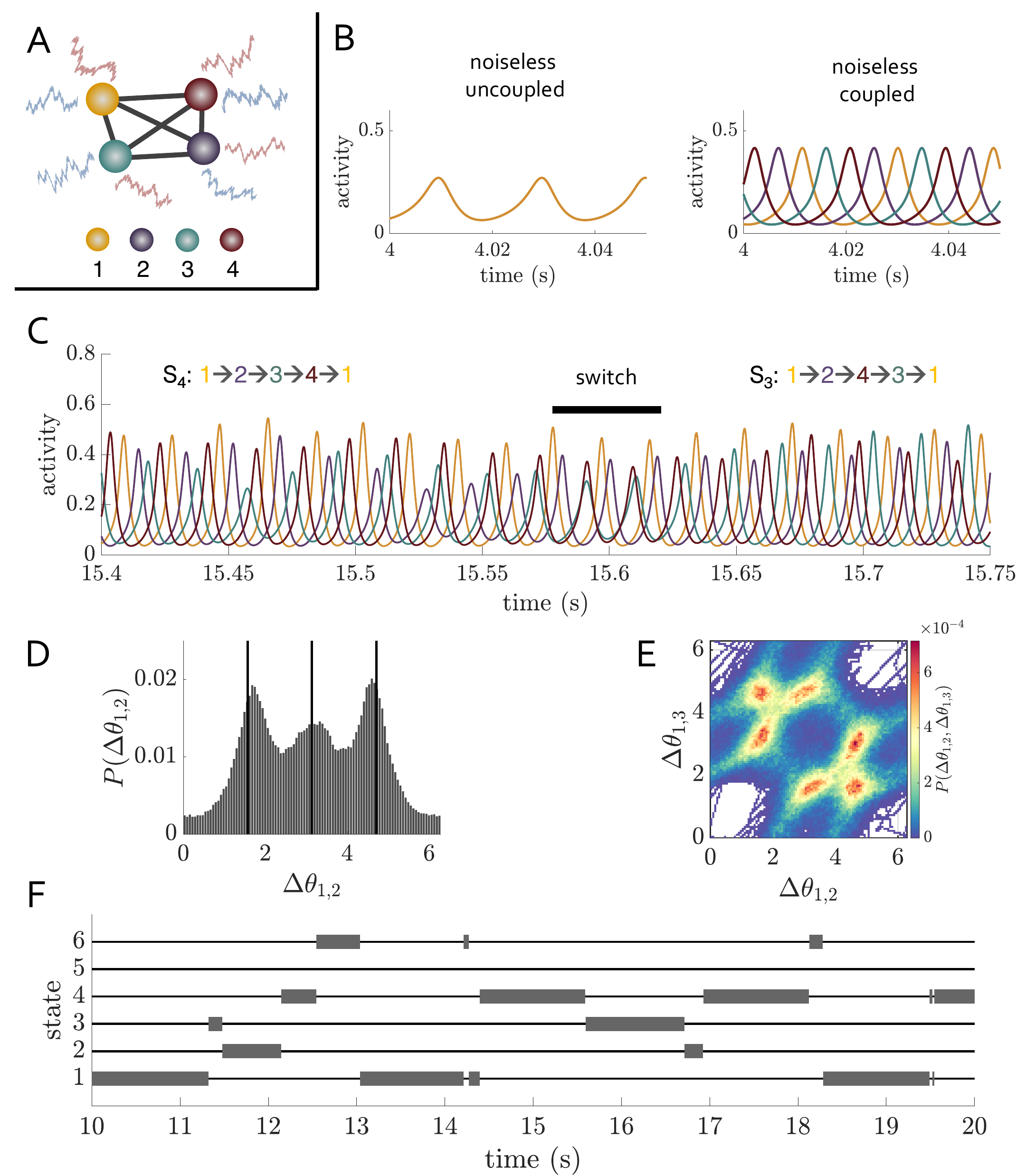}
	\caption[Baseline phase-locking behavior of a 4-area network with stochastic background inputs.]{\textbf{Baseline phase-locking behavior of a 4-area network with stochastic background inputs.} \textbf{(A)} We study a 4-area neural mass network, where the excitatory and inhibitory subpopulations within each area receive independent stochastic background drives generated from an OU process (blue and red inputs in the schematic). For this analysis, we set $\overline{P_{E}} = 1.3$, $\overline{P_{I}} = 0$, $G_{EE} = 0.25$, and $T_{D} = 2.5$ ms. When we make the background inputs stochastic, we use parameters $\alpha = 10$ and $\sigma = 0.2$. \textbf{(B)} \textit{Left:} Time-series of an uncoupled, deterministic WC unit. \textit{Right:} Example time-series of four all-to-all coupled deterministic WC units, where in this case, the system is in state $\mathbf{S_{6}}$ characterized by the ordering $1 \rightarrow 3 \rightarrow 4 \rightarrow 2 \rightarrow 1$. \textbf{(C)} A snapshot of the network's activity when the neural populations are driven by a noisy environment. Note the spontaneous switch from collective state $\mathbf{S_4}$ to collective state $\mathbf{S_3}$ at $t\approx 15.6$ seconds. \textbf{(D)} The distribution $P(\Delta \theta_{1,2})$ of the pairwise phase differences between area 1 and area 2 across a long simulation of the system. \textbf{(E)} The joint distribution $P(\Delta \theta_{1,2}, \Delta \theta_{1,3})$ of $\Delta \theta_{1,2}$ and $\Delta \theta_{1,3}$ across a long simulation. \textbf{(F)} A 10-second-long segment of the system that shows which of the six collective states is occupied (indicated by the thick gray bars) as a function of time.}
	\label{f:4node_baseline_stochastic}
\end{FPfigure}

We conclude our analysis of the model's baseline dynamical regimes by studying how stochastically fluctuating background inputs affect collective activity patterns in the 4-area network (Fig.~\ref{f:4node_baseline_stochastic}A). In particular, we examine a scenario in which an isolated, deterministic WC unit intrinsically oscillates, and when the four regions are coupled, the network phase-locks into one of the six multistable states discussed in the previous section (Fig.~\ref{f:4node_baseline_stochastic}B). We acheive this by setting the input to the excitatory subpopulations at $\overline{P_{E,j}} = 1.3$ for $j \in \{1,...,4\}$, the input to the inhibitory subpopulations at $\overline{P_{I,j}} = 0$ for $j \in \{1,...,4\}$, the network coupling strength to $G_{EE} = 0.25$, and the interareal time delay to $T_{D} = 2.5$ ms. As for the 2-area networks, noise is incorporated by letting the inputs to the E and I subpopulations of each region -- $P_{E,j}(t)$ and $P_{I,j}(t)$ -- evolve according to independent OU processes. In what follows, we fix the rate of mean reversion to $\alpha = 10$ and the noise strength to $\sigma = 0.2$. (We will briefly study the consequences of changing these two quantities at the end of this section). Because a large amount of data is required to accurately map out the system's dynamics in the stochastic regime, the following analyses are based on a very long simulation of 22 minutes in length.

With the chosen parameters, the regional activity becomes slightly more irregular over time, exhibiting both variation in oscillation amplitude and instantaneous frequency (Fig.~\ref{f:4node_baseline_stochastic}C). In addition, while approximate phase-locking is still observed within certain time-segments, the relative phase relations between consecutive units become less rigid and fluctuate with some spread around the set of values $\Delta \theta \in \{-\pi/2, \pi, \pi/2\}$ observed in the noiseless case. Nonetheless, it is still perfectly possible to define looser versions of six collective states that emerge in the noiseless limit, and to track the temporal evolution of these patterns. For example, a simple way of defining the network state as a function of time is to just consider the instantaneous ordering of the phases (or activity peaks) of each unit. 

Applying this simple state-extraction method to the activity time-series illustrated in Fig.~\ref{f:4node_baseline_stochastic}C, we observe an example of spontaneous state-switching. At the beginning of the depicted time-window, the network is in a noisy version of state $\mathbf{S_{4}}$ characterized by the phase ordering $1\rightarrow 2 \rightarrow 3 \rightarrow 4 \rightarrow 1$. But at $t \approx 15.6$ seconds, the green and red areas exchange places, inducing a change in the collective activity pattern to a noisy version of state $\mathbf{S_{3}}$ characterized by the new ordering $1\rightarrow 2 \rightarrow 4 \rightarrow 3 \rightarrow 1$. In this way, the stochastically fluctuating environment that drives the network can induce random transitions in the system's \textit{collective} phase-locking pattern \cite{Palmigiano2017:FlexibleInformation, Witt2013:Controlling}.

The fact that the network undergoes random transitions between noisy versions of the collective states that exist in the absence of noise can perhaps be better recognized by examining distributions of the instantaneous phase relations across a very long simulation. We focus first on the distribution $P(\Delta \theta_{1,2})$ between a given pair of units 1 and 2, which exhibits three clear peaks located near the set of phase differences that arise in the deterministic limit  (Fig.~\ref{f:4node_baseline_stochastic}D). The presence of these modes indicates that even with noise, there are still preferred dynamical configurations that the system moves between as it evolves. It is even more enlightening to estimate the joint distribution $P(\Delta \theta_{1,2}, \Delta \theta_{1,3})$ of $\Delta \theta_{1,2}$ and $\Delta \theta_{1,3}$. Doing so reveals six ``hotspots" of particularly high likelihood (Fig.~\ref{f:4node_baseline_stochastic}E). Moreover, these frequently visited patterns correspond to both of the phase relations simultaneously occupying values close to what they would be if the system were in one of the six phase-locked patterns from the deterministic simulations. The hotspots in the joint distribution are thus signatures of the multistable attractors present in the noiseless setting. In the presence of stochastic inputs, these states become blurred but still discernible versions of their noiseless selves, and they become metstable, with the network spontaneously transitioning into and out of them.

Finally, we show in Fig.~\ref{f:4node_baseline_stochastic}F the state occupation as a function of time across a relatively long 10-second window. For this analysis, we use the simple state-extraction method described above based on the ordering of the units' phases. This figure illustrates the evolution of the network's collective activity patterns and the spontaneous transitions between configurations. Note that for the chosen parameter set, the system often dwells in a particular state for on the order of $\sim$1 second. Hence, although the network does switch between different states, this switching occurs on a relatively slow time-scale relative to the oscillation period.

\begin{figure}
	\centering
	\includegraphics[width=\textwidth]{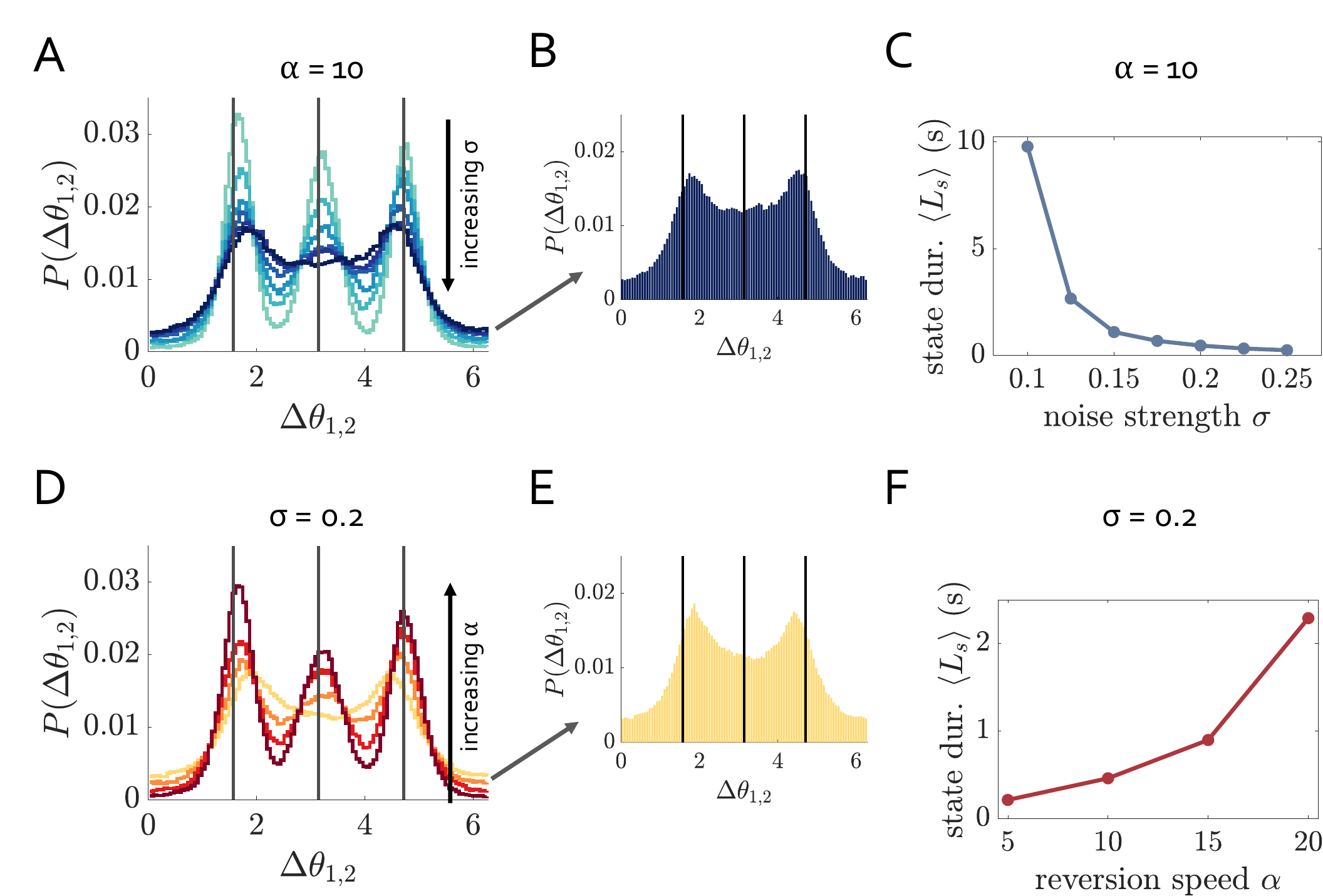}
	\caption[Effect of varying the noise parameters on the dynamics of a 4-area network with stochastic background inputs.]{\textbf{Effect of varying the noise parameters on the dynamics of a 4-area network with stochastic background inputs.} For this analysis, we set $\overline{P_{E}} = 1.3$, $\overline{P_{I}} = 0$, $G_{EE} = 0.25$, and $T_{D} = 2.5$ ms. \textbf{(A)} For $\alpha=10$, the distribution of phase differences $P(\Delta \theta_{1,2})$ between area 1 and area 2 for varying $\sigma$ ($\sigma = \{0.125,0.15,0.175,0.2, 0.225, 0.25\}$, increasing as the curves go from light to dark). \textbf{(B)} Close-up of $P(\Delta \theta_{1,2})$ for $\alpha = 10$ and $\sigma = 0.25$. \textbf{(C)} The mean state duration $\langle L_{s} \rangle$ as a function of the noise strength $\sigma$. \textbf{(D)} For $\sigma=0.2$, the distribution of phase differences $P(\Delta \theta_{1,2})$ between area 1 and area 2 for varying $\alpha$ ($\alpha = \{5,10,15,20\}$, increasing as the curves go from light to dark). \textbf{(E)} Close-up of $P(\Delta \theta_{1,2})$ for $\sigma= 0.2$ and $\alpha = 5$. \textbf{(F)} The mean state duration $\langle L_{s} \rangle$ as a function of the rate of mean reversion $\alpha$.}
	\label{f:4node_baseline_stochastic2}
\end{figure}

We conclude this section by briefly studying the effects of the noise parameters $\alpha$ and $\sigma$. As for the 2-area network, increasing the noise strength $\sigma$ (Fig.~\ref{f:4node_baseline_stochastic2}A) or decreasing the rate of mean reversion $\alpha$ (Fig.~\ref{f:4node_baseline_stochastic2}D) both have the effect of reducing the height and broadening the width of the peaks in the phase difference distribution $P(\Delta \theta_{1,2})$ (note that the conclusions are similar for any pair of units $i \neq j$). Hence, these two parameters control the tightness of phase-locking in the network and the ``blurriness" of the collective states relative to the deterministic limit. Importantly, if $\sigma$ is too large (Fig.~\ref{f:4node_baseline_stochastic2}B) or if $\alpha$ is too small (Fig.~\ref{f:4node_baseline_stochastic2}E), then the noise itself destroys the dynamical behavior of interest. In particular, if the network is overly noisy, then it is not possible to discern three clear peaks in the phase difference distributions (for both large $\sigma$ and low $\alpha$, the original mode at $\Delta\theta_{1,2} \approx \pi$ is washed out). This, in turn, indicates the degradation of the phase-locking patterns that we wish to analyze. Going forward, we always work in a parameter regime where the stochastic network exhibits clear signatures of the collective states observed in the noiseless scenario.

We also examine the average duration of collective states as a function of the noise parameters $\alpha$ and $\sigma$. Accordingly, we extract the temporal evolution of the network's state across a long simulation using the naive method that considers only the relative ordering of the units' phases at a given moment. From this state ``vector", we calculate a set of state durations $\{L_{s}\}$ by computing the lengths of time for which the network uninterruptedly remains in the same state before switching to a new configuration. The mean state duration $\langle L_{s} \rangle$ is then calculated as the average of this set. Similar to the 2-area circuit, we find here that the mean duration $\langle L_{s} \rangle$ decreases with increasing $\sigma$ (Fig.~\ref{f:4node_baseline_stochastic2}C) or decreasing $\alpha$ (Fig.~\ref{f:4node_baseline_stochastic2}F).

\subsection{Investigating the effects of local perturbations on collective dynamical states}

In the previous sections, we mapped out the behavior of 2-area and 4-area model brain circuits under baseline conditions, focusing on parameter regimes yielding multistable phase-locking. We now turn to an investigation of how those collective states respond to different local perturbations. This analysis is motivated by the question of how a circuit's functional state -- determined by its phase-locking pattern -- can be rapidly controlled via modulatory inputs, bypassing the need for potentially slow and/or costly structural modifications. To that end, the rest of the results section is organized as follows. In Sec.~\ref{s:pulsed_stim}, we study how the 2- and 4-area networks respond to brief input pulses applied to one area, and we consider both the deterministic and stochastic situations. In Sec.~\ref{s:AC_stim}, we then examine how both circuits respond to rhythmic inputs applied to a single region, and we again separately study both the deterministic and stochastic settings. Finally, in Sec.~\ref{s:parameter_variation}, we consider the robustness of various results to changes in the baseline model parameters.

\subsection{Pulse perturbations can induce state-switching in multiarea networks}
\label{s:pulsed_stim}

In this section, we study the response of the neural mass networks to brief stimulation pulses applied to only a single region in the larger system. In what follows, we establish circumstances where pulse perturbations are ineffective at modifying collective states, but also those where the perturbation can successfully induce switching between different functional connectivity patterns, hence enabling rapid state-selection in brain circuits that exhibit collective multistability.

\subsubsection{2-area networks: Deterministic limit}
\label{s:2node_deterministic_pulse}

\begin{FPfigure}
	\centering
	\includegraphics[width=\textwidth]{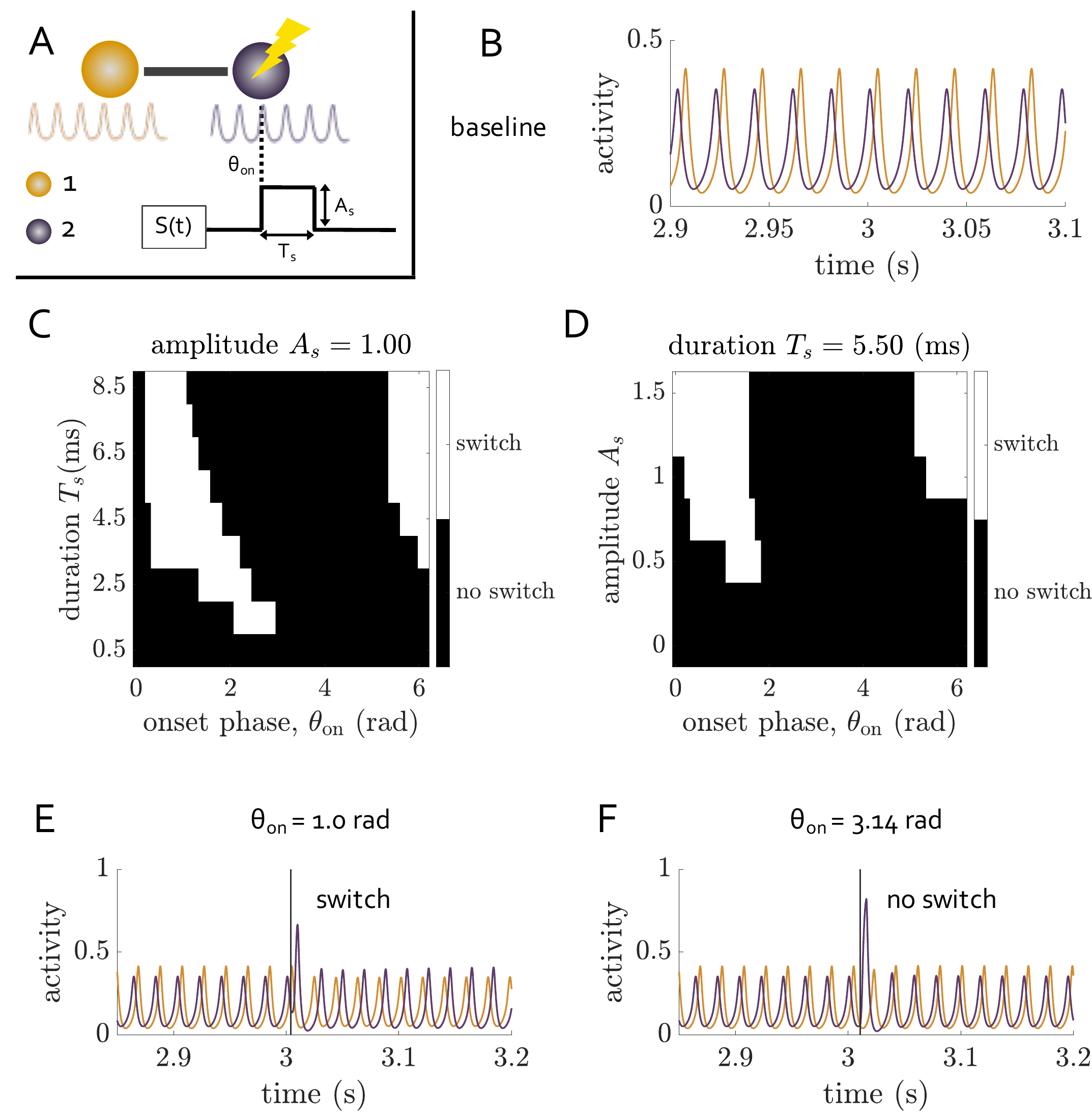}
	\caption[Response of a deterministic 2-area network to brief pulse perturbations.]{\textbf{Response of a deterministic 2-area network to brief pulse perturbations.} \textbf{(A)} Schematic of the model setup. We study the response of a deterministic 2-area brain circuit to a brief input pulse of amplitude $A_{s}$ and duration $T_{s}$. The perturbation targets only one of the two regions, which in this case is the area that leads in phase. Throughout, we set $P_{E} = 1.35$, $G_{EE} = 0.2$, and $T_{D} = 1.5$ ms. \textbf{(B)} Activity time-series of the two regions under baseline conditions. The network locks into collective state $\mathbf{S_{2}}$, where area 2 leads area 1 in phase. \textbf{(C)} Surface plot that shows whether an external input pulse causes a switch in the phase-locking pattern (white area) as a function of the onset phase $\theta_{\mathrm{on}}$ and the pulse duration $T_{s}$. \textbf{(D)} Surface plot that shows whether an external input pulse causes a switch in the phase-locking pattern (white area) as a function of the onset phase $\theta_{\mathrm{on}}$ and the pulse amplitude $A_{s}$. \textbf{(E)} Activity time-series that shows the effects of a pulse perturbation applied to the purple area at the time denoted by the black line (pulse parameters are $A_{s}=1$, $T_{s} = 5.5$ ms, $\theta_{\mathrm{on}} = 1$ rad). The external input causes the purple area to undergo a phase-delay, which switches the phase-locking pattern. \textbf{(F)} Activity time-series that shows the effects of a pulse perturbation applied to the purple area at the time denoted by the black line (pulse parameters are $A_{s}=1$, $T_{s} = 5.5$ ms, $\theta_{\mathrm{on}} = \pi$ rad). The external input causes the purple area to undergo a slight phase-advance, and the phase-locking pattern is maintained.}
	\label{f:2node_deterministic_pulseStim}
\end{FPfigure}

Here we start by considering 2-area networks that operate in the limit of deterministic background inputs, and with paramters chosen such that out-of-phase locking emerges in the collective dynamics (see Sec.~\ref{s:2area_deterministic}). Concretely, we consider parameters $P_{E,j} = 1.35$ for $j \in \{1,2\}$, $G_{EE} = 0.2$, and $T_{D} = 1.5$ ms. Fig.~\ref{f:2node_deterministic_pulseStim}B shows a segment of the activity time-series that illustrates one of the two possible multistable states. As described in Sec.~\ref{s:pulse_input_description}, a local pulse perturbation is enacted by stimulating the excitatory subpopulation of one of the areas in the circuit with a square wave pulse of amplitude $A_{s} > 0$, duration $T_{s} > 0$, and at an onset phase $\theta_{\mathrm{on}}$ of the ongoing oscillation (Fig.~\ref{f:2node_deterministic_pulseStim}A).

For the 2-area circuits, we are particularly interested in the question of when the perturbation can induce a switch in the network's collective phase-locking state by altering the lead-lag relationship. To examine this, we proceed as follows. We first run a baseline simulation without a perturbation applied. In this case, the system quickly flows towards one of its stable phase-locking configurations. We then run a second simulation using the same initial conditions, and again allow the system ample time to fully settle into the same attractor. But, for the second simulation, we then activate an external perturbation to the leading area at a time when its oscillation phase is equal to $\theta_{\mathrm{on}}$. This process is repeated anew for 50 onset phases $\theta_{\mathrm{on}}$ spaced evenly in the interval $[0,2\pi]$ in order to examine the dependence on the timing of the stimulation relative to the perturbed area's ongoing rhythm. Note that because the perturbation is transient in nature, the system will eventually reequilibriate into one of its two stable collective states. The key point is thus to determine whether applying the perturbation causes the phase difference $\Delta \theta$ to switch sign after the network relaxes.

Our results indicate that for sufficiently strong and long enough pulse amplitudes and durations, appropriately-timed external stimulation pulses can cause a switching of the network's phase-locking configuration (Figs.~\ref{f:2node_deterministic_pulseStim}C,D). In particular, Fig.~\ref{f:2node_deterministic_pulseStim}C shows a surface plot of the state-switching region as a function of the pulse duration $T_{s}$ and onset phase $\theta_{\mathrm{on}}$ for fixed amplitude $A_{s} = 1.0$, and Fig.~\ref{f:2node_deterministic_pulseStim}D shows the same information as a function of the pulse amplitude $A_{s}$ and onset phase $\theta_{\mathrm{on}}$ for fixed duration $T_{s} = 5.5$ ms. From these analyses, we observe that there are specific ranges of onset phases for which a transition in the collective state occur. Because we consider a perturbation of the phase-leader, the transition areas correspond to the stimulated region undergoing a phase-delay relative to the unperturbed region. The range of onset phases that lead to a successful state switch are determined by the phase response properties of the system \cite{Smeal2010:PhaseResponse, Canavier2015:PhaseResetting,Battaglia2012:DynamicEffective,Witt2013:Controlling, Lisitsyn2019:Causally}.

In our convention, $\theta = 0$ corresponds to the peaks of the oscillation cycles, and so we see that for smaller amplitudes and durations, successful switching occurs on the falling side of the oscillation before the trough ($\theta = \pi$) is reached. For increasing pulse duration and amplitude, onset phases that lead to a state change tend to shift more towards $\theta = 0$, and perturbations that arrive just before or at the oscillation peak can also lead to state-switching. Fig.~\ref{f:2node_deterministic_pulseStim}E and Fig.~\ref{f:2node_deterministic_pulseStim}F show segments of the two units’ activity time-series just prior to and just after application of the stimulation pulse. Panel E is an example of a successful state-switch. Before the pulse, the purple unit leads the yellow unit in phase, and then at the time of the stimulation (indicated by the black bar), a new peak is initiated in the purple unit's oscillation. That stimulation-induced effect subsequently induces a phase-delay of the purple area's activity, which in turn allows the yellow area to become the phase-leader. Alternatively, panel F depicts an example of an unsuccessful state switch. While the amplitude and duration of the stimulation are kept the same, the onset phase here is such that the pulse begins near the trough of the purple area's cycle. This scenario leads to a transient phase-advance for the purple unit, and it remains the phase-leader after the pulse is turned off.

In sum, we find that if the external pulse is activated within a ``critical" portion of the receiving area's oscillation, then the induced transient change in activity at the perturbed site can move the network from one of its stable phase-locked attractors to the other. Hence, it is possible to control the lead-lag relationships using localized stimulation. This result is consistent with the findings of other studies that have used models based on interneuron-mediated gamma rhythms \cite{Battaglia2012:DynamicEffective,Palmigiano2017:FlexibleInformation,Lisitsyn2019:Causally,Witt2013:Controlling}. 

\subsubsection{4-area networks: Deterministic limit}
\label{s:4area_deterministic_pulseStim}

\begin{FPfigure}
	\centering
	\includegraphics[width=1\textwidth]{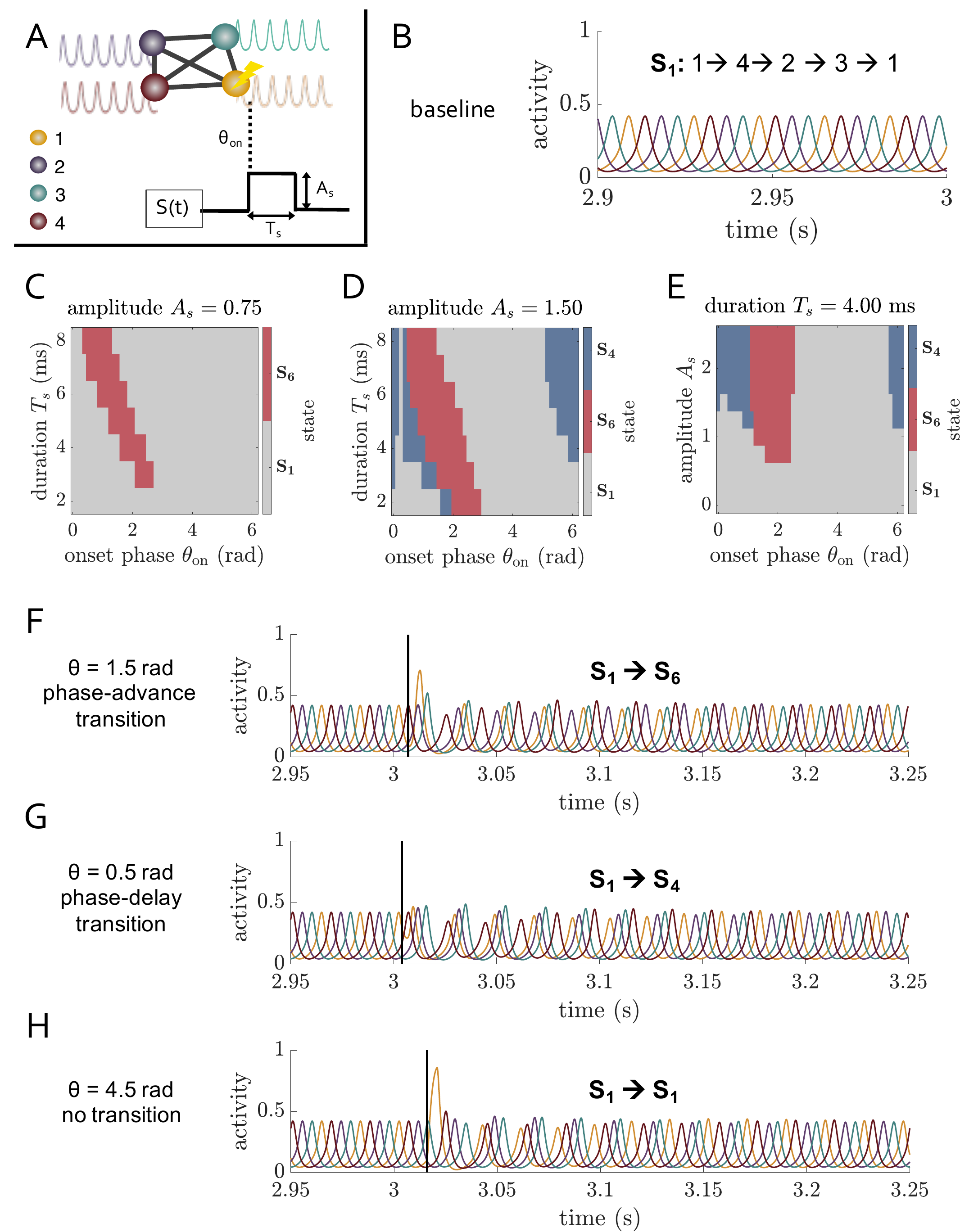}
	\caption[Response of a deterministic 4-area network to brief pulse perturbations.]{\textbf{Response of a deterministic 4-area network to brief pulse perturbations.} \textbf{(A)} Schematic of the model setup. We study the response of a deterministic 4-area brain circuit to a brief input pulse of amplitude $A_{s}$ and duration $T_{s}$. The perturbation targets only one of the two regions, which in this case is area 1 (yellow). Throughout, we set the network parameters to $P_{E} = 1.325$, $G_{EE} = 0.2$, and $T_{D} = 2.5$ ms. \textbf{(B)} Activity time-series of the four areas under baseline conditions. The network locks into collective state $\mathbf{S_{1}}$. \textbf{(C)} Surface plot showing the phase-locking state that the network equilibrates to after a pulse input of varying duration and onset phase $\theta_{\mathrm{on}}$ for fixed amplitude $A_{s} = 0.75$. \textbf{(D)} Surface plot showing the phase-locking state that the network equilibrates to after a pulse input of varying duration and onset phase $\theta_{\mathrm{on}}$ for fixed amplitude $A_{s} = 1.5$. \textbf{(E)} Surface plot showing the phase-locking state that the network equilibrates to after a pulse input of varying amplitude and onset phase $\theta_{\mathrm{on}}$ for fixed duration $T_{s} = 4.0$ ms. \textbf{(F)} Activity time-series showing the effects of a pulse perturbation applied to the yellow area at the time denoted by the black line (pulse parameters are $A_{s} = 1.25$, $T_{s} = 5$ ms, $\theta_{\mathrm{on}} = 1.5$ rad). The external input causes the yellow area to phase-advance, such that the state switches from $\mathbf{S_{1}} \rightarrow \mathbf{S_{6}}$. \textbf{(G)} Activity time-series showing the effects of a pulse perturbation applied to the yellow area at the time denoted by the black line (pulse parameters are $A_{s} = 1.25$, $T_{s} = 5$ ms, $\theta_{\mathrm{on}} = 0.5$ rad). The external input causes the yellow area to phase-delay, such that the state switches from $\mathbf{S_{1}} \rightarrow \mathbf{S_{4}}$. \textbf{(H)} Activity time-series showing the effects of a pulse perturbation applied to the yellow area at the time denoted by the black line (pulse parameters are $A_{s} = 1.25$, $T_{s} = 5$ ms, $\theta_{\mathrm{on}} = 4.5$ rad). Though the external input induces a large transient effect at the stimulated site, it does not lead to a reconfiguration of the phase-locking pattern.}
	\label{f:4node_deterministic_pulseStim}
\end{FPfigure}

Having studied the simplest possible network composed of 2 coupled regions, we now examine the response of the deterministic, 4-area network to local input pulses. For this analysis, we set the baseline network parameters to $P_{E,j} = 1.325$ for $j \in \{1,...,4\}$, $G_{EE} = 0.2$, and $T_{D} = 2.5$ ms. These choices yield oscillatory population activity at each model brain region, and the collective dynamics of the circuit as a whole exhibits multistable phase-locking, where a given initial condition leads to one of six possible collective states (see Sec.~\ref{s:4area_deterministic} and Fig.~\ref{f:4node_baseline} for details). We show an example of a baseline network activity pattern here in Fig.~\ref{f:4node_deterministic_pulseStim}B, where the network locks into state $\mathbf{S_{1}}$ characterized by the phase-ordering $1 \rightarrow 4 \rightarrow 2 \rightarrow 3 \rightarrow 1$. As before, we introduce a local perturbation to the system by injecting a square wave pulse input of duration $T_{s}$, amplitude $A_{s}$, and onset phase $\theta_{\mathrm{on}}$ into the excitatory subpopulation of one neural mass (Sec.~\ref{s:pulse_input_description}), which here is area 1 (see Fig.~\ref{f:4node_deterministic_pulseStim}A for a schematic of the model setup).

We wish to understand whether a local perturbation can be used to shift the network's phase-locking configuration from $\mathbf{S_{1}} \rightarrow \mathbf{S_{j\neq 1}}$, bypassing the need to alter structural connections in order to alter the networks' functional state. Accordingly, we begin in a similar manner to the 2-node case, and first allow the dynamics to evolve in the absence of a perturbation. For the chosen initial conditions, the network settles into collective state $\mathbf{S_{1}}$ (Fig.~\ref{f:4node_deterministic_pulseStim}A). We then run a new simulation wherein the dynamics are prepared in the same initial state, but after $t = 3$ seconds, a pulse perturbation of amplitude $A_{s}$ and duration $T_{s}$ is delivered to area 1, arriving at a time when area 1's phase is equal to a particular value $\theta_{\mathrm{on}}$. This latter step is repeated for a range of pulse amplitudes and durations, and for 50 onset phases distributed uniformly across a full oscillation cycle of the receiving area. To determine if the phase-locking pattern can be reconfigured by such external inputs, we then compare the state of the network after the transient effects of the perturbation die way to the original state from the baseline condition.

The complete results of our analysis are depicted in Fig.~\ref{f:4node_deterministic_pulseStim}C--H. Figs.~\ref{f:4node_deterministic_pulseStim}C and D show surface plots of the state ID $\mathbf{S_{i}}$ as a function of the onset phase $\theta_{\mathrm{on}}$ and the pulse duration $T_{s}$ for fixed amplitude $A_{s} = 0.75$ and $A_{s} = 1.5$, respectively. Fig.~\ref{f:4node_deterministic_pulseStim}E shows the state ID instead for fixed duration $T_{s} = 4$ ms and varying onset phase and amplitude. We first note that below a certain pulse amplitude and duration, the perturbation does not alter the functional state of the network, regardless of the onset phase. This fact is illustrated by the gray region (indicating a trivial $\mathbf{S_{1}} \rightarrow  \mathbf{S_{1}}$ transition) as one scans the lower portions of Figs.~\ref{f:4node_deterministic_pulseStim}C,E horizontally. But, if the strength and/or duration of the external input are strong enough and/or long enough, then an appropriately-timed focal perturbation can indeed switch the collective state of the network into a new configuration (colored areas in Figs.~\ref{f:4node_deterministic_pulseStim}C,D,E). 

Examining Figs.~\ref{f:4node_deterministic_pulseStim}C,D,E, the first additional point we make is that for lower stimulation strengths (Fig.~\ref{f:4node_deterministic_pulseStim}C), the observed non-trivial state transition corresponds to $\mathbf{S_{1}} \rightarrow  \mathbf{S_{6}}$ (light red region). Looking back at the set of collective states and their IDs (Fig.~\ref{f:4node_baseline}E), we see that $\mathbf{S_{1}}$ transitions to $\mathbf{S_{6}}$ via region 1 shifting one slot ahead of its position in $\mathbf{S_{1}}$ (i.e. the yellow area 1 swaps places with the green area 3). We show an example of a pulse input to region 1 that induces the $\mathbf{S_{1}} \rightarrow  \mathbf{S_{6}}$ state switch in Fig.~\ref{f:4node_deterministic_pulseStim}F. Note that the precisely timed perturbation causes area 1 (yellow) to undergo a phase-advance that then alters the phase-locking pattern in the long-term. 

The second fact we point out is that for larger pulse amplitudes and/or durations (Figs.~\ref{f:4node_deterministic_pulseStim}D,E), two state transitions become unlocked for different sets of onset phases: $\mathbf{S_{1}} \rightarrow  \mathbf{S_{6}}$ (light red regimes) and $\mathbf{S_{1}} \rightarrow  \mathbf{S_{4}}$ (blue regimes). We already know that the former corresponds to the stimulated site phase-advancing by one position. Again referring to the dictionary of collective states laid out in Fig.~\ref{f:4node_baseline}E, we see that the latter transformation $\mathbf{S_{1}} \rightarrow  \mathbf{S_{4}}$ corresponds to the opposite scenario, where the perturbed area 1 falls back one space and comes to sit behind the red area 4. This state-switch occurs for onset phases nearer to the peak of the ongoing rhythm. An example of a perturbation-induced switch from pattern $\mathbf{S_{1}}$ to $\mathbf{S_{4}}$ is depicted in Fig.~\ref{f:4node_deterministic_pulseStim}G, where one can see that area 1 undergoes a phase-delay that leads to the change in the functional state. For completeness, we also show an example of an onset phase that does \textit{not} lead to a change in the system's state, despite the external stimulation causing a significant transient change in the perturbed regions' activity (Fig.~\ref{f:4node_deterministic_pulseStim}H). Finally, while we considered the case of applying the perturbation to area 1, the symmetry of the system implies that the results would be analogous for a different choice of the stimulated site, but some of the allowed state transitions would be distinct.

\subsubsection{2-area networks: Stochastic scenario}
\label{s:2area_stochastic_pulse}

As we have noted previously, incorporating stochasticity into the dynamical model is an important step towards increasing the model's applicability to real neural dynamics. In this section, we thus move on to analyze 2-area circuits subject to noisy background drive (Sec.~\ref{s:stochastic_backgroundDrive}), but that operate in regimes characterized by spontaneous switching between out-of-phase configurations (see Sec.~\ref{s:2area_baseline_stochastic} for analyses of the system's baseline activity patterns in the presence of stochastic background drive). In such cases, it is still possible to extract noisy, transient versions of collective states, and the question we ask here is whether phase-locking can still be effectively controlled via brief, local input pulses in these more realistic regimes. In what follows, we analyze four different working points that are meant to demonstrate conditions under which rapid state-selection via local perturbations remains possible in the presence of noise, as well as when control capabilities begin to break down.

Before describing the effects of pulse inputs at different working points, we explain the general protocol we use to analyze the stochastic scenario. As before, we are concerned with whether or not the external input can switch the network's lead-lag relation. But, because noise results in spontaneous state-switching and causes the phase-locking to become imperfect, assessing these effects becomes slightly more involved. For all parameter sets, we begin by running a long (150-second) simulation without perturbations applied; from the resulting time-series, we are able to resolve and characterize the network's favored activity patterns. In order to extract preferred collective states, we compute the phase difference $\Delta \theta$ between the two units at each time point across the simulation, and bin the set of absolute values $\{|\Delta \theta|\}$ into 100 bins of equal width; the preferred phase difference $|\Delta \theta^{*}|$ is defined as that corresponding to the peak of this histogram. The next step is to determine candidate windows for applying a perturbation. When it comes to this step, first note that we specifically consider the effect of applying a perturbation during windows in which the phase difference is near one of the preferred configurations $\pm \Delta \theta^{*}$ for at least one cycle. In other words, the question is: given that the system is close to one of its favored states prior to the perturbation, does a local pulse input change the sign of the phase difference and hence switch the lead-lag configuration. Note that requiring the two units to be locked around a certain $\Delta \theta$ for at least one cycle ensures that we can properly compare the effects of stimulating one of the two regions at different phases of its ongoing oscillation. Also note that we could consider a more constrained criteria in which the system must remain in a particular configuration for longer than one oscillation cycle, but in what follows, we consider the least stringent selection method. In particular, candidate perturbation windows are extracted by finding time segments during which the absolute phase difference $|\Delta \theta|$ satisfies $|\Delta \theta| \in [\Delta\theta^{*} - \delta, |\Delta\theta^{*}|+\delta]$ for at least one oscillation cycle, where $\delta$ is a threshold that determines how far $\Delta \theta$ can vary around the peak value $|\Delta \theta^{*}|$. The subsequent analyses all assume a default value of $\delta = \pi/6$, though we do also briefly examine the effect of varying this quantity (see Fig.~\ref{f:081620201420_pulseStim_varyDelta}). 

Of the full set of candidate windows, we select 100 of them at random in which to apply perturbations. Specifically, for each window, we run a new simulation that uses the exact same initial conditions and noise realization, but we perturb the phase-leading area with a square wave pulse of amplitude $A_{s}$, duration $T_{s}$, and onset phase $\theta_{\mathrm{on}}$ (Sec.~\ref{s:modeling_perturbations}). For a given set of pulse parameters, onset phase, and time-window, the effect of the external input on the networks' collective state is assessed by determining \textit{(1)} if the sign of the phase difference switches within $n_{s}$ oscillation cycles after the perturbation onset, and \textit{(2)} if the sign of the phase difference remains reversed for a subsequent $n_{m}$ cycles after the initial switch. If both of these criteria are met, then we conclude that the input pulse is able to successfully control the network activity pattern (for the given $n_{s}$ and $n_{m}$). To determine the overall efficacy of a perturbation with fixed parameters on inducing a lasting state-transition, we repeat the aforementioned process for each of the 100 randomly selected windows, and from these trials, we compute the fraction that lead to a successful alteration of the collective phase relation. In what follows, we will often refer to this fraction as the ``switching probability" (or ``transition probability") $P_{\mathrm{switch}}$, but note that because $P_{\mathrm{switch}}$ is computed from a finite number of samples, it is only an estimate of the true value. 

Importantly, the collective state will switch on its own due to the stochastic environment. It is therefore helpful to compare the perturbation-induced switching probability to the chance of observing a \textit{spontaneous} change in the sign of the phase difference without applying the perturbation. To do this, we consider the same set of 100 time-windows, and compute the fraction of those windows for which both \textit{(1)} the sign of the phase difference switches within $n_{s}$ oscillation cycles on its own, and \textit{(2)} the new lead-lag relation is maintained for a subsequent $n_{m}$ cycles after the initial switch. This procedure allows us to ask whether the estimated likelihood of observing a state-transition due to the perturbation itself exceeds the estimated probability of the system transitioning on its own, due to noise. In the remainder of the text, we often refer to the estimated chance level of observing a state transition as the ``spontaneous switching probability" or ``spontaneous transition probability".

Note that if given enough time, the lead-lag relation will change. Therefore, we are interested in the question of whether local perturbations can be harnessed to achieve ``on-demand" or immediate state-selection. To this end, we consider a default value $n_{s} = 2$. Moreover, for coherent oscillatory activity to be functionally relevant, it is also necessary that a particular phase relationship be maintained for at least a few oscillation cycles; for this reason, we consider a default value $n_{m} = 5$. Unless stated otherwise, the reader can assume in the coming text that the above default values are used to compute the switching probabilities $P_{\mathrm{switch}}$.

We now present our key results regarding the effects of local pulse perturbations on state-switching in stochastic 2-area networks. We refer the reader to Sec.~\ref{s:2area_baseline_stochastic} for a complete analysis of the baseline behaviors of the 2-area networks with noisy background drive, as here we only provide a recap of the key points. To begin, we consider a baseline working point of $\overline{P_{E,j}} = 1.35$ for $j \in \{1,2\}$, $T_{D} = 1.5$ ms, $G_{EE} = 0.2$, $\alpha = 10$, and $\sigma = 0.2$ (Working Point 1; Fig.~\ref{f:081220202120_pulseStim}). For these parameters but in the absence of noise ($\sigma \rightarrow 0$), we saw that the phase-locking pattern could be controlled via appropriately-timed external input pulses (Fig.~\ref{f:2node_deterministic_pulseStim}C--F). When the background drive is instead made stochastic, we observe that the baseline distribution of phase differences $P[\Delta \theta]$ develops two broad but still well-defined out-of-phase peaks that indicate preferred phase relations that the system spontaneously transitions between over time (Fig.~\ref{f:081220202120_pulseStim}A). In particular, a given lead-lag relationship (characterized by whether the sign of the phase difference between the two areas is positive or negative) lasts for on average $\langle L_{s} \rangle \approx 378$ ms before switching (Fig.~\ref{f:081220202120_pulseStim}B). Thus, for this operating point, it is possible to extract time-windows during which the system exists in blurred versions of the noiseless collective states. In particular, the vertical lines in the $P[\Delta \theta]$ histogram (Fig.~\ref{f:081220202120_pulseStim}A) show the thresholds used to select the candidate time-windows for applying pulse perturbations; note that this criteria extracts time-windows during which $\Delta \theta$ is near one of its most probable out-of-phase configurations.

\begin{FPfigure}
	\centering
	\includegraphics[width=1\textwidth]{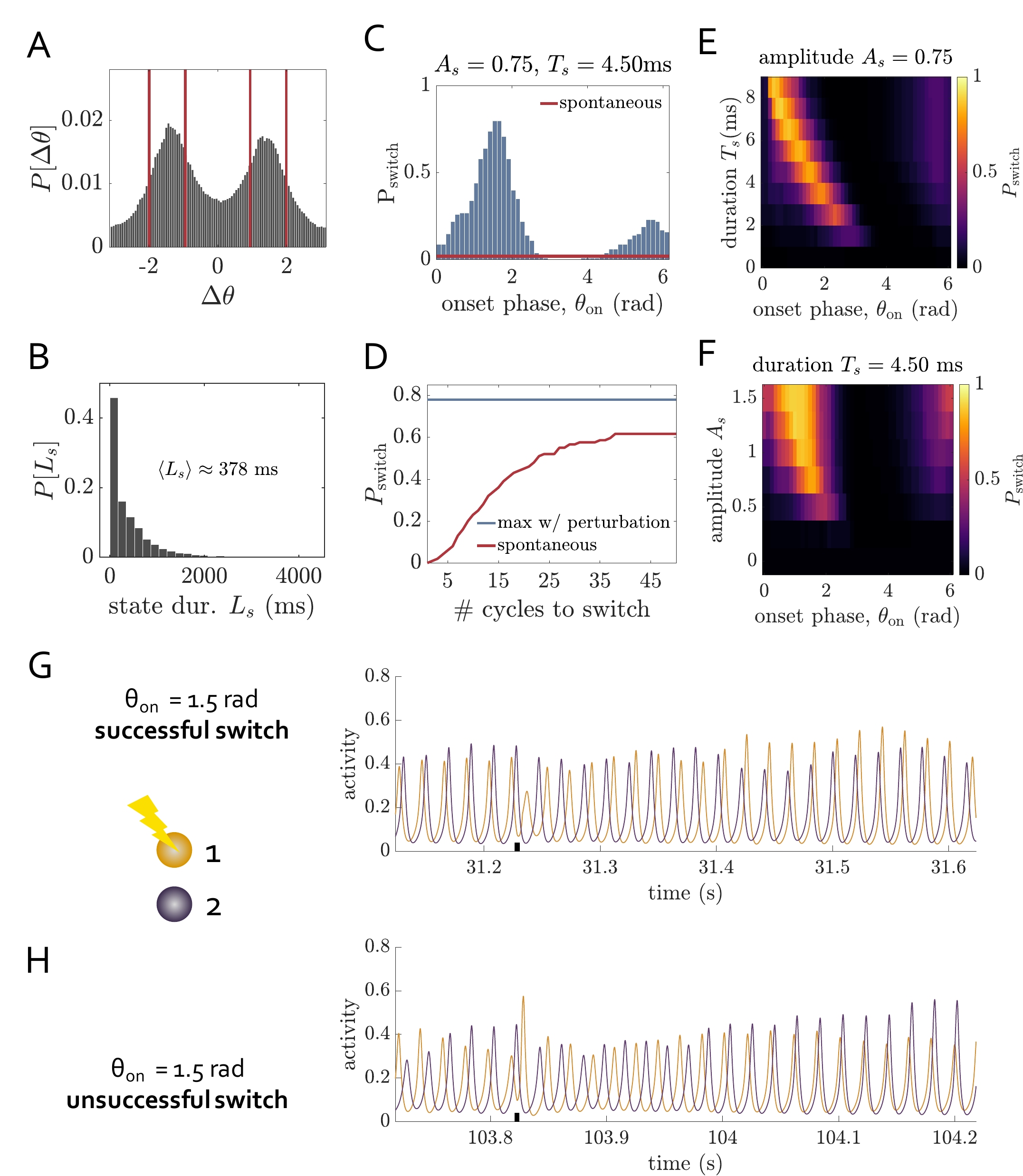}
	\caption[Response of a stochastic 2-area network to brief perturbation pulses applied to the phase-leader: Working Point 1.]{\textbf{Response of a stochastic 2-area network to brief perturbation pulses applied to the phase-leader: Working Point 1.} The key network parameters are: $\overline{P_{E}} = 1.35$, $T_{D} = 1.5$ ms, $G_{EE} = 0.2$, $\alpha = 10$, $\sigma = 0.2$. \textbf{(A)} The baseline distribution of phase differences $\Delta \theta$ between the two coupled areas across a long simulation. The red lines around each peak indicate the range of $\Delta \theta$ values used to select candidate windows for perturbation; we consider a width $\pm \delta = \pi/6$ around each peak. \textbf{(B)} The distribution of state durations $L_{s}$ with the mean value $\langle L_{s} \rangle$ printed inside the panel. \textbf{(C)} The heights of the blue bars show the estimated switching probabilities $P_{\mathrm{switch}}$ for pulse inputs applied at different onset phases $\theta_{\mathrm{on}}$. A successful state transition is said to occur if the sign of the interareal phase relation switches within 2 cycles and the new configuration lasts for at least 5 more cycles following the initial switch. The pulse has amplitude $A_{s} = 0.75$ and duration $T_{s} = 4.5$ ms. The red horizontal line indicates the estimated likelihood of observing the same behavior spontaneously, i.e., in the absence of a perturbation. \textbf{(D)} Across all onset phases $\theta_{\mathrm{on}}$, the blue line indicates the maximum switching probability $\mathrm{max}[P_{\mathrm{switch}}]$. A successful state transition is said to occur if the sign of the interareal phase relation switches within 1 cycle and the new configuration lasts for at least 5 more cycles following the initial switch. The perturbation has amplitude $A_{s} = 0.75$ and duration $T_{s} = 4.5$ ms. The red curve shows the estimated likelihood of observing a spontaneous switch that lasts for at least 5 periods and that occurs within a varying number of oscillation cycles. \textbf{(E)} The estimated switching probability $P_{\mathrm{switch}}$ as a function of the onset phase $\theta_{\mathrm{on}}$ and the pulse duration $T_{s}$. The pulse amplitude is fixed at $A_{s} = 0.75$, and a successful state transition is said to occur if the sign of the interareal phase relation switches within 2 cycles and the new configuration lasts for at least 5 more cycles following the initial switch. \textbf{(F)} The switching probability $P_{\mathrm{switch}}$ as a function of the onset phase $\theta_{\mathrm{on}}$ and the pulse amplitude $A_{s}$. The pulse duration is fixed at $T_{s} = 4.5$ ms, and a successful state transition is said to occur if the sign of the interareal phase relation switches within 2 cycles and the new configuration lasts for at least 5 more cycles following the initial switch. \textbf{(G)} Activity time-series from a trial where a successful state-switch is induced by applying a pulse perturbation to the yellow area at the time denoted by the black bar (pulse parameters are $A_{s} = 0.75$, $T_{s} = 4.5$ ms, $\theta_{\mathrm{on}} = 1.5$ rad). \textbf{(H)} Activity time-series from a different trial that shows an unsuccessful state-switch with the same pulse parameters.}
	\label{f:081220202120_pulseStim}
\end{FPfigure}

Although the system is now noisy, we find that for this working point, suitable external inputs that target the phase-leader can still very effectively control the phase-locking pattern. As an example, Fig.~\ref{f:081220202120_pulseStim}C shows the switching probability $P_{\mathrm{switch}}$ as a function of the onset phase $\theta_{\mathrm{on}}$ for pulses of fixed duration $T_{s} = 4.5$ ms and amplitude $A_{s} = 0.75$. Note that there are specific ranges of onset phases for which the perturbation-induced likelihood of triggering a state transition is well above the chance level. Indeed, the maximum switching probability reaches a value of near 0.8. To see more clearly how a local perturbation can induce a state-switch, we show an example of the activity time-series for a successful transition in Fig.~\ref{f:081220202120_pulseStim}G. However, despite the high switching likelihood, the stochastic nature of the system still renders some perturbations ineffective, even if they arrive at an onset phase that would yield a switch in the deterministic limit. Fig.~\ref{f:081220202120_pulseStim}H depicts an example of an unsuccessful trial that uses the exact same pulse parameters as before. First note that, due to fluctuations in the oscillation amplitudes and phase-locking, the transient effect of the external input is noticeably different in comparing Figs.~\ref{f:081220202120_pulseStim}G and H. Moreover, while the perturbation can alter the sign of the phase difference for the couple of cycles right after its onset, this effect decays within a 5-cycle period for this time-window.

To better appreciate the influence of focal input pulses on collective phase-locking states, it is helpful to more carefully contrast the perturbation-induced effects to what would be expected spontaneously. Specifically, we compute the spontaneous switching probability for varying $n_{s}$ (holding $n_{m} = 5$) and compare the resulting curve to the maximum perturbation-induced transition probability across all onset phases for the smallest possible $n_{s} = 1$. This analysis reveals that the chance of observing a spontaneous state change within an allowed waiting time of up to 50 cycles is still well below the maximum switching probability triggered by a perturbation within only a single cycle (Fig.~\ref{f:081220202120_pulseStim}D). Hence, if the system is currently in a well-defined collective state near the peak of $P[\Delta \theta]$, then the efficacy of the perturbation-induced effects that occur immediately will not be matched in the spontaneous dynamics even if one waited tens of oscillation cycles. Crucially, this observation allows us to conclude that although state-control may be overall less efficient in the stochastic regime, perturbations are nonetheless able to elicit state switching on very fast time scales with efficacies that far exceed chance levels. In this way, the functional benefits of multistable collective phase-locking and modulatory control inputs continue to hold under more realistic conditions.

The dependence of the transition probability $P_{\mathrm{switch}}$ on the pulse parameters $A_{s}$ and $T_{s}$ and the onset phase $\theta_{\mathrm{on}}$ is summarized in Figs.~\ref{f:081220202120_pulseStim}E,F. As expected, these plots resemble blurred versions of their counterparts from the noiseless system (Fig.~\ref{f:2node_deterministic_pulseStim}C,D). For this reason, we do not unpack each feature again, and instead refer the reader to Sec.~\ref{s:2area_stochastic_pulse} for a detailed discussion. Here we just highlight that for strong enough and/or long enough external pulses, the switching probability peaks at values well above chance for specific ranges of the onset phase. These bands correspond to parameter combinations where focal perturbations successfully initiate rapid state transitions that would occur only very rarely on their own.

When the background inputs are made stochastic, there are even more knobs to tune, and it is impossible to completely explore the influence of the network's baseline dynamics on the capacity for functional state control. Here we choose to examine one important aspect of the system, which is the signal-to-noise ratio. There are a number of ways this could be changed, but here we briefly consider the effect of keeping the network coupling, delay, and noise parameters the same, but lowering the mean background drive to $\overline{P_{E,j}} = 1.30$ for $j \in {1,2}$ (Working Point 2). As discussed in Sec.~\ref{f:uncoupled_WC}, this process has the effect of lowering the amplitude of regional oscillations at baseline, in turn making the network more susceptible to noise. Indeed, compared to the previous working point, lowering $\overline{P_{E}}$ leads to decreased peak heights in the phase difference distribution $P[\Delta \theta]$ (Fig.~\ref{f:081620201420_pulseStim}A) and decreased state durations (Fig.~\ref{f:081620201420_pulseStim}B). Nevertheless, it is still clear that the network has preferred dynamical configurations, allowing us to again assess the outcomes of external input pulses on collective states. 

Proceeding with the same analyses as before, we first note that when the signal-to-noise ratio is decreased, the likelihood of inducing fast, lasting state-switches still remains well above chance levels if the input is received during certain phases of the ongoing oscillation (Figs.~\ref{f:081620201420_pulseStim}C,E,F). However, it is also apparent that the perturbation-induced switching probabilities undergo a global decrease compared to the previous working point (compare Figs.~\ref{f:081620201420_pulseStim}C,E,F to Figs.~\ref{f:081220202120_pulseStim}C,E,F). Thus, not surprisingly, the collective states are overall more difficult to control when the system is inherently more noisy. Additionally, because baseline state-switching is also faster, the likelihood of inducing an immediate state-switch with a perturbation will eventually be matched (and exceeded) in the spontaneous dynamics within a few tens of cycles (Fig.~\ref{f:081620201420_pulseStim}D). Even so, it remains a crucial point that despite the heightened noise, perturbations can still significantly accelerate transitions in the collective state. This fact allows for enhanced functional flexibility and significant probabilities of state reconfiguration within just one oscillation cycle (Figs.~\ref{f:081620201420_pulseStim}G,H). We also find that the perturbation-induced effects remain well above chance levels even when the window selection width $\delta$ is doubled (see Fig.~\ref{f:081620201420_pulseStim_varyDelta}).

\begin{FPfigure}
	\centering
	\includegraphics[width=1\textwidth]{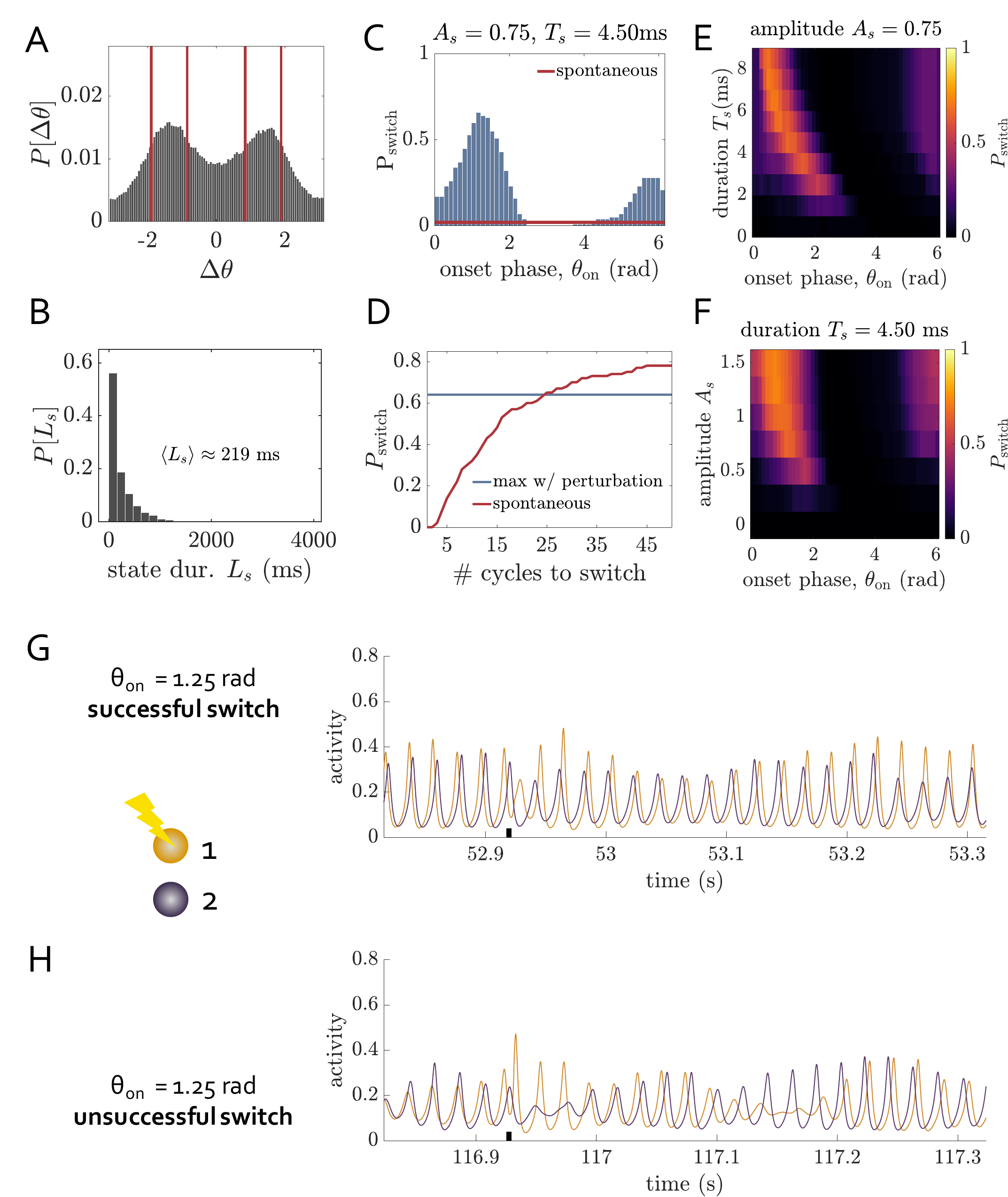}
	\caption[Response of a stochastic 2-area network to brief perturbation pulses applied to the phase-leader: Working Point 2.]{\textbf{Response of a stochastic 2-area network to brief perturbation pulses applied to the phase-leader: Working Point 2.} The key network parameters are: $\overline{P_{E}} = 1.30$, $T_{D} = 1.5$ ms, $G_{EE} = 0.2$, $\alpha = 10$, $\sigma = 0.2$. \textbf{(A)} The baseline distribution of phase differences $\Delta \theta$ between the two coupled areas across a long simulation. The red lines around each peak indicate the range of $\Delta \theta$ values used to select candidate windows for perturbation; we consider a width $\pm \delta = \pi/6$ around each peak. \textbf{(B)} The distribution of state durations $L_{s}$ with the mean value $\langle L_{s} \rangle$ printed inside the panel. \textbf{(C)} The heights of the blue bars show the estimated switching probabilities $P_{\mathrm{switch}}$ for pulse inputs applied at different onset phases $\theta_{\mathrm{on}}$. A successful state transition is said to occur if the sign of the interareal phase relation switches within 2 cycles and the new configuration lasts for at least 5 more cycles following the initial switch. The pulse has amplitude $A_{s} = 0.75$ and duration $T_{s} = 4.5$ ms. The red horizontal line indicates the estimated likelihood of observing the same behavior spontaneously, i.e., in the absence of a perturbation. \textbf{(D)} Across all onset phases $\theta_{\mathrm{on}}$, the blue line indicates the maximum switching probability $\mathrm{max}[P_{\mathrm{switch}}]$. A successful state transition is said to occur if the sign of the interareal phase relation switches within 1 cycle and the new configuration lasts for at least 5 more cycles following the initial switch. The perturbation has amplitude $A_{s} = 0.75$ and duration $T_{s} = 4.5$ ms. The red curve shows the estimated likelihood of observing a spontaneous switch that lasts for at least 5 periods and that occurs within a varying number of oscillation cycles. \textbf{(E)} The estimated switching probability $P_{\mathrm{switch}}$ as a function of the onset phase $\theta_{\mathrm{on}}$ and the pulse duration $T_{s}$. The pulse amplitude is fixed at $A_{s} = 0.75$, and a successful state transition is said to occur if the sign of the interareal phase relation switches within 2 cycles and the new configuration lasts for at least 5 more cycles following the initial switch. \textbf{(F)} The estimated switching probability $P_{\mathrm{switch}}$ as a function of the onset phase $\theta_{\mathrm{on}}$ and the pulse amplitude $A_{s}$. The pulse duration is fixed at $T_{s} = 4.5$ ms, and a successful state transition is said to occur if the sign of the interareal phase relation switches within 2 cycles and the new configuration lasts for at least 5 more cycles following the initial switch. \textbf{(G)} Activity time-series from a trial where a successful state-switch is induced by applying a pulse perturbation to the yellow area at the time denoted by the black bar (pulse parameters are $A_{s} = 0.75$, $T_{s} = 4.5$ ms, $\theta_{\mathrm{on}} = 1.25$ rad). \textbf{(H)} Activity time-series from a different trial that shows an unsuccessful state-switch with the same pulse parameters.}
	\label{f:081620201420_pulseStim}
\end{FPfigure}

We next consider whether pulse inputs can still be used to modulate collective phase-locking for a quantitatively different dynamical regime where in the absence of noise, the system does not have attractors that correspond to out-of-phase states. In particular, we set the parameters to $\overline{P_{E,j}} = 1.25$ for $j \in \{1,2\}$, $T_{D} = 1.5$ ms, $G_{EE} = 0.3$, $\alpha = 10$, and $\sigma = 0.25$ (Working Point 3). As discussed in  Sec.~\ref{s:2area_baseline_stochastic}, here the background drive is such that the deterministic network actually locks into an in-phase configuration (Fig.~\ref{f:2node_baseline_stochastic_lowDrive}A). However, as we saw in Sec.~\ref{s:2area_baseline_stochastic} (and see again here) the addition of appropriate background noise can lead to preferred out-of-phase configurations (Fig.~\ref{f:102220202020_pulseStim}A) and fast switching between lead-lag relations (Fig.~\ref{f:102220202020_pulseStim}B). Although out-of-phase locking arises here for a somewhat different reason than the previously examined operating points, the two clear peaks in the distribution $P(\Delta \theta)$ allow us to extract windows that correspond to well-defined collective states in which we can test the effects of perturbations. 

Continuing as before, we find that there remains a band of onset phases (for strong and long enough pulses) in which the perturbation-induced switching probability exceeds the spontaneous level (Figs.~\ref{f:102220202020_pulseStim}C). Fig.~\ref{f:102220202020_pulseStim}G shows the activity time series just before and after a successful state change. But, notice that the peak transition probabilities across a range of pulse amplitudes and durations are quite drastically reduced, reaching levels of only about 0.25 (Figs.~\ref{f:102220202020_pulseStim}E,F). An example of an unsuccessful trial that uses the exact same pulse parameters is shown in Fig.~\ref{f:102220202020_pulseStim}H; here, the influence of the noise takes over after the perturbation is turned off, and prevents the triggered reverse in the lead-lag relationship from lasting more than just 2-3 oscillation cycles (thus resulting in failed state-control). Nonetheless, although overall performance is reduced in this regime, we still wish to underscore that brief external inputs are still more effective than chance at eliciting very rapid reconfigurations of functional connectivity. That is, approximately 25\% of the time, carefully-timed pulses can lead to a sustained (lasting 5 cycles) change in the sign of the phase relation within one period, while one would need to wait about six cycles in the absence of a modulatory input (Fig.~\ref{f:102220202020_pulseStim}C).

\begin{FPfigure}
	\centering
	\includegraphics[width=1\textwidth]{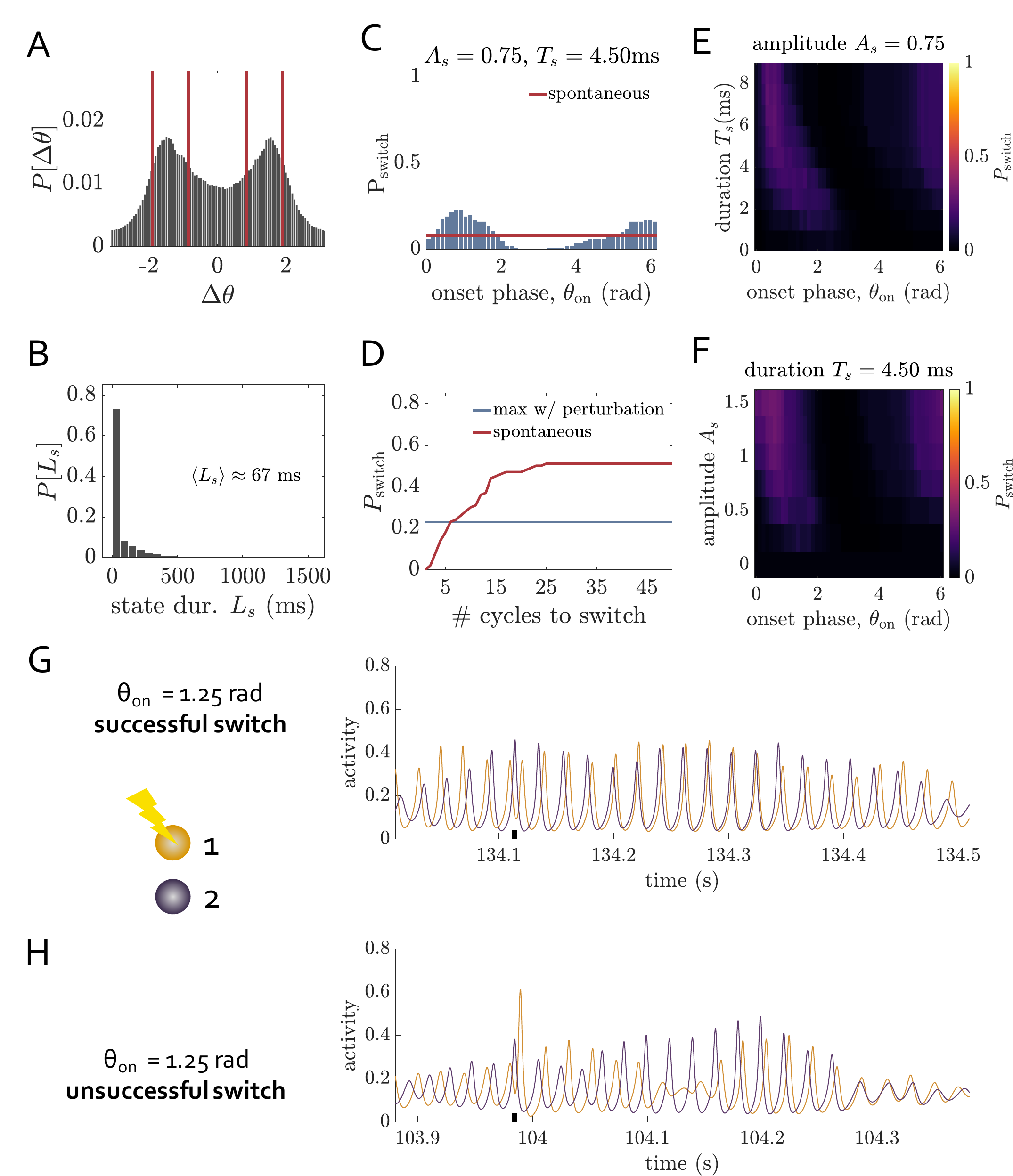}
	\caption[Response of a stochastic 2-area network to brief perturbation pulses applied to the phase-leader: Working Point 3.]{\textbf{Response of a stochastic 2-area network to brief perturbation pulses applied to the phase-leader: Working Point 3}. The key network parameters are: $\overline{P_{E}} = 1.25$, $T_{D} = 1.5$ ms, $G_{EE} = 0.3$, $\alpha = 10$, $\sigma = 0.25$. \textbf{(A)} The baseline distribution of phase differences $\Delta \theta$ between the two coupled areas across a long simulation. The red lines around each peak indicate the range of $\Delta \theta$ values used to select candidate windows for perturbation; we consider a width $\pm \delta = \pi/6$ around each peak. \textbf{(B)} The distribution of state durations $L_{s}$ with the mean value $\langle L_{s} \rangle$ printed inside the panel. \textbf{(C)} The heights of the blue bars show the estimated switching probabilities $P_{\mathrm{switch}}$ for pulse inputs applied at different onset phases $\theta_{\mathrm{on}}$. A successful state transition is said to occur if the sign of the interareal phase relation switches within 2 cycles and the new configuration lasts for at least 5 more cycles following the initial switch. The pulse has amplitude $A_{s} = 0.75$ and duration $T_{s} = 4.5$ ms. The red horizontal line indicates the estimated likelihood of observing the same behavior spontaneously, i.e., in the absence of a perturbation. \textbf{(D)} Across all onset phases $\theta_{\mathrm{on}}$, the blue line indicates the maximum switching probability $\mathrm{max}[P_{\mathrm{switch}}]$. A successful state transition is said to occur if the sign of the interareal phase relation switches within 1 cycle and the new configuration lasts for at least 5 more cycles following the initial switch. The perturbation has amplitude $A_{s} = 0.75$ and duration $T_{s} = 4.5$ ms. The red curve shows the estimated likelihood of observing a spontaneous switch that lasts for at least 5 periods and that occurs within a varying number of oscillation cycles. \textbf{(E)} The estimated switching probability $P_{\mathrm{switch}}$ as a function of the onset phase $\theta_{\mathrm{on}}$ and the pulse duration $T_{s}$. The pulse amplitude is fixed at $A_{s} = 0.75$, and a successful state transition is said to occur if the sign of the interareal phase relation switches within 2 cycles and the new configuration lasts for at least 5 more cycles following the initial switch. \textbf{(F)} The estimated switching probability $P_{\mathrm{switch}}$ as a function of the onset phase $\theta_{\mathrm{on}}$ and the pulse amplitude $A_{s}$. The pulse duration is fixed at $T_{s} = 4.5$ ms, and a successful state transition is said to occur if the sign of the interareal phase relation switches within 2 cycles and the new configuration lasts for at least 5 more cycles following the initial switch. \textbf{(G)} Activity time-series from a trial where a successful state-switch is induced by applying a pulse perturbation to the yellow area at the time denoted by the black bar (pulse parameters are $A_{s} = 0.75$, $T_{s} = 4.5$ ms, $\theta_{\mathrm{on}} = 1.25$ rad). \textbf{(H)} Activity time-series from a different trial that shows an unsuccessful state-switch with the same pulse parameters.}
	\label{f:102220202020_pulseStim}
\end{FPfigure}

Understanding when meaningful modulation of collective states breaks down is also important. While functional state control could become ineffective for multiple reasons, here we show that one scenario occurs when the background drive is decreased even further. In particular, we consider the same parameters as used in Fig.~\ref{f:102220202020_pulseStim}, with the exception that we set $\overline{P_{E,j}} = 1.20$ for $j \in \{1,2\}$ (Working Point 4). The baseline dynamics for this situation were studied in detail in Sec.~\ref{s:2area_baseline_stochastic} and Fig.~\ref{f:2node_baseline_stochastic_lowDrive}. The key point is that for this working point, even when the network is coupled, it is mostly the noise itself that induces robust regional oscillations and subsequent out-of-phase locking. Nonetheless, we selected parameters such that the peak values of the phase difference distribution $P[\Delta \theta]$ are similar between the $\overline{P_{E}} = 1.25$ and $\overline{P_{E}} = 1.20$ working points (compare Figs.~\ref{f:102220202020_pulseStim}A and ~\ref{f:102220201220_pulseStim}A). In this way, we control for significant variation in the noisiness of the collective states at different operating points (though do note that collective states are still shorter lived when the background drive is lower; Fig.~\ref{f:102220201220_pulseStim}B). Interestingly, although sharp peaks still manifest in $P[\Delta \theta]$, the switching probabilities drop to $<$ 10\% and do not robustly exceed the spontaneous switching level (Fig.~\ref{f:102220201220_pulseStim}C). Fig.~\ref{f:102220201220_pulseStim}D shows an example of the activity time-series for an unsuccessful trial; due to the influence of noise over the system's dynamics, precise and lasting state-switching is difficult to induce in this regime.

\begin{figure}[h!]
	\centering
	\includegraphics[width=0.99\textwidth]{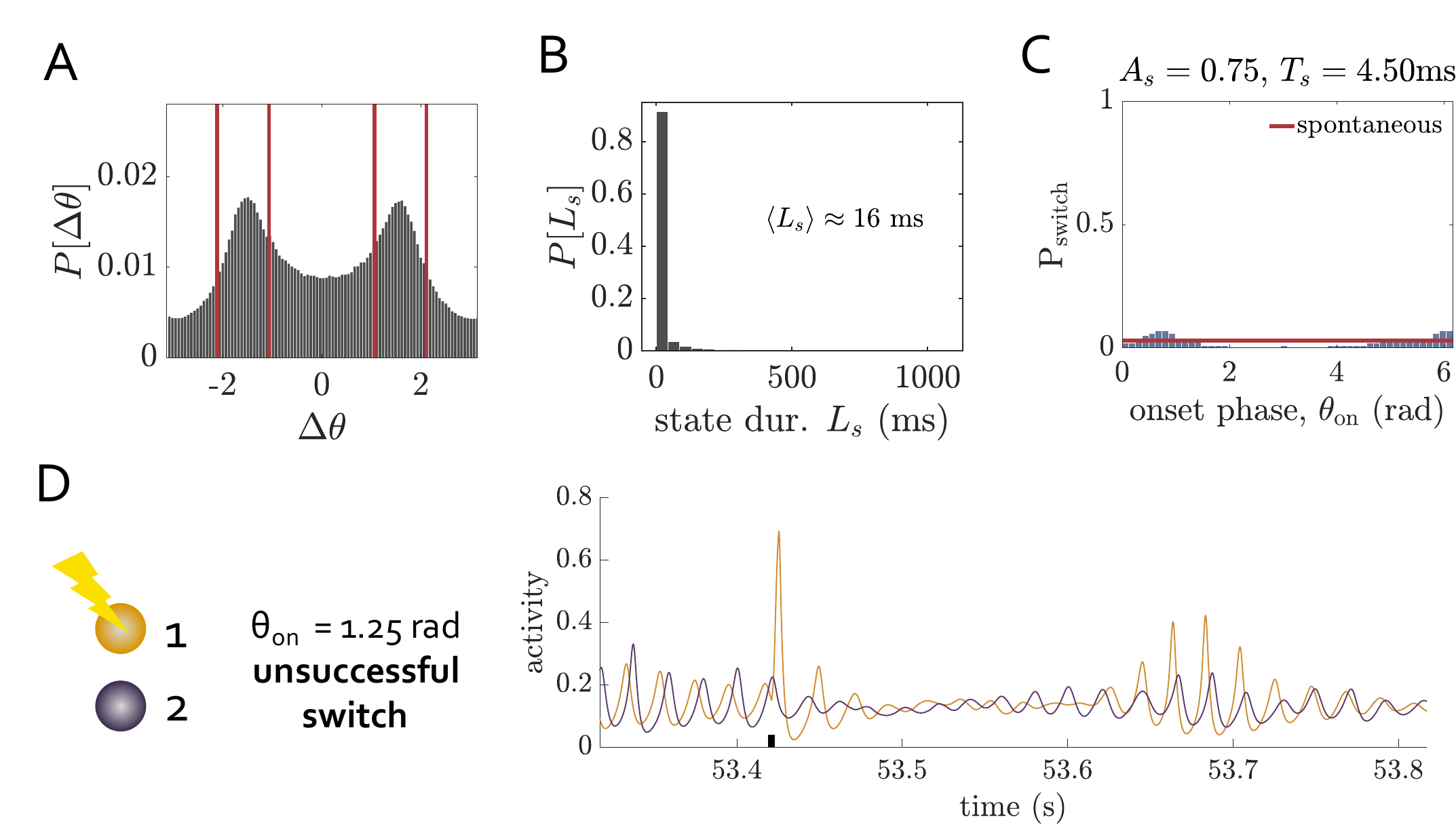}
	\caption[Response of a stochastic 2-area network to brief perturbation pulses applied to the phase-leader: Working Point 4.]{\textbf{Response of a stochastic 2-area network to brief perturbation pulses applied to the phase-leader: Working Point 4.} The key network parameters are: $\overline{P_{E}} = 1.20$, $T_{D} = 1.5$ ms, $G_{EE} = 0.3$, $\alpha = 10$, $\sigma = 0.25$. \textbf{(A)} The baseline distribution of phase differences $\Delta \theta$ between the two coupled areas across a long simulation. The red lines around each peak indicate the range of $\Delta \theta$ values used to select candidate windows for perturbation; we consider a width $\pm \delta = \pi/6$ around each peak. \textbf{(B)} The distribution of state durations $L_{s}$ with the mean value $\langle L_{s} \rangle$ printed inside the panel. \textbf{(C)} The heights of the blue bars show the estimated switching probabilities $P_{\mathrm{switch}}$ for pulse inputs applied at different onset phases $\theta_{\mathrm{on}}$. A successful state transition is said to occur if the sign of the interareal phase relation switches within 2 cycles and the new configuration lasts for at least 5 more cycles following the initial switch. The pulse has amplitude $A_{s} = 0.75$ and duration $T_{s} = 4.5$ ms. The red horizontal line indicates the estimated likelihood of observing the same behavior spontaneously, i.e., in the absence of a perturbation. \textbf{(D)} Across all onset phases $\theta_{\mathrm{on}}$, the blue line indicates the maximum switching probability $\mathrm{max}[P_{\mathrm{switch}}]$. A successful state transition is said to occur if the sign of the interareal phase relation switches within 1 cycle and the new configuration lasts for at least 5 more cycles following the initial switch. The perturbation has amplitude $A_{s} = 0.75$ and duration $T_{s} = 4.5$ ms. The red curve shows the estimated likelihood of observing a spontaneous switch that lasts for at least 5 periods and that occurs within a varying number of oscillation cycles. \textbf{(E)} Activity time-series showing a failed state-switch after applying a pulse perturbation to the yellow area at the time denoted by the black bar (pulse parameters are $A_{s} = 0.75$, $T_{s} = 4.5$ ms, $\theta_{\mathrm{on}} = 1.25$ rad).}
	\label{f:102220201220_pulseStim}
\end{figure}

\subsubsection{4-area networks: Stochastic scenario}
\label{s:4area_stochastic_pulse}

In the preceding subsection, we identified conditions under which local perturbations could modulate collective states in 2-area networks in the presence of noise. Building upon those analyses, we now ask whether perturbation-induced switching between different transiently-stable phase-locking patterns can still be achieved in the slightly more complex case of a 4-area network driven by stochastic background inputs.

We characterized the baseline behavior of 4-region brain circuits in the presence of noisy background drive in Sec.~\ref{s:4area_stochastic_baseline}. There we saw that depending on the combination of the network and noise parameters, the system could naturally cycle through approximate forms of the multistable phase-locking patterns than manifest in the noiseless limit. Those are the regimes of interest for our current investigation, where we ask whether targeted external inputs can still control noisy collective states that spontaneously transition among one another at baseline. In order to investigate this question, the first step is to define a protocol that can be used to identify time windows during which the network resides in dynamical configurations reminiscent of the perfectly phase-locked states seen in the deterministic scenario (i.e, those in Fig.~\ref{f:4node_baseline}). 

To begin, we run a long, 150-second simulation of the dynamics without any perturbations. To extract time periods during which network activity is near one of the six multistable states $\mathbf{S_{j}}$ that arise in the deterministic case, we make use of what the known phase relations are for each of those configurations. Specifically, in the absence of noise, each collective state corresponds to the four phases being distributed uniformly around the unit circle in a particular order (see Fig.~\ref{f:4node_baseline}). Thus, the phase difference between a given pair of regions will always take on one of three values, such that $\Delta \theta_{i,j} \in \{-\pi/2, \pi, \pi/2\}$ for $i \neq j$. For example, in state $\mathbf{S_{1}}$, we have: $\Delta \theta_{1,2} = \pi$, $\Delta \theta_{1,3} = -\pi/2$, $\Delta \theta {1,4} = \pi/2$, $\Delta \theta{2,3} = \pi/2$, $\Delta \theta{2,4} = -\pi/2$, and $\Delta \theta_{3,4} = \pi$. Similar phase relations can be defined for the five other unique states. To determine if the system exists in a blurred version of state $\mathbf{S_{j}}$ when noise is present, we allow for some spread around the six noiseless phase differences. That is, if $\Delta \theta_{1,2}^{*,j}$ is the phase difference between unit $1$ and $2$ in the noiseless version of state $\mathbf{S_{j}}$, then we would say that $\Delta \theta_{1,2}$ from the stochastic simulation should fall within $\Delta \theta_{1,2}^{*,j} - 2\delta < \Delta \theta_{1,2} < \Delta \theta_{1,2}^{*,j} + 2\delta $ (and analogously for the other five phase relations at the same time) in order for the system to be considered in a noisy version of $\mathbf{S_{j}}$. In what follows, we use a default of $\delta = \pi/6$. 

To assess whether pulse perturbations have a meaningful effect on the circuit's dynamics, we next need to collect a set of candidate windows for applying the external input. The candidate set is determined by finding all windows during which the system remains in the same collective state for at least one cycle of the perturbed area's baseline oscillation. We then select 100 of those candidate windows at random for further analysis. In this way, we ask: given that the network is near one of its preferred attractors when it receives a perturbation, can the external input induce a change in the collective state? To answer this question, for each selected window we run a new simulation where the noise realization and initial conditions are kept the same, but an input pulse of amplitude $A_{s}$ and duration $T_{s}$ is applied to region 1 at different phases $\theta_{\mathrm{on}}$ of its ongoing oscillation. By continuing to let the dynamics evolve after receiving the perturbation, we analyze its effect on the network's state by determining \textit{(1)} if the perturbation caused the collective state to change to a new configuration within $n_{s}$ oscillation cycles after the perturbation onset, and if so, \textit{(2)} whether the new state is maintained for a subsequent $n_{m}$ cycles after the initial switch. Note that to determine a state transition, we only require that the relative ordering of the phases changes, but we do not put any further constraints on what the values of the phase differences must be afterwards. In what follows, we use default values $n_{s} = 2$ and $n_{m} = 5$; hence, we are interested in the ability to \textit{quickly} shift the network activity pattern and have the alteration last for a functionally-relevant length of time. 

To summarize the overall effect of an input pulse with certain parameters, we compute the fraction of the 100 selected windows for which the perturbation induces a phase-advancing (stimulated site moves ahead of the area initially preceding it) or phase-delaying (stimulated site moves behind the area initially lagging it) state transition. Note that these were the two types of state-transitions that we observed in studying the response of the noiseless 4-area circuit to brief input pulses (Sec.~\ref{s:4area_stochastic_pulse}). In what follows, we denote these two fractions as $P_{\mathrm{advance}}$ and $P_{\mathrm{delay}}$, and often refer to them as the ``phase-advancing transition probabilities" and the ``phase-delaying transition probabilities".  The final step is to compare the perturbation-induced switching probabilities to the chance of observing the same effects in the spontaneous dynamics. To make this comparison, we consider the same time windows used for the perturbation analysis, and calculate the fraction for which the phase-ordering transitions to a configuration that corresponds to a phase-advance or phase-delay of area 1. As before, a successful trial is one in which the switch occurs within $n_{s}$ cycles from the time the perturbation would have been applied and where the new configuration lasts for at least $n_{m}$ cycles afterward.

We are now prepared to study the effects of brief input pulses on state-switching in 4-area networks driven by a noisy environment. We consider a baseline operating point $\overline{P_{E,j}} = 1.3$ for $j \in \{1,...,4\}$, $T_{D} = 2.5$ ms, $G_{EE} = 0.2$, $\alpha = 5$, and $\sigma = 0.1$ (Working Point 1). With these parameters, one can easily resolve three clear peaks in the distribution $P(\Delta \theta_{1,2})$ of the pairwise phase differences between area 1 and area 2 (Fig.~\ref{f:4node_stochastic_pulseStim_100320201120}), and these peaks are centered close to what the stable phase relations would be in the absence of noise. (Note that the results are similar for any pair of units $i \neq j$). As discussed thoroughly during our characterization of the baseline dynamics of stochastic 4-region circuits, the presence of these modes signifies that the system dwells in and spontaneously moves between variants of the multistable collective states that are attractors in the noiseless version of the network (please see Sec.~\ref{s:4area_stochastic_baseline} for complete details). Using the protocol outlined above, we can thus extract time windows during which the network resides in one of those configurations, and test whether an external stimulation signal can cause a reconfiguration of the activity pattern. For context, we note that for the chosen working point, it takes an average of $ \langle L_{s} \rangle \approx 740$ ms for the relative ordering of the phases to change spontaneously. Hence, it is pertinent to understand whether external inputs could bring about more rapid state shifts.

Our analyses of the stochastic network indicate that aptly-timed perturbations can still bring about state transitions with efficacies well above chance levels (Fig.~\ref{f:4node_stochastic_pulseStim_100320201120}). To see this, first consider Fig.~\ref{f:4node_stochastic_pulseStim_100320201120}D, which shows the estimated phase-advancing transition probability $P_{\mathrm{advance}}$ as a function of the onset phase $\theta_{\mathrm{on}}$ (the pulse amplitude is $A_{s} = 1.5$ and the pulse duration is $T_{s} = 4$ ms). Of note is that for inputs received within a certain band of oscillation phases, $P_{\mathrm{advance}}$ peaks at levels that reach $\approx$ 0.5. In contrast, the likelihood that the same transitions would be observed spontaneously is zero; this is because when the network resides near one of its collective attractors, it takes much longer than just a couple of oscillation cycles for the background noise to elicit a state switch. Compounding on this, there would be no guarantee that the next state observed in the natural dynamics should be the one that corresponds to region 1 undergoing a 1-hop phase-advance. Thus, although the perturbation-induced phase-advancing transition probability is only about 50\% at its maximum, this is still quite significant given the nature of the spontaneous dynamics. Fig.~\ref{f:4node_stochastic_pulseStim_100320201120}B offers an example of network activity just before and after a perturbation to the yellow area that induces a successful phase-advance state transition. Before the external pulse arrives, the relative phase-order is $1 \rightarrow 3 \rightarrow 4 \rightarrow 2 \rightarrow 1$. When region 1 receives the stimulation, its oscillation undergoes a transient shifting that pushes it ahead of region 2. This in turn establishes the new phase-order $1 \rightarrow 2 \rightarrow 3 \rightarrow 4 \rightarrow 1$. Although the new state is transient and the corresponding phase relations imperfect, the effect of the perturbation may still be enough to briefly alter or disrupt a computation that was potentially enabled by the previous collective pattern. 

In the deterministic limit, strong and long enough pulses could additionally trigger state transitions wherein the stimulated site underwent a phase-delay relative to its neighboring oscillation. Here we find that this effect can also be induced in the noisy system with likelihoods significantly above chance (Fig.~\ref{f:4node_stochastic_pulseStim_100320201120}G). As before, the stimulation must be carefully timed relative to the targeted area's rhythm, but if this is the case, then the phase-delay transition probability $P_{\mathrm{delay}}$ can also reach maximum values near 50\% (at least for the shown pulse parameters $A_{s} = 1.5$ and $T_{s} = 4$ ms). An example of a successful trial in which this transition occurs is depicted in Fig.~\ref{f:4node_stochastic_pulseStim_100320201120}C; the external input causes a second peak in the yellow area's oscillation that then leads to an overall 1-hop delay and change in the phase-order to the new state $1 \rightarrow 4 \rightarrow 2 \rightarrow 3 \rightarrow 1$.

We also study how the estimated state transition probabilities $P_{\mathrm{advance}}$ and $P_{\mathrm{delay}}$ vary as a function of the pulse amplitude and duration (Figs.~\ref{f:4node_stochastic_pulseStim_100320201120}E,F,H,I). It is again no surprise that these heat maps look like probabilistic versions of Figs.~\ref{f:4node_deterministic_pulseStim}D,E. While we do see that there is some variation of the maximum values of the phase-advancing and phase-delaying transition probabilities as a function of the pulse amplitude and duration, the onset phase is critical to triggering reliable state switches. Importantly, we find that $P_{\mathrm{advance}}$ and $P_{\mathrm{delay}}$ can exceed 0.3 -- which is well above the spontaneous switching probability -- across a range of $A_{s}$-$T_{s}$ combinations, so long as the timing of the pulse is within an optimal range.

\begin{FPfigure}
	\centering
	\includegraphics[width=1\textwidth]{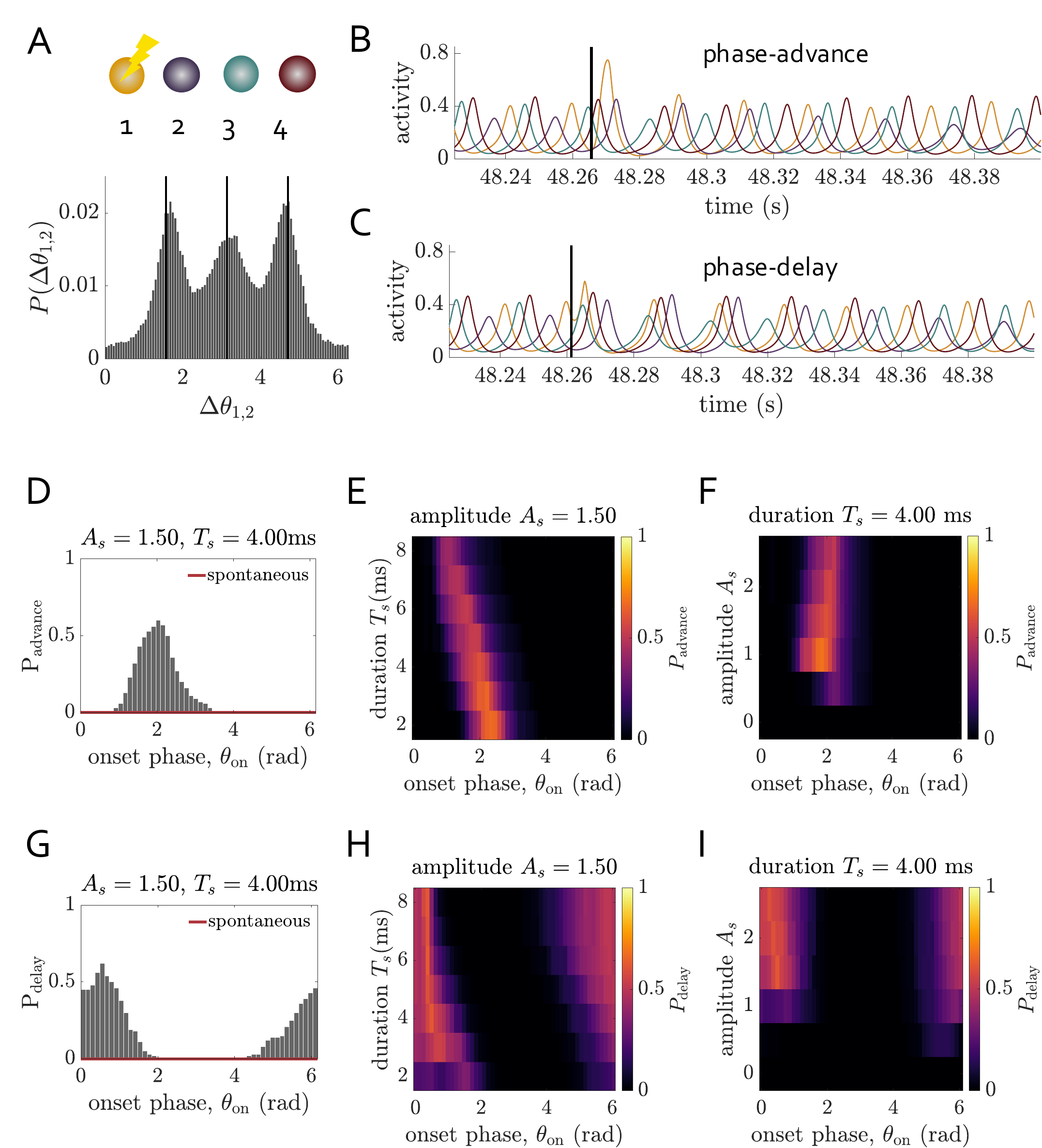}
	\caption[Response of a stochastic 4-area network to brief perturbation pulses applied to area 1: Working Point 1.]{\textbf{Response of a stochastic 4-area network to brief perturbation pulses applied to area 1: Working Point 1}. The key network parameters are: $\overline{P_{E}} = 1.3$, $T_{D} = 2.5$ ms, $G_{EE} = 0.2$, $\alpha = 5$, $\sigma = 0.1$. \textbf{(A)} The baseline distribution $P(\Delta \theta_{1,2})$ of phase differences $\Delta \theta_{1,2}$ between areas 1 and 2 across a long simulation. The black lines indicate the values of the stable phase differences in the absence of noise. \textbf{(B)} Activity time-series from a trial where a successful phase-advance transition is induced by applying a pulse perturbation to the yellow area at the time denoted by the black bar (pulse parameters are $A_{s} = 1.5$, $T_{s} = 4.0$ ms, $\theta_{\mathrm{on}} = 2.0$ rad). \textbf{(C)} Activity time-series from a trial where a successful phase-delay transition is induced by applying a pulse perturbation to the yellow area at the time denoted by the black bar (pulse parameters are $A_{s} = 1.5$, $T_{s} = 4.0$ ms, $\theta_{\mathrm{on}} = 0.5$ rad). \textbf{(D)} The estimated phase-advancing transition probability $P_{\mathrm{advance}}$ induced by pulse inputs applied at different onset phases $\theta_{\mathrm{on}}$. A successful state transition occurs if the ordering of the phases switches to the correct configuration within 2 cycles and the new ordering lasts for at least 5 more cycles following the initial switch. The red horizontal line indicates the estimated likelihood of observing the same behavior spontaneously, i.e., in the absence of a perturbation. The pulse parameters are $A_{s} = 1.5$, $T_{s} = 4.0$ ms. \textbf{(E)} The estimated phase-advancing transition probability $P_{\mathrm{advance}}$ as a function of the pulse onset phase $\theta_{\mathrm{on}}$ and the pulse duration $T_{s}$, for fixed amplitude $A_{s} = 1.5$. A successful state transition occurs if the ordering of the phases switches to the correct configuration within 2 cycles and the new ordering lasts for at least 5 more cycles following the initial switch. \textbf{(F)} The estimated phase-advancing transition probability $P_{\mathrm{advance}}$ as a function of the pulse onset phase $\theta_{\mathrm{on}}$ and the pulse amplitude $A_{s}$, for fixed duration $T_{s} = 4.0$ ms. A successful state transition occurs if the ordering of the phases switches to the correct configuration within 2 cycles and the new ordering lasts for at least 5 more cycles following the initial switch. \textbf{(G)} The estimated phase-delaying transition probability $P_{\mathrm{delay}}$ induced by pulse inputs applied at different onset phases $\theta_{\mathrm{on}}$. A successful state transition occurs if the ordering of the phases switches to the correct configuration within 2 cycles and the new ordering lasts for at least 5 more cycles following the initial switch. The red horizontal line indicates the estimated likelihood of observing the same behavior spontaneously, i.e., in the absence of a perturbation. The pulse parameters are $A_{s} = 1.5$, $T_{s} = 4.0$ ms. \textbf{(H)} The estimated phase-delaying transition probability $P_{\mathrm{delay}}$ as a function of the pulse onset phase $\theta_{\mathrm{on}}$ and the pulse duration $T_{s}$, for fixed amplitude $A_{s} = 1.5$. A successful state transition occurs if the ordering of the phases switches to the correct configuration within 2 cycles and the new ordering lasts for at least 5 more cycles following the initial switch. \textbf{(I)} The estimated phase-delaying transition probability $P_{\mathrm{delay}}$ as a function of the pulse onset phase $\theta_{\mathrm{on}}$ and the pulse amplitude $A_{s}$, for fixed duration $T_{s} = 4.0$ ms. A successful state transition occurs if the ordering of the phases switches to the correct configuration within 2 cycles and the new ordering lasts for at least 5 more cycles following the initial switch.}
	\label{f:4node_stochastic_pulseStim_100320201120}
\end{FPfigure}

Though we do not have the computational resources to exhaustively explore how the circuit's baseline parameters affect the ability of local stimulation to control network states, we do study one direction of interest, which is the effect of enhancing the noise strength. In particular, we increase $\sigma$ to 0.125 while keeping all other parameters identical (Working Point 2). Crucially, increasing the noise in this manner does not completely wipe-out the structure of the collective activity patterns that we wish to investigate in the first place, but it does cause a smearing and reduction of the height of the peaks in the distribution $P(\Delta \theta_{1,2})$ (Fig.~\ref{f:4node_stochastic_pulseStim_111020201320}A; note that the conclusions are similar for any pair of units $i \neq j$). As determined in our analysis of the baseline network dynamics (Sec.~\ref{s:4area_stochastic_baseline}), increasing $\sigma$ also reduces the time the system tends to spend in one phase ordering before switching. In particular, here $\langle L_{s} \rangle$ decreases to $\approx$ 428 ms. Nonetheless, we can still use our methodology to select time-windows when the collective dynamics are in a well-defined state, and then test whether perturbations can robustly kick the activity pattern out of that configuration.

Carrying out the same set of analyses as for the previous working point, we find qualitatively similar results for the higher-noise setting (Fig.~\ref{f:4node_stochastic_pulseStim_111020201320}B--I). In particular, well-timed stimulation pulses can induce both rapid phase-advance and phase-delay state transitions with efficacies that exceed the spontaneous levels. As expected, we do also observe an overall reduction in $P_{\mathrm{advance}}$ and $P_{\mathrm{delay}}$ across the considered range of pulse parameters when compared to the less noisy regime. This finding simply quantifies the fact that state modulation becomes more difficult when the network operates in a more stochastic background environment. Nonetheless, we can still conclude that there is a range of noise levels for which perturbations can continue to have a meaningful influence over system dynamics. Though we do not follow-up on these points further, it may be possible to slowly increase the noise level until the perturbation-induced transition probabilities are reduced to near chance levels. Indeed, it may be interesting to see if this crossover regime occurs before a complete merging of the three peaks in the baseline phase difference distribution.

The final point we wish to make is that the efficacy of state-control also depends on the criteria used to select the perturbation windows. As the width parameter $\delta$ is increased, the maximum state transition probabilities across all onset phases tend to decrease for a given fixed combination of the pulse amplitude and duration (see Fig.~\ref{f:4node_stochastic_pulseStim_100320201120_varyDelta}). In other words, when network phase relations are more variable across the set of perturbation windows, it becomes more difficult to induce fast and sustained reconfigurations of the collective phase ordering. One reason this might occur is that if the set of phase relations is slightly different across the perturbation windows, then the optimal range of onset phases for inducing a state switch may also be slightly different from window to window. In turn, the state transition probabilities at a fixed $\theta_{\mathrm{on}}$ are likely to be smaller for larger $\delta$. It is also generally reasonable to expect that perturbations applied during noisier time-windows will simply have less reliable effects than those applied in the middle of a time segment when the network is very close to one of the attractors from the noiseless system. Indeed, whether or not a perturbation yields a lasting change to the circuit's activity pattern also depends on the evolution of the system after the perturbation is turned off. In this way, stimulation effects will be cleaner if the stimulation is enacted in the midst of a time window when the stochastic system more closely resembles the deterministic one. Decreasing $\delta$ is more likely to select perturbation windows where that is the case. Importantly, windows where the phase configuration is tighter also constitute times when the network is very unlikely to change on its own, and are thus arguably when a perturbation may be most needed.

\begin{FPfigure}
	\centering
	\includegraphics[width=1\textwidth]{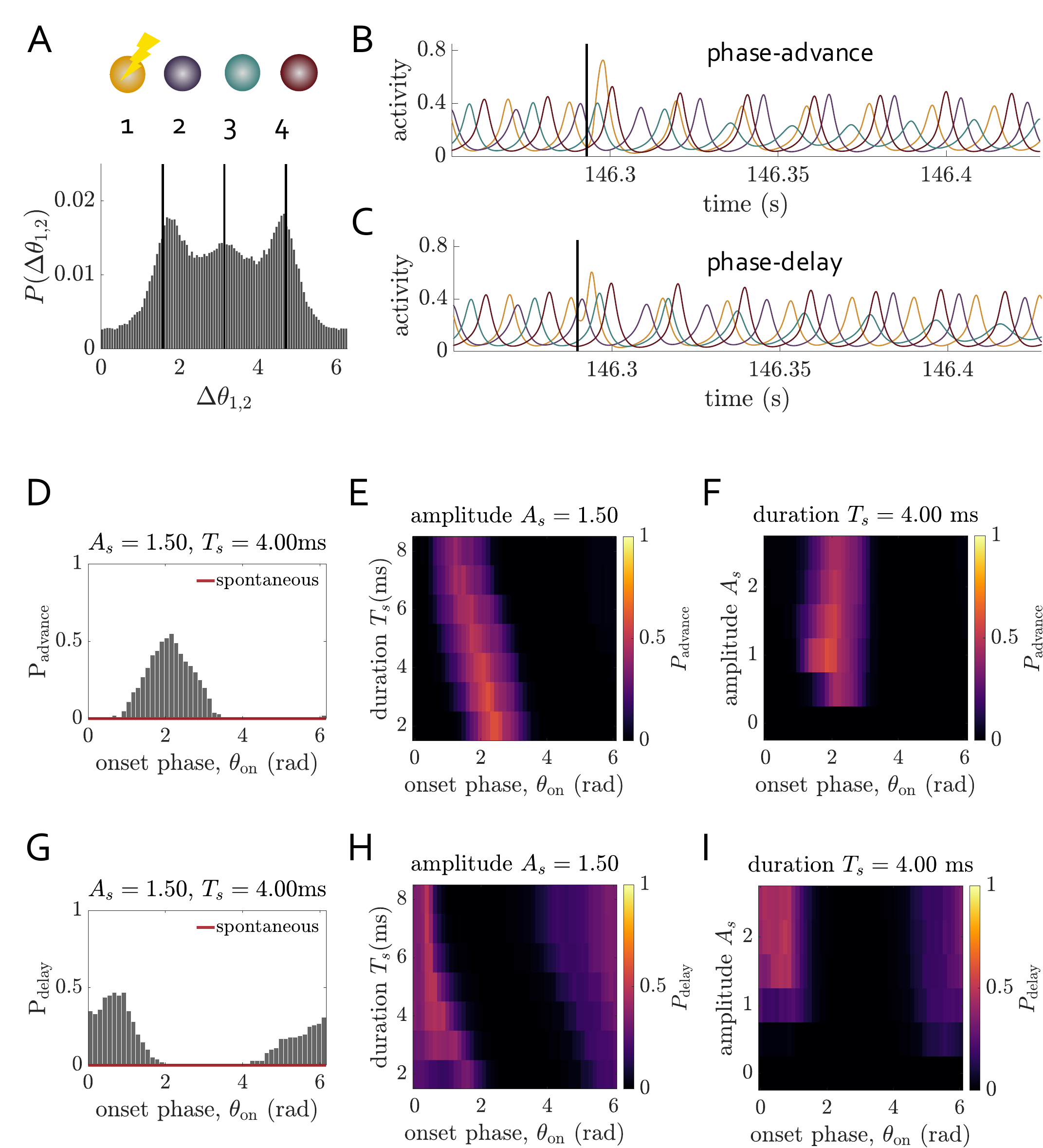}
	\caption[Response of a stochastic 4-area network to brief perturbation pulses applied to area 1: Working Point 2.]{\textbf{Response of a stochastic 4-area network to brief perturbation pulses applied to area 1: Working Point 2.} The key network parameters are: $\overline{P_{E}} = 1.3$, $T_{D} = 2.5$ ms, $G_{EE} = 0.2$, $\alpha = 5$, $\sigma = 0.125$. \textbf{(A)} The baseline distribution $P(\Delta \theta_{1,2})$ of phase differences $\Delta \theta_{1,2}$ between areas 1 and 2 across a long simulation. The black lines indicate the values of the stable phase differences in the absence of noise. \textbf{(B)} Activity time-series from a trial where a successful phase-advance transition is induced by applying a pulse perturbation to the yellow area at the time denoted by the black bar (pulse parameters are $A_{s} = 1.5$, $T_{s} = 4.0$ ms, $\theta_{\mathrm{on}} = 1.8$ rad). \textbf{(C)} Activity time-series from a trial where a successful phase-delay transition is induced by applying a pulse perturbation to the yellow area at the time denoted by the black bar (pulse parameters are $A_{s} = 1.5$, $T_{s} = 4.0$ ms, $\theta_{\mathrm{on}} = 0.6$ rad). \textbf{(D)} The estimated phase-advancing transition probability $P_{\mathrm{advance}}$ induced by pulse inputs applied at different onset phases $\theta_{\mathrm{on}}$. A successful state transition occurs if the ordering of the phases switches to the correct configuration within 2 cycles and the new ordering lasts for at least 5 more cycles following the initial switch. The red horizontal line indicates the estimated likelihood of observing the same behavior spontaneously, i.e., in the absence of a perturbation. The pulse parameters are $A_{s} = 1.5$, $T_{s} = 4.0$ ms.  \textbf{(E)} The estimated phase-advancing transition probability $P_{\mathrm{advance}}$ as a function of the pulse onset phase $\theta_{\mathrm{on}}$ and the pulse duration $T_{s}$, for fixed amplitude $A_{s} = 1.5$. A successful state transition occurs if the ordering of the phases switches to the correct configuration within 2 cycles and the new ordering lasts for at least 5 more cycles following the initial switch. \textbf{(F)} The estimated phase-advancing transition probability $P_{\mathrm{advance}}$ as a function of the pulse onset phase $\theta_{\mathrm{on}}$ and the pulse amplitude $A_{s}$, for fixed duration $T_{s} = 4.0$ ms. A successful state transition occurs if the ordering of the phases switches to the correct configuration within 2 cycles and the new ordering lasts for at least 5 more cycles following the initial switch. \textbf{(G)} The estimated phase-delaying transition probability $P_{\mathrm{delay}}$ induced by pulse inputs applied at different onset phases $\theta_{\mathrm{on}}$. A successful state transition occurs if the ordering of the phases switches to the correct configuration within 2 cycles and the new ordering lasts for at least 5 more cycles following the initial switch. The red horizontal line indicates the estimated likelihood of observing the same behavior spontaneously, i.e., in the absence of a perturbation. The pulse parameters are $A_{s} = 1.5$, $T_{s} = 4.0$ ms. \textbf{(H)} The estimated phase-delaying transition probability $P_{\mathrm{delay}}$ as a function of the pulse onset phase $\theta_{\mathrm{on}}$ and the pulse duration $T_{s}$, for fixed amplitude $A_{s} = 1.5$. A successful state transition occurs if the ordering of the phases switches to the correct configuration within 2 cycles and the new ordering lasts for at least 5 more cycles following the initial switch. \textbf{(I)} The estimated phase-delaying transition probability $P_{\mathrm{delay}}$ as a function of the pulse onset phase $\theta_{\mathrm{on}}$ and the pulse amplitude $A_{s}$, for fixed duration $T_{s} = 4.0$ ms. A successful state switch occurs if the ordering of the phases switches to the correct configuration within 2 cycles and the new ordering lasts for at least 5 more cycles following the initial switch.}
	\label{f:4node_stochastic_pulseStim_111020201320}
\end{FPfigure}

\subsection{Rhythmic stimulation can induce state-morphing in multiarea networks}
\label{s:AC_stim}

In the previous section we studied the effects of brief, targeted perturbation pulses on the collective dynamics of multiarea brain circuits. We found that under certain conditions, these external inputs could trigger a complete reorganization of the phase-locking state to a topologically distinct functional configuration. We turn next to an examination of how network activity is modulated by sustained rhythmic stimulation that again targets a single region. As before, we begin by understanding the simplest scenarios of 2-area and 4-area circuits in the deterministic limit, and then move to the more intricate case of stochastic background drive. In what follows, our analyses will reveal that, depending on the nature of the external input and/or the baseline dynamics, the focal stimulation will either lead to a breakdown of the collective phase-locking states or will enable more fine-tuned control of the interareal phase relationships. 

\subsubsection{2-area networks: Deterministic limit}
\label{s:2node_deterministic_AC}

\begin{figure}
	\centering
	\includegraphics[width=\textwidth]{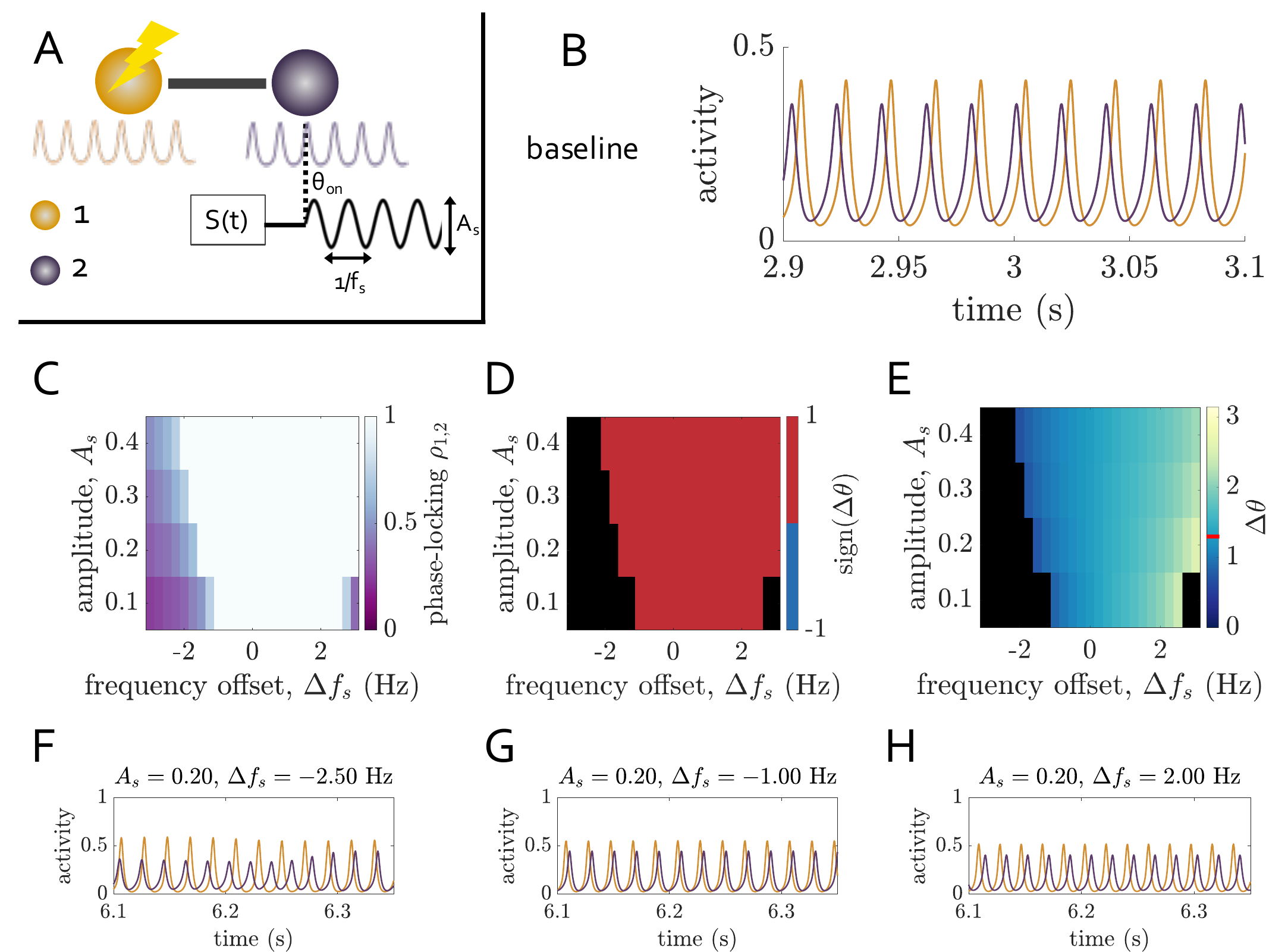}
	\caption[Response of a deterministic 2-area network to rhythmic stimulation.]{\textbf{Response of a deterministic 2-area network to rhythmic stimulation.} \textbf{(A)} Schematic of the model setup. We study the response of a deterministic 2-area brain circuit to a sinusoidal input of amplitude $A_{s}$ and frequency $f_{s}$. The perturbation targets only one of the two regions. The network parameters are fixed to $P_{E} = 1.35$, $G_{EE} = 0.2$, and $T_{D} = 1.5$ ms. \textbf{(B)} Activity time-series of the two regions under baseline conditions. The network locks into collective state $\mathbf{S_{2}}$, where area 2 leads area 1 in phase. \textbf{(C)} The phase-locking value $\rho_{1,2}$ between the two areas as a function of the stimulation frequency offset $\Delta f_{s}$ and the stimulation amplitude $A_{s}$. \textbf{(D)} The sign of the phase difference between the two regions' activity as a function of the stimulation frequency offset $\Delta f_{s}$ and the stimulation amplitude $A_{s}$. \textbf{(E)} The value of the phase difference between the two regions' activity as a function of the stimulation frequency offset $\Delta f_{s}$ and the stimulation amplitude $A_{s}$. The red bar on the color scale indicates the value of the phase difference at baseline. \textbf{(F)} Activity time-series showing the effects of stimulating the yellow area with $A_{s} = 0.2$ and $\Delta f_{s} = -2.5$ Hz. \textbf{(G)} Activity time-series showing the effects of stimulating the yellow area with $A_{s} = 0.2$ and $\Delta f_{s} = -1.0$ Hz. \textbf{(H)} Activity time-series showing the effects of stimulating the yellow area with $A_{s} = 0.2$ and $\Delta f_{s} = 2.0$ Hz.}
	\label{f:2node_deterministic_ACStim}
\end{figure}

We start by investigating the ability of focal, rhythmic stimulation to control the dynamical state of 2-region networks subject to deterministic background drive. As before, we set model parameters such that regional activity oscillates in time and such that the network phase-locks into a particular lead-lag configuration at baseline. For the following analysis, we use $P_{E,j} = 1.35$ for $j \in \{1,2\}$, $G_{EE} = 0.2$, and $T_{D} = 1.5$ ms. Fig.~\ref{f:2node_deterministic_ACStim}B shows a brief segment of the baseline network activity for these parameters; for this example, the system falls into state $\mathbf{S_{2}}$, for which area 2 leads area 1 in phase. The external stimulation is introduced by injecting a sinusoidal drive of amplitude $A_{s}$ and frequency $f_{s}$ into the excitatory pool of one region at an onset phase $\theta_{\mathrm{on}}$ (see Sec.~\ref{s:rhythmic_input_description} for full details). Here we assume that the stimulation targets area 1 (Fig.~\ref{f:2node_deterministic_ACStim}A).

We begin by simulating the network dynamics for 2 seconds with no stimulation applied, which allows the system to settle into a phase-locked attractor (here state $\mathbf{S_{2}}$). After the system has relaxed to an equilibrium, the rhythmic stimulation is initiated the next time the phase of the receiving area's intrinsic oscillation is equal to $\mathbf{\theta_{on}}$, and it remains present for the rest of the simulation. To assess the dependence on the parameters of the external input, we repeat this protocol for 50 onset phases evenly spaced in the range $[0,2\pi]$, and for a range of stimulation frequencies and amplitudes.

The effects of the external input on the network's collective dynamical state are assessed in two ways. We first determine whether circuit activity remains phase-locked in the presence of the stimulation by computing the PLV between the two regions across a 2-second window that begins 2 seconds after the stimulation is first turned on. This is important to understand, as we are interested in when coherent dynamics exist in the first place. If the network does remain in a locked configuration, we then measure the value and sign of the phase difference $\Delta \theta$ between the two areas over the same time segment. Both of these latter two quantities are then compared against their values in the baseline scenario. Note that in what follows, we quantify various results in terms of the frequency offset $\Delta f_{s} = f_{s} - f_{o}$ between the stimulation $f_{s}$ and intrinsic $f_{o}$ oscillation frequency. 

Fig.~\ref{f:2node_deterministic_ACStim}C shows the phase-locking value $\rho_{1,2}$ between the two units in the network for different stimulation amplitudes and frequency offsets. Note first that the system remains phase-locked for a larger range of $\Delta f_{s}$ as the amplitude increases. Also observe that it is generally easier for the network to remain in a coherent state for positive values of $\Delta f_{s}$ than for negative ones. This fact implies that phase-locking is more easily maintained when the perturbed unit is driven at frequencies slightly greater than its baseline frequency compared to frequencies slightly less than its baseline frequency. Although we do not study a wider set of $\Delta f_{s}$ values here, we would expect the PLV to eventually drop off for large enough positive values as well. Importantly, outside of the locking region indicated by $\rho_{1,2} = 1$, the external stimulation actually destroys the functional connectivity states that we set out to examine in the first place. An example of such a breakdown of coherent activity is shown in Fig.~\ref{f:2node_deterministic_ACStim}F. While the ability to steer a multiarea circuit into an entirely new dynamical regime with focal inputs is important and interesting, we leave further investigation of this direction to future work.

We next consider the direction of the phase relation between the areas inside the phase-locked region. Recall that prior to the stimulation, the network exists in state $\mathbf{S_{2}}$, where area 2 leads area 1 ($\Delta \theta < 0$). When area 1 receives the localized external input, however, we find that the lead-lag configuration is consistently altered such that region 1 leads region 2 ($\Delta \theta > 0$), and that this result is independent of the stimulation amplitude, frequency offset, and onset phase (Fig.~\ref{f:2node_deterministic_ACStim}D). In other words, when the rhythmic stimulation is ongoing, its presence breaks the symmetry of the system and causes the perturbed area to become the persistent phase-leader in parameter ranges where phase-locking is maintained. Perhaps of more interest is the fine-tuned modulation of the phase difference enabled by varying the frequency offset $\Delta f_{s}$. This effect is depicted in Fig.~\ref{f:2node_deterministic_ACStim}E, which shows the phase difference $\Delta \theta$ between the two areas as a function of the stimulation amplitude and frequency offset. Here we observe that increasing $\Delta f_{s}$ widens the phase difference, whereas decreasing $\Delta f_{s}$ has the opposite effect. Hence, for a given amplitude, a large range of phase relations is possible via manipulation of the frequency offset. Examples of the activity time-series for negative and positive values of $\Delta f_{s}$ are shown in Figs.~\ref{f:2node_deterministic_ACStim}G and H, respectively.

To summarize, we find that for the 2-area circuit, locally-applied rhythmic stimulation first breaks the symmetry of the system and biases the dynamics into a specific phase ordering. Second, within that constraint, the spatial configuration of the areas' oscillations can be varied by tuning the stimulation frequency. We refer to this latter type of regulation of the collective phase-locking state as ``state-morphing".

\subsubsection{4-area networks: Deterministic limit}
\label{s:4node_deterministic_AC}

\begin{FPfigure}
	\centering
	\includegraphics[width=\textwidth]{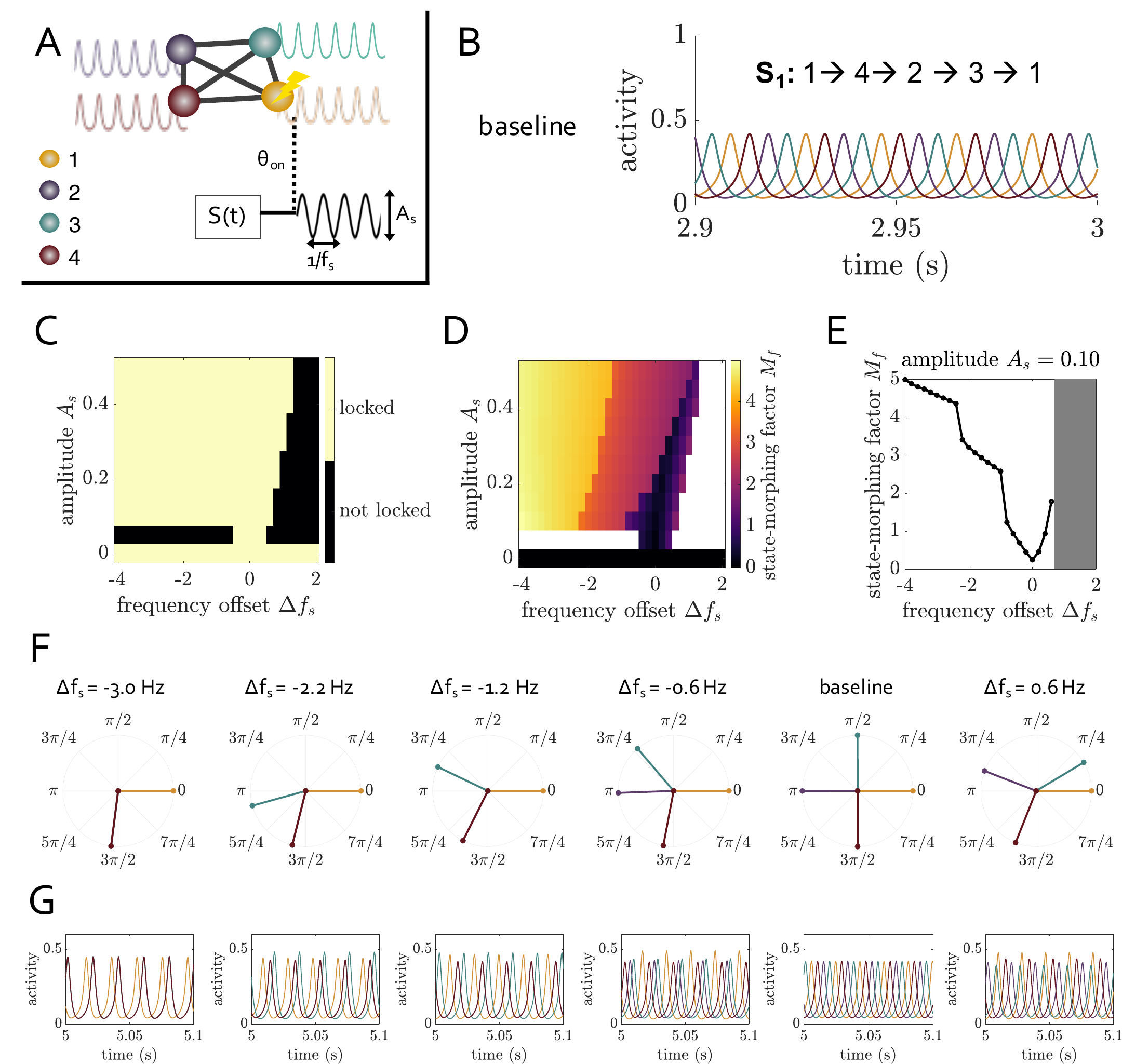}
	\caption[Response of a deterministic 4-area network to rhythmic stimulation.]{\textbf{Response of a deterministic 4-area network to rhythmic stimulation.} \textbf{(A)} Schematic of the model setup. We study the response of a deterministic 4-area brain circuit to a sinusoidal input of amplitude $A_{s}$ and frequency $f_{s}$. The perturbation targets only one of the four regions, which in this case is area 1 (yellow). The network parameters are: $P_{E} = 1.325$, $G_{EE} = 0.2$, and $T_{D} = 2.5$ ms. \textbf{(B)} Activity time-series of the four regions under baseline conditions. The network locks into collective state $\mathbf{S_{1}}$. \textbf{(C)} The yellow area indicates where the network is collectively phase-locked as a function of the stimulation frequency offset $\Delta f_{s}$ and the stimulation amplitude $A_{s}$. \textbf{(D)} The state-morphing factor $M_{f}$ as a function of the stimulation frequency offset $\Delta f_{s}$ and the stimulation amplitude $A_{s}$. The white areas correspond to parameters where the circuit is not phase-locked. \textbf{(E)} The state-morphing factor $M_{f}$ as a function of the stimulation frequency offset $\Delta f_{s}$ for a fixed stimulation amplitude $A_{s} = 0.1$. In the gray area, network phase-locking breaks down due to the stimulation. \textbf{(F)} Relative phase configurations of the network for different frequency offsets $\Delta f_{s}$ for a fixed amplitude $A_{s} = 0.1$. \textbf{(G)} Snapshots of the activity time-series for different frequency offsets $\Delta f_{s}$ for a fixed amplitude $A_{s} = 0.1$.}
	\label{f:4node_deterministic_ACStim}
\end{FPfigure}

In this section we examine how the collective states of the deterministic 4-area network respond to sustained, focally-applied rhythmic stimulation. For the baseline parameters of the circuit, we use the same ones as employed in Sec.~\ref{s:4area_deterministic_pulseStim}: $P_{E,j} = 1.325$ for $j \in \{1,...,4\}$, $G_{EE} = 0.2$, and $T_{D} = 2.5$ ms. For these values, population activity phase-locks, and depending on the initial conditions, the lead-lag relations correspond to one of six unique collective states (see Sec.~\ref{s:4area_deterministic}, Fig.~\ref{f:4node_baseline}E for details). In what follows, the system is prepared in state $\mathbf{S_{1}}$, and is then perturbed from that starting position. Fig.~\ref{f:4node_deterministic_ACStim}B shows a snapshot of the network's activity pattern at baseline, from which one observes the phase-ordering $1 \rightarrow 4 \rightarrow 2 \rightarrow 3 \rightarrow 1$. As for the 2-area circuit, system dynamics are perturbed via a sinusoidal input to the excitatory subpopulation of area 1 (yellow unit), and we study the effects of the frequency $f_{s}$ of the external signal, its amplitude $A_{s}$, and the phase of the ongoing oscillation $\theta_{\mathrm{on}}$ at which the stimulation is initiated (see Sec.~\ref{s:modeling_perturbations} for complete details). A schematic of the system setup is depicted in Fig.~\ref{f:4node_deterministic_ACStim}A.

To understand how sustained rhythmic input may reshape or modulate the networks' phase-locked activity patterns, we first map the conditions under which coherent dynamics are maintained in the presence of the stimulation. To determine these conditions, we scan over a range of stimulation frequencies and amplitudes, and from those simulations, compute the phase-locking value between all regions to determine if the network remains in a locked state for each parameter combination. More precisely, for a given stimulation frequency and amplitude, system dynamics are first allowed to settle into state $\mathbf{S_{1}}$ across a 2-second window during which no external input is applied. After this relaxation time period, the stimulation is activated the next time the oscillation phase of the receiving area is $\approx \pi$, after which the stimulation remains on for the remainder of the simulation. We then compute the PLV between all region pairs across a 2-second window that begins 2 seconds after the stimulation onset, again in order to allow network dynamics to reequilibrate. 

The ``locking region" of the circuit (i.e., combinations of the frequency offset $\Delta f_{s}$ and amplitude $A_{s}$ for which the mean PLV over all region pairs exceeds a high threshold of 0.99) is presented in Fig.~\ref{f:4node_deterministic_ACStim}C. Interestingly, this analysis reveals that while the network can indeed remain in a phase-locked state for certain external inputs, the locking region is skewed towards lower values of $\Delta f_{s}$. In other words, network-wide locking can be maintained at lower stimulation amplitudes when the stimulation frequency is slightly \textit{smaller} than the baseline oscillation speed. Note that this does not mean the stimulated site itself cannot lock to higher frequency input, but that the network as a whole cannot remain in a coherent configuration if the external rhythm is too fast. Hence, the leftward skew in Fig.~\ref{f:4node_deterministic_ACStim}C is really a property of the entire coupled system, rather than of a single WC unit. Understanding this phenomenon on a deeper level would be an interesting direction for future work, though we do not explore it further here. A second albeit expected point is that, as the stimulation amplitude is increased, the range of frequency offsets across which phase-locking is maintained expands. This is especially noticeable for $\Delta f_{s} > 0$. 

Having delineated the locking-region for the network under focal, rhythmic stimulation, we are now in a position to examine if and how the actual phase relations among the neural populations are modulated by the external drive. In general, we find that the rhythmic input can have two distinct effects on the collective state. First, if the stimulation is turned on at particular onset phases of the ongoing oscillation, then it can induce a change in the phase-ordering in a manner similar to the state-switching that occurs for the input pulses studied in Sec.~\ref{s:4area_deterministic_pulseStim} (not shown). Note that this effect occurs due to a transient shifting of the regional oscillations as they first react to the turning-on of the stimulation. Importantly, though, even if the onset phase is such that no switching is induced, the sustained rhythmic input still causes a separate modulation of the phase-locking pattern. Specifically, it induces a shifting -- or ``morphing" -- of the spatial configuration of the phase relations relative to their geometric arrangement at baseline (Fig.~\ref{f:4node_deterministic_ACStim}F). That is, once a phase-ordering is established (whether or not it is different from that which existed prior to the stimulation onset), the external drive further regulates the actual phase differences between each pair of areas such that the $\Delta \theta_{ij}$'s between consecutive areas $i$ and $j\neq i$ are no longer necessarily equal to $\pm \pi/2$. We again refer to this more fine-tuned realignment of the spatial layout of the oscillation phases within a particular phase-ordering as ``state-morphing''.

To quantify the extent of state-morphing induced by a rhythmic external input to one region in the circuit, we define a quantity that we term the ``state-morphing factor", $M_{f}$. In doing so, we first reiterate that ``state-switching" refers to a change in the ordering of the phases relative to one another (i.e. it corresponds to a topological rearrangement of the functional state that depends on the area labels). In contrast, we would like our notion of state-morphing to be a measure of how much the \textit{geometrical} organization of the phases differs between baseline and the case of oscillatory external drive in a manner that is independent of the original node labeling/ordering. To define such a quantity, we let $\mathbf{S_{j}^{s}}$ for $j \in \{1,...,6\}$ label the state of the system under focal stimulation (after transient effects have been discarded), where $j \in \{1,...,6\}$ indicates the ordering of the units' activity peaks (or phases). $\mathbf{S_{j}^{s}}$ has a corresponding baseline state $\mathbf{S_{j}}$ with the same circular ordering of the phases. In addition to this topological information, we also tabulate the relative spatial positions of the phases in each state $\mathbf{S_{j}^{s}}$ and $\mathbf{S_{j}}$ relative to the same reference area (e.g., the stimulated region 1). In particular, because the areas are phase-locked under both baseline and stimulation conditions, we perform this tabulation of the phase positions in the reference frame that rotates with the same angular velocity as the system under study and oriented such that region 1 is always located at a phase of zero radians in that frame. Doing this for the stimulation condition, we can then define a set of phases $\{\theta^{s'}_{i,j}\}$ $\in [0,2\pi]$ for $i \in \{1,...,4\}$ that indicate the angular positions of the four areas in state $\mathbf{S_{j}^{s}}$ relative to $\theta^{s'}_{1,j} = 0$. The same procedure can be carried out for the corresponding baseline state $\mathbf{S_{j}}$ with the same phase-ordering, yielding another set of relative phases $\{\theta^{'}_{i,j}\}$ $\in [0,2\pi]$ for $i \in \{1,...,4\}$, again relative to $\theta^{'}_{1,j} = 0$. Once the two sets of phases are computed, we calculate the state morphing factor (for fixed $j$) as

\begin{equation}
M_{f} = \sum_{i=1}^{N}|\theta^{s'}_{i,j} - \theta^{'}_{i,j}|.
\label{eq:morphing_factor}
\end{equation}
~\\
This quantity summarizes the extent of the spatial shifting of the phase configuration under focal stimulation relative to the arrangement at baseline with the same topological ordering of the phases.

In Fig.~\ref{f:4node_deterministic_ACStim}D, we show the state-morphing factor (within the phase-locked region) as a function of the stimulation amplitude $A_{s}$ and the frequency offset $\Delta f_{s}$. This analysis indicates that the extent of state-morphing is primarily modulated by the frequency offset, which is especially clear by examining values of $\Delta f_{s} < 0$ for fixed amplitude. In particular, we find that as $\Delta f_{s}$ decreases from zero, the amount of morphing of the collective state increases. This behavior indicates that the frequency of external input is a control parameter that tunes how much the geometric configuration of the phases changes relative to the baseline configuration. Also note, though, that near a frequency of $\Delta f_{s} = 0$, for example, that state-morphing can be induced simply by increasing the amplitude of the stimulation. Thus, both stimulation parameters can play a role in reorganizing the collective phase-locking state.

Going back to the dependence of $M_{f}$ on the frequency offset $\Delta f_{s}$, we observe that especially for smaller amplitudes, there are three main morphing regimes highlighted by the three main blocks of color in Fig.~\ref{f:4node_deterministic_ACStim}D. To see these regimes even more clearly, we consider a cross-section of the state morphing factor for fixed amplitude $A_{s} = 0.1$ but varying $\Delta f_{s}$ (Fig.~\ref{f:4node_deterministic_ACStim}E). Here one can see that for $\Delta f_{s} < 0$, there are three clusters that each contain points with relatively similar values of $M_{f}$, and the clusters themselves are separated by steeper vertical ``jumps" as the frequency offset decreases. We also see that as $\Delta f_{s}$ increases in the positive direction, the state-morphing factor also increases, up until the point that phase-locking breaks down altogether. 

In order to see more clearly how the interareal phase relations are modulated for different levels of the state-morphing factor, we examine the relative phases of each area (Fig.~\ref{f:4node_deterministic_ACStim}F) and their activity time-series (Fig.~\ref{f:4node_deterministic_ACStim}G) for different values of the frequency offset $\Delta f_{s}$ at a fixed amplitude $A_{s} = 0.1$. We observe that for small, negative frequency offsets all four regions remain separated by a non-zero phase difference. However, the phase lag between the stimulated area 1 (yellow) and the area that leads it (region 3, green) increases slightly from baseline, and the stimulated region also slightly increases its phase-lead over region 4 (red). If the frequency offset is decreased slightly further, we enter the second $M_{f}$ cluster. In this regime, the area that initially lagged the stimulated region (area 4, red) and the area that was initially anti-phase with the stimulated region (area 2, purple) merge with one another such that they become in-phase and collectively further behind the stimulated site. Simultaneously, the phase difference between area 3 (green) and the stimulated area increases. Decreasing $\Delta f_{s}$ slightly more within this same regime further modulates the spatial arrangement of the oscillatory activity peaks, such that area 1, which receives the external input, becomes a global phase-leader. When the frequency offset is made yet smaller, the circuit enters the third state-morphing cluster present in Fig.~\ref{f:4node_deterministic_ACStim}E. Here, all units except the stimulated region merge into a single in-phase cluster, which itself lags behind the stimulated site. We can also consider slightly positive frequency offsets at this same stimulation amplitude (before phase-locking breaks down). In this case, we see that the most dramatic effect is that the stimulated area moves closer to the region that leads it in phase; the other phase relations are also modulated slightly as well.

In sum, we see that focal, rhythmic stimulation morphs the collective phase-locking state, inducing shifts in the relative spatial arrangement of the population oscillations. We comment on potential functional roles of this type of state control in the Discussion section.

\subsubsection{2-area networks: Stochastic scenario}
\label{s:2node_stochastic_AC}

\begin{figure}[h!]
	\centering
	\includegraphics[width=\textwidth]{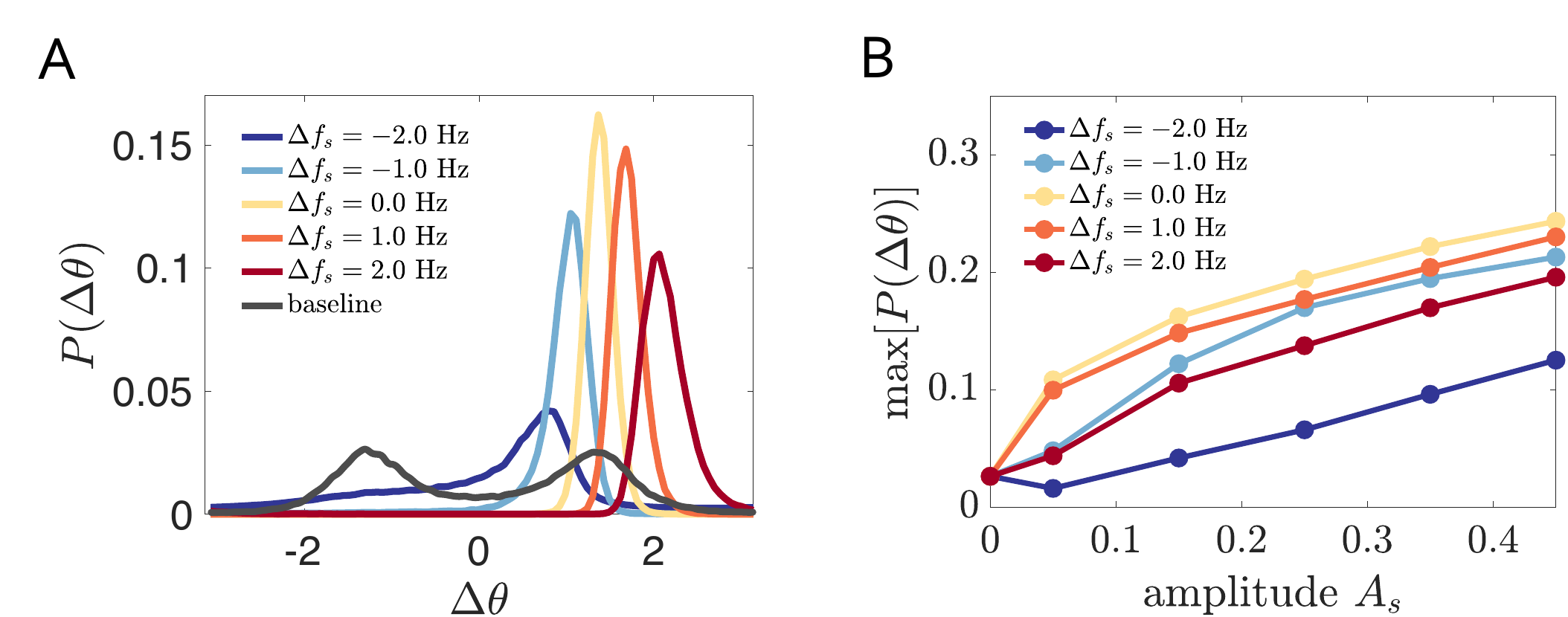}
	\caption[Response of a stochastic 2-area network to rhythmic stimulation.]{\textbf{Response of a stochastic 2-area network to rhythmic stimulation.} The key network parameters are: $\overline{P_{E}} = 1.35$, $T_{D} = 1.5$ ms, $G_{EE} = 0.2$, $\alpha = 15$, $\sigma = 0.15$. \textbf{(A)} Distribution $P(\Delta \theta)$ of the phase difference between the two areas in the network at baseline (gray curve), and for different values of the frequency offset $\Delta f_{s}$ (colored curves). The stimulation amplitude is fixed to $A_{s} = 0.25$ and is applied to area 1. Note that $\Delta \theta > 0$ corresponds to a configuration in which area 1 leads area 2 in phase. \textbf{(B)} The maximum value of the phase difference distribution, $\max{[P(\Delta \theta)]}$, as a function of the stimulation amplitude $A_{s}$ and for different frequency offsets $\Delta f_{s}$ (different colors).}
	\label{f:071120201216_2area_ACstim_stochastic}
\end{figure}

In the previous two subsections, we examined the response of collective phase-locking states to rhythmic stimulation when the dynamics were deterministic. Here we continue this analysis for the more realistic scenario in which the model brain circuits are driven by a stochastically fluctuating background environment (Sec.~\ref{s:stochastic_backgroundDrive}). We begin with the 2-area system, and consider the baseline parameters $\overline{P_{E,j}} = 1.35$ for $j \in \{1,2\}$, $T_{D} = 1.5$ ms, $G_{EE} = 0.2$, $\alpha = 15$, and $\sigma = 0.15$. As detailed thoroughly in Sec.~\ref{s:2area_baseline_stochastic} and Fig.~\ref{f:2node_baseline_stochastic}, introducing noise causes both lead-lag configurations to be sampled as the system evolves. A consequence of this behavior is the appearance of two out-of-phase peaks in the distribution of interareal phase differences $P[\Delta \theta]$ (see the gray curve in Fig.~\ref{f:071120201216_2area_ACstim_stochastic}A, which was derived from a long baseline simulation of 22 minutes in length). 

To assess the effects of focally-applied rhythmic input on the circuit's collective dynamics, we examine how it alters the distribution $P(\Delta \theta)$ relative to the baseline shape. To do this, we stimulate area 1 in the same way as before, and again vary both the stimulation amplitude $A_{s}$ and frequency offset $\Delta f_{s}$. Note that because the system is now stochastic, the oscillation frequency can vary from cycle to cycle, but the stimulation frequency can still be compared to the frequency that corresponds to the peak of the power spectra computed across the entirety of the baseline simulation. For each combination of $A_{s}$ and $\Delta f_{s}$, we run a simulation equal in length to the baseline version (i.e. 22 minutes in duration), and then recompute $P(\Delta \theta)$. 

In Fig.~\ref{f:071120201216_2area_ACstim_stochastic}A, we plot the distributions of phase differences $P(\Delta \theta)$ for different values of the frequency offset $\Delta f_{s}$ and for a fixed stimulation amplitude $A_{s} = 0.25$. As with the noiseless case (Fig.~\ref{f:2node_deterministic_ACStim}D,E) we find that the stimulation breaks the symmetry of $P(\Delta \theta)$ present at baseline and causes one of the two lead-lag directions to become preferred. Specifically, under the influence of rhythmic input, the two equally tall peaks disappear in favor of asymmetric distributions with boosted maxima at values $\Delta \theta > 0$. In this way, stimulating region 1 causes that area to preferentially become the phase-leader. Moreover, akin to the deterministic results, the frequency offset $\Delta f_{s}$ controls the location of the peak of $P[(Delta \theta)$. Hence, fine-tuning of the collective phase relation -- albeit in a probabilistic manner -- can be obtained by modulating a property of the external input.

For the stochastic scenario, we additionally observe that the peak height of $P(\Delta \theta)$ varies with $\Delta f_{s}$. In particular, stimulating with $\Delta f_{s} = 0$ Hz (such that the stimulation frequency is the same as the baseline frequency) yields the tallest and narrowest phase difference distributions. Shifting the frequency offset to the left or right of zero leads to broader and shorter distributions (less consistent phase-locking) with the effect being more drastic for negative frequency offsets $\Delta f_{s} < 0$. These results can be understood in the context of the noiseless case as well, where we saw that network phase-locking broke down more easily when $\Delta f_{s}$ was shifted away from zero in the negative direction than \textit{vice versa} (Fig.~\ref{f:2node_deterministic_ACStim}C). Finally, we observe that for all frequency offsets, the peak of the phase difference distribution $P(\Delta \theta)$ increases with the stimulation amplitude $A_{s}$ (Fig.~\ref{f:071120201216_2area_ACstim_stochastic}B). As we should expect, this behavior indicates that a stronger input leads to tighter phase relations.

\subsubsection{4-area networks: Stochastic scenario}
\label{s:4node_stochastic_AC}

\begin{figure}
	\centering
	\includegraphics[width=\textwidth]{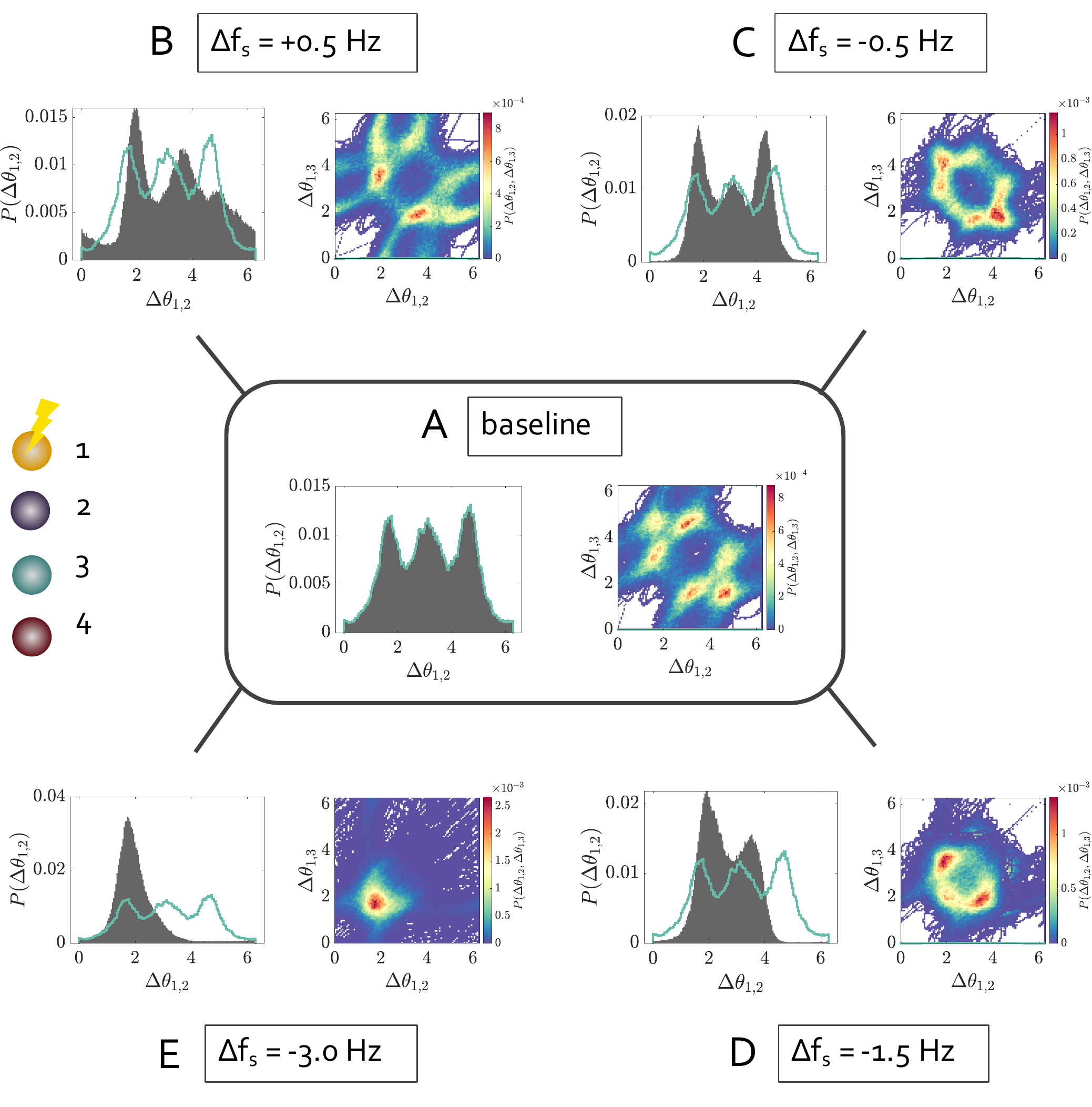}
	\caption[Response of a stochastic 4-area network to rhythmic stimulation.]{\textbf{Response of a stochastic 4-area network to rhythmic stimulation.} The key network parameters are: $\overline{P_{E}} = 1.325$, $T_{D} = 2.5$ ms, $G_{EE} = 0.2$, $\alpha = 15$, $\sigma = 0.2$. \textbf{(A)} The distribution $P(\Delta \theta_{1,2})$ of the phase difference between areas 1 and 2 (Left) and the joint distribution $P(\Delta \theta_{1,2}, \Delta \theta_{1,3})$ of $\Delta \theta_{1,2}$ and $\Delta \theta_{1,3}$ (Right) under baseline conditions. \textbf{(B--E)} The gray histograms in the left panels show the distribution $P(\Delta \theta_{1,2})$ of the phase difference between areas 1 and 2 for a stimulation amplitude $A_{s} = 0.1$ and for varying frequency offsets $\Delta f_{s} = \{0.5,-0.5-1.5,-3.0\}$ Hz. For comparison, the green curves show the baseline distribution. The right panels show the joint distribution $P(\Delta \theta_{1,2}, \Delta \theta_{1,3})$ of $\Delta \theta_{1,2}$ and $\Delta \theta_{1,3}$ for a stimulation amplitude $A_{s} = 0.1$ and varying frequency offsets $\Delta f_{s} = \{0.5,-0.5-1.5,-3.0\}$ Hz.} 
	\label{f:4node_stochastic_ACStim}
\end{figure}

Continuing methodically, in this section we examine the response of stochastic 4-area networks to targeted, rhythmic stimulation. To carefully compare with the deterministic case, we use the same baseline parameters as in Sec.~\ref{s:4node_deterministic_AC} ($\overline{P_{E,j}} = 1.325$ for $j \in \{1,...,4\}$, $T_{D} = 2.5$ ms, $G_{EE} = 0.2$). In terms of the noise, we set $\alpha = 15$ and $\sigma = 0.2$. A detailed characterization of the baseline dynamics of the circuit with stochastic inputs can be found in Sec.~\ref{s:4area_stochastic_baseline}. Here we simply remind the reader that the noise causes the network to spontaneously sample collective activity patterns that resemble approximate forms of the six multistable states from the deterministic limit. This results in a trimodal distribution $P(\Delta \theta_{1,2})$ of the phase differences between a given pair of units 1 and 2 in the network (Fig.~\ref{f:4node_stochastic_ACStim}A,Left) and the emergence of six high density areas in the joint distribution $P(\Delta \theta_{1,2}, \Delta \theta_{1,3})$ of two unique pairs of phase relations (Fig.~\ref{f:4node_stochastic_ACStim}A,Right). 

We model rhythmic stimulation identically to the deterministic case, which is by injecting a sinusoidal input of amplitude $A_{s}$ and frequency offset $\Delta f_{s}$ into the excitatory subpopulation of area 1 (see Sec.~\ref{s:AC_stim} for details). In the absence of noise, a key finding of our analysis was that varying the frequency offset $\Delta f_{s}$ resulted in a morphing of the collective state wherein a given phase-ordering was maintained but the precise values of the pairwise phase differences shifted relative to the baseline configuration (Sec.~\ref{s:4node_deterministic_AC}, Fig.~\ref{f:4node_deterministic_ACStim}). Here, we thus focus on how changing $\Delta f_{s}$ modulates the distributions $P(\Delta \theta_{1,2}$ and $P(\Delta \theta_{1,2}, \Delta \theta_{1,3}$ relative to their baseline shapes (in what follows, note that the conclusions are the same for any $P(\Delta \theta_{1,j \neq 1}$ and $P(\Delta \theta_{1, j \neq 1}, \Delta \theta_{1,k \neq j}$). To do this, we run long simulations (15 minutes in total length) for different values of the stimulation frequency offset, holding the stimulation amplitude fixed at a value $A_{s} = 0.1$. This procedure allows us to properly sample the phase-locking statistics of the network's dynamics under different stimulation conditions. Note that the frequency offset is computed as the difference between the stimulation frequency and the peak frequency from the baseline power spectrum determined from the entire length of the simulation. The results of our analysis are presented in Fig.~\ref{f:4node_stochastic_ACStim}; the center panel corresponds to baseline conditions and the four panels around the edge correspond to stimulating area 1 with rhythmic input at four different values of $\Delta f_{s}$. We unpack each case one-by-one.

To begin, we look more closely at $\Delta f_{s} = 0.5$ Hz, when the stimulation frequency is slightly higher than baseline (Fig.~\ref{f:4node_stochastic_ACStim}B). With this external input, the distribution $P(\Delta \theta_{1,2})$ exhibits three peaks of varying height, and the peak locations are also shifted relative to the peak locations at baseline. Six splotches can also still be resolved in the joint distribution $P(\Delta \theta_{1,2}, \Delta \theta_{1,3})$ (though some of them are more blurred along certain directions) indicating that there are still six different collective states. Based on the way the modes shift, we can begin to see that the peaks in the $\Delta \theta_{1,2}$ histogram are signatures of the three distinct phase relations that exist between the stimulated node and the other areas in the network in the deterministic version of the system. In particular, the left peak shifts closer to $\pi$, meaning that in a given collective state, the stimulated area will phase-lead one of the regions by a larger amount relative to baseline; this same effect can be seen for one particular phase-ordering in column 6 of Fig.~\ref{f:4node_deterministic_ACStim}F, where the yellow and red areas move apart. With noise, the system will spontaneously visit different phase orderings, but in each case, there will always be one region that tends to lag further behind the stimulated area than it would in the corresponding baseline configuration with the same phase ordering. Similarly, we see that the middle and right peaks both shift towards $\Delta \theta_{1,2} = 2\pi$, indicating that the stimulated area narrows the gap but still lags behind the two other regions. This behavior is akin to the green and purple areas shifting towards the stimulated area in the deterministic example. In sum, we thus still observe the state morphing effects induced by rhythmic external input when the circuit is stochastic. One important point, though, is that while the noiseless results might lead us to conclude that all three peaks should be of approximately equal height in the noisy system, this is clearly not the case. Rather, we see that the peak heights decrease from left to right, suggesting that some of the phase relations are more stable and robust under noise than others. This manifests as a stretching of four of the six hotspots in the joint distribution as well. One reason peak asymmetry might occur is that for $\Delta f_{s} = 0.5$ Hz, the collective dynamics (in the deterministic system) are close to an instability, in that increasing the frequency offset a bit further leads to the onset of a different dynamical regime in which phase-locking breaks down. In the non-phase-locked condition, the system may spend longer amounts of time near certain phase differences as it evolves. It may be the case that -- due to noise -- the network exhibits signatures of this non-phase-locked regime as well. A complete understanding of this situation requires further analysis.

When the frequency offset is instead decreased to $\Delta f_{s} = -0.5$ Hz, we again see evidence of spontaneous switching between morphed versions of the six collective baseline states (Fig.~\ref{f:4node_stochastic_ACStim}C). In particular, $P(\Delta \theta_{1,2})$ remains trimodal, but the left and right peaks move towards $\pi$, indicating slightly increased absolute phase differences with the stimulated node relative to baseline. These modulations are relatively consistent with the picture in Fig.~\ref{f:4node_deterministic_ACStim}F, Column 4, where the green and red phases shift further away from the stimulated site, whereas the purple area remains near its location prior to stimulation. Note also that the precision of the two out-of-phase peaks becomes tighter under stimulation.

Continuing to decrease the frequency offset to $\Delta f_{s} = -1.5$ Hz results in even more drastic state morphing (Fig.~\ref{f:4node_stochastic_ACStim}C) that can yet again be understood or predicted from the corresponding deterministic scenario. For these conditions, the phase difference distribution $P(\Delta \theta_{1,2})$ exhibits only two peaks, each of which are also located at different positions than any of the three peaks in the baseline histogram. In addition, the left mode is significantly higher than the right mode. This situation corresponds loosely to the regime in the noiseless case where two of the non-stimulated areas merge to become in-phase with one another and collectively further lag the stimulated region, while the third area increases its lead over the stimulated area (Fig.~\ref{f:4node_deterministic_ACStim}F, Column 3). There are thus two distinct phase relations regarding the stimulated site. In the stochastic circuit, this fact manifests as the distribution $P(\Delta \theta_{1,2})$ becoming bimodal: an area $j\neq 1$ (e.g., $j = 2$) will either lag the stimulated region in a given collective state (left peak) or lead the stimulated region (right peak). It is also more likely for a non-stimulated area to be part of the in-phase cluster than not, which explains the higher amplitude of the left mode relative to the right one. Also note that the joint distribution now exhibits three high-density areas, which are signatures of the now only three distinct collective states that the system spontaneously transitions among (either areas 2 and 3, 2 and 4, or 3 and 4 will comprise the in-phase group). Moreover, because it is more likely for one area to be leading the stimulated site and the other lagging it, the off-diagonal hotspots are more populated than the on-diagonal one. 

If the stimulation frequency is reduced yet further relative to the peak baseline frequency (e.g. $\Delta f_{s} = -3.0$ Hz), the state morphing progresses further as well. Here we see that under stimulation, the phase difference distribution $P(\Delta \theta_{1,2})$ has only a single mode, with a position indicating that the stimulated area becomes a global phase-leader. Indeed, without noise, these stimulation parameters yield a state in which the three unperturbed nodes are drawn into a single in-phase cluster, which itself lags behind the activity of the stimulated site. As expected, in the stochastic network, the joint distribution $P(\Delta \theta_{1,2}, \Delta \theta_{1,3})$ now shows only a single well-populated region centered around the location where the two considered phase relations are approximately equal. This configuration corresponds to a smeared version of the single collective state that exists in the deterministic limit (Fig.~\ref{f:4node_deterministic_ACStim}F, Column 1).

In total, the results of this section indicate that when the 4-area circuit is subject to stochastic background drive, targeted, rhythmic stimulation can still lead to approximate forms of collective state-morphing.

\section{Discussion}
\label{s:ch3_discussion}

In this report we used computational modeling to investigate how coherence-based functional connectivity states in multiarea brain networks can be controlled via targeted stimulation. We worked specifically with model networks that yield multistable phase-locking patterns in the absence of external perturbations and noise. Then, in the context of simple 2-area and 4-area circuits, we explored functional state modulation via locally-applied \textit{(i)} brief stimulation pulses, and \textit{(ii)} sustained rhythmic inputs. When dynamics were deterministic, we first established that short-lived perturbation pulses injected into a single region can induce transitions between different phase-locking patterns. Consistent with previous studies that have used models of interneuron-based gamma synchrony \cite{Battaglia2012:DynamicEffective,Witt2013:Controlling,Palmigiano2017:FlexibleInformation,Lisitsyn2019:Causally}, in the 2-region network, this state-switching amounted to changing the sign of the collective phase relation. In the 4-region system, we further showed that two distinct state-transitions are possible, which involve either a phase-advance or a phase-delay of the perturbed site. In addition to reconfiguring the global organization of collective states, we found that focal sinusoidal input can also regulate the spatial structure of the network's phase relations, allowing for an additional type of state modulation. Lastly, the most crucial component of this study was an examination of how stochastic fluctuations in the background inputs to the network affect state control. At baseline, incorporating this type of realistic noise caused the multistable activity patterns of the interareal circuits to instead become only transiently stable. Nevertheless, we showed that the aforementioned control strategies could still be implemented successfully (in a probabalistic manner) if noise levels are not too high. Finally, we illustrated that state-switching modulations become infeasible in a regime where regional oscillations and out-of-phase locking are purely noise-driven, highlighting that the efficacy of collective state-control depends on how the states themselves are generated.

\subsection{Connections to past work and ideas}

It is crucial to acknowledge the connections between our investigation and a number of inspiring studies that have previously posited and explored the themes upon which our analysis rests. First, although functional interactions may take many forms, we focused here on those generated via the coherence of oscillatory activity from distributed neural populations. Hence, our study relied on the well-studied communication-through-coherence hypothesis postulating that effective communication between two neuronal groups arises when their intrinsic population rhythms lock with an appropriate phase relation \cite{Fries2015:Rhythms,Fries:2005a,Bastos2015:Communication,Maris2016:DiversePhase,Varela2001:TheBrainweb,Siegel2012:Spectral}. A number of experimental studies have lent support to this idea \cite{Womelsdorf2007:ModulationOfNeuronal,Roberts2013:Robust,Schoffelen2005:Neuronal,Hipp2011:Oscillatory,Bosman2012:AttentionalStimulus,Gregoriou2009:HighFrequency,Grothe2012:Switching}, and it has also been explored in modeling efforts that use both reduced neural mass models \cite{Battaglia2012:DynamicEffective,Perez-Cervera:2020aa} and more detailed spiking models \cite{Witt2013:Controlling,Palmigiano2017:FlexibleInformation,Lisitsyn2019:Causally,Barardi2014:PhaseCoherence,Tiesinga2010:Mechanisms,Quax2017:TopDown}.

A critical concept at the heart of this study is that of multistability, wherein multiple stable states coexist in a system for a given set of fixed parameters. Multistability has been hypothesized to allow neural systems -- across many spatial and temporal scales -- to switch between a variety of different functions as the requirements of the system change \cite{Kelso2012:Multistability}. This capability may be invaluable when the underlying structure or parameters of the system at hand cannot be changed on functionally-relevant timescales. Importantly, multistability could be an inherent property of a single dynamical element, or it could be an emergent, collective property of a multi-component system. The type of multistability in this work falls under the latter of those two forms. In particular, a ``state" here was considered to be a particular pattern of network phase-locking, and that pattern as a whole was the multistable entity (rather than the nature of the activity in a single brain area). Other types of multistability may also manifest in brain dynamics and have been examined as well (see, e.g., Refs. \cite{Freyer2011:Biophysical,Deco2012:OngoingCortical,Freyer2012:Canonical}).

It is essential to emphasize that this study is not the first to propose multistable phase-locking and collective state control as mechanisms that could underlie flexible interareal communication in brain networks. On the contrary, our analysis is strongly inspired by and builds directly upon the ideas presented in a collection of studies by \citet{Battaglia2012:DynamicEffective}, \citet{Witt2013:Controlling}, and \citet{Palmigiano2017:FlexibleInformation}. Because the work presented here strongly draws from these past efforts, it is important to highlight their contributions and the relationships to the current study. 

We begin with \citet{Battaglia2012:DynamicEffective}, which, using both spiking simulations and a rate model, employed a causal analysis of information transfer for different multistable phase-locking states of 2-area and 3-area cortical circuits. This study showed that the different lead-lag relationships afforded by different multistable states differentially regulated the directionality of effective connectivity in the network. Hence, a direct connection between out-of-phase oscillatory coherence and flexible information flow was established. Moreover, the authors numerically and semi-analytically studied state-switching in the 2-area networks via pulse currents. The study in \citet{Witt2013:Controlling} also examined out-of-phase coherence in 2-area circuits using a spiking model, and focused specifically on formulating a detailed, biologically-plausible model of optogenetic stimulation. In particular, they developed a closed-loop stimulation protocol capable of carefully controlling the functional connectivity state of the network. We also bring specific attention to the work of \citet{Palmigiano2017:FlexibleInformation}. While the previously mentioned studies examined mainly a strongly synchronized regime, \citet{Palmigiano2017:FlexibleInformation} used a biophysically-detailed spiking model of coupled cortical areas to study out-of-phase locked states in a significantly more realistic transient synchrony regime. In particular, the authors used information-theoretic analyses to demonstrate that highly stochastic networks that exhibit only weak synchrony and short bursts of interareal coherence generate flexible information routing states dictated by collective phase relations during brief, phase-locked windows. 

The main goal of the present work was to perform a numerical investigation of how collective states in multi-regional brain circuits can be controlled by local stimulation, and we attempted to build upon the previously noted studies mainly on this front. For example, in addition to brief pulse perturbations, we also examined the consequences of sustained rhythmic stimulation for modulating coherence-based functional connectivity states. Moreover, we also focused on studying the effects of background noise and the baseline dynamical regime of the interareal networks on the efficacy of collective state control via local inputs. 

It is also important to note that, although we were interested here in multistable phase-locking in the context of interareal communication in large-scale brain circuits, an earlier-proposed functional role of multistable phase-locking relates to central pattern generators (CPGs) in the nervous system. CPGs are specialized neuronal circuits that can inherently produce rhythmic, phase-locked activity patterns thought to underlie repetitive motor functions such as breathing or walking \cite{Marder:2001aa,Marder1996:Principles}. Past studies have shown that some biophysical models of these cellular networks are capable of producing multistable states \cite{Canavier:1997aa,Canavier1999:Control,Golubitsky:1998aa,Wojcik:2014aa,Wojcik:2011aa}, and have explored the ability of transient inputs to multistable, deterministic CPGs to induce switching between patterns that may generate different motor behaviors (e.g., those that correspond to different gaits of a quadruped \cite{Canavier1999:Control} or those that correspond to cat locomotion \textit{vs.} paw-shaking \cite{Parker:2018aa}). Despite the difference in the spatiotemporal scales and models relevant to studying CPGs \textit{vs.} long-range communication in multiarea brain networks, similar dynamical mechanisms and control strategies can apply in both cases.

\subsection{Broader functional implications}

In this section we briefly expand on some potential broader implications of this work related to the operating principles of interareal brain networks. First, as initially suggested and discussed in previous studies \cite{Battaglia2012:DynamicEffective,Palmigiano2017:FlexibleInformation}, the emergence of multistable phase-locking may be useful from a functional standpoint by enabling a single structural network to generate a multiplicity of effective connectivity states. In this way, the system would not necessarily need to rewire its structural connections -- which may be impossible on short time-scales or energetically expensive -- to support different information routing pathways. However, if the system is deterministic (or even if the intrinsic dynamics strongly enough outweigh any background noise) having multistability alone is not enough for computational flexibility. In such a limit, once a particular phase-locking pattern stabilizes, it would never (or rarely) change. Even in regimes where stochastic background fluctuations induce spontaneous switching between the circuit's collective attractors, proper network function may still require selecting certain states at specific times. In either of these scenarios, what is needed is a means of controlling the collective phase relations.

We thus set out to investigate the potential of targeted external inputs to reconfigure multistable phase-locking patterns in multiregional model brain networks. Importantly, these modulatory signals could represent exogenously-generated sensory inputs, modulation from other brain areas, or artifical stimulation. Though one could study arbitrarily complex control signals, we considered two approachable types -- brief pulse stimuli and sustained rhythmic stimulation -- that may nonetheless be relevant for and/or utilized by neural systems. Moreover, we were interested here in what is attainable via inputs that target only one area. While ``control" in the brain might often require simultaneous manipulation of several areas, there is appeal in considering strategies that may demand less complex calibration and coordination. 

We began by considering collective state modulation in noiseless interareal circuits. For both the 2-area and 4-area networks, we found that brief, focal input pulses -- if strong enough and long enough, and applied at an appropriate phase of the receiving oscillation -- could cause a rearrangement of the circuit's phase-locking pattern. Importantly, similar effects have been reported previously in other studies that have used alternative neural population models \cite{Battaglia2012:DynamicEffective,Witt2013:Controlling}. Asumming that different phase-locking patterns correspond to distinct information routing states, these past studies have noted that a significant functional implication of being able to induce such state-switching would be the ability to directly control the direction of interareal information transfer. In this way, a single focal pulse can not only transiently affect the activity of the stimulated region, but could also induce a \textit{global} alteration of interareal communication patterns, which may in turn allow the system to perform a different function.

A key component of this study was the analysis of functional state modulation in the presence of realistic stochastic background drive. As in the more complex models presented in Refs. \cite{Witt2013:Controlling,Palmigiano2017:FlexibleInformation}, incorporating noise causes the network to spontaneously transition between the collection of phase-locked configurations that are simultaneously stable in the absence of noise. In the presence of some stochasticity, the notion of perturbation-induced state-switching thus becomes probabilistic and its effectiveness must be measured relative to the chance of observing the same result spontaneously. Indeed, if given long enough, the stochastic networks will sample all of their preferred states on their own. However, critical to the functional capabilities of brain networks is also how quickly they are able to enter or leave particular states. 

For both the 2-area and 4-area networks with stochastic baseline inputs, we found that stimulation pulses -- if applied during windows of reasonable interareal coherence -- could indeed significantly accelerate state switching. Thus, the local control signal has a consequential effect precisely when it is needed most; that is, when the collective state would take significantly longer to change on its own. It is also important to note, though, that the efficacies of perturbation-induced state switching decreased with decreasing signal-to-noise levels. This effect is certainly undesirable, but because the networks rarely exited a given phase-locked state within just a couple of cycles, the perturbations continued to be useful for state regulation (albeit to a lesser extent) even when the baseline activity patterns were more volatile. One other point to comment on is that, even if the phase configuration after a perturbation cannot stabilize near one of the noiseless attractors of the network, this may not automatically imply that the perturbation would have no functional consequence. Rather, it may be useful to simply prevent communication or information transfer in particular directions. This type of state control might still be achieved if the perturbation can momentarily reconfigure the network's lead-lag relations, even if the post-perturbation state is short-lived or does not exhibit tight phase-locking. In sum, our findings suggest that even in a stochastic setting, brief input pulses can be harnessed to rapidly reconfigure phase-locking patterns in interareal networks. In this way, well-timed, focal perturbations could serve as a mechanism for de-selecting/selecting certain collective states at particular moments, which may in turn enable controlled, rapid switching between different computations performed by the circuit.

Also crucial to understand is when state control can break down. In the context of our model, we showed that one scenario in which brief pulse inputs failed to control the network state occurred when the baseline drive was reduced such that regional oscillations and transient phase-locking were purely noise-driven. In particular, we studied a parameter set where the 2-area network did not exhibit robust regional oscillations in the absence of noise, but with stochastic baseline inputs, exhibited bursts of strongly rhythmic activity and two out-of-phase peaks in its phase difference distribution. Importantly, consideration of this working point first showed how the presence of out-of-phase peaks in the phase difference distribution do not in and of themselves indicate collective multistability in the deterministic limit. For the scenario discussed here, the asymmetrical phase relations instead arose entirely due to fluctuations in the level of background drive to each unit. Nonetheless, it was still possible to extract time windows wherein the network was approximately phase-locked with a preferred phase relation, and to test the effects of pulse stimuli on controlling the network state during those epochs. In doing so, we found that state-switching could not be reliably induced at levels significantly above chance. Hence, while some features of the baseline dynamics appeared to have similar structure to other, controllable working points, the low-drive working point was not controllabe via brief stimulation pulses. This finding may seem surprising initially. However, if asymmetric phase-locking is a result \textit{only} of noise rather than underlying multistability, then an only transient input will not be able to induce a reliable and lasting effect. In addition, the fluctuating and relatively low amplitude of the regional oscillations and very rapid dissolution of phase-locking further hinder the ability to control collective states in the low-drive regime. One interesting implication of these findings is the suggestion that, by observing the response to focal perturbations, one may be able to discern the dynamical nature or operating point of collective neural activity states \textit{in vivo}, for example, via optogenetic stimulation \cite{Cardin:2010aa}.

We have thus far focused on the functional implications of being able to induce state-switching -- a global reorganization of the collective lead-lag relationships in an interareal network -- via pulse inputs. In the second portion of our study, we also explored the influence of targeted rhythmic stimulation, finding that a different type of state-modulation became possible in this case. For the 2-area networks with and without noise, the first consequence of oscillatory drive (with appropriate amplitude and frequency) was a symmetry-breaking effect: it caused the stimulated area to become the global phase-leader. A possible functional implication of this effect is that interareal communication would be prevented (or would be significantly more unlikely) in one direction and boosted in the other (given that the sign of the phase relation governs the direction of information flow \cite{Battaglia2012:DynamicEffective, Palmigiano2017:FlexibleInformation}). Additionally, we found that tuning the frequency of the input signal could enable a more fine-tuned shifting of the phase difference to larger or smaller values. This more tailored control may be important if the system requires a slight adjustment of a particular information transmission pathway in order to more accurately perform a certain operation in a perturbed environment, for example.

The long-lasting effects of rhythmic stimulation on the collective states in 4-area circuits were slightly more complex, but they could still be characterized as a form of state-morphing: a modulation of the spatial arrangement of the phase relations relative to the corresponding uniform, baseline configuration. If the stimulation frequency was close to the baseline frequency, we found that the pairwise phase relations shifted relative to one another -- with different pairs becoming closer together or farther apart -- but all phase differences remained non-zero. This kind of fine-tuned adjusting of the collective phase-locking state again has functional implications within the context of the CTC framework, where interareal phase relations are thought to play a role in regulating the efficiency of interareal communication \cite{Fries2015:Rhythms,Fries:2005a,Bastos2015:Communication,Maris2016:DiversePhase}. For example, it may be beneficial to have a means of re-weighting (boosting or depressing) the strengths of a hard-wired circuit's effective routing channels -- without reversing the direction of information flow entirely -- in order to optimally process and transfer information depending on the broader context. We also found that morphing effects could be made more extreme by altering the stimulation frequency further. For example, two areas initially separated by a non-zero phase-lag could be pushed together, such that they merged into a single in-phase cluster. This type of modulation might allow for more symmetric information flow between the areas that coalesce into a single in-phase group, while maintaing directed information flow between the rest of the network elements that exhibit out-of-phase locking.

\subsection{Limitations and future work}

This investigation has a number of limitations, and therefore leaves open many important directions to explore in forthcoming studies. First, in our use of the Wilson-Cowan model, we have studied a computationally tractable but highly idealized realization of neural population activity. On the one hand, the relative simplicity of the WC model permitted a thorough mapping of networks' baseline dynamics in the presence of stochastic background inputs, as well as in-depth analyses of functional state manipulation at a variety of working points, for different networks, and for different perturbations. Moreover, it may even be possible to make some analytical progress in the context of this model in future work. On the other hand, though, only more biologically detailed models allow for \textit{(i)} linking microscopic, spiking-level neuronal activity to population-level phase-locking \cite{Palmigiano2017:FlexibleInformation} and collective state control \cite{Witt2013:Controlling}, and \textit{(ii)} an understanding of how macroscopic functional states relate to interareal communication and information flow at the level of individual spike patterns \cite{Battaglia2012:DynamicEffective}.

A second limitation of this study is the simplicity of the anatomical network connectivities. Specifically, we opted to examine canonical structural motifs that exhibited clean, multistable phase-locking in the deterministic limit and hence allowed us to focus on and parse the effects of dynamical perturbations (both with and without background noise). It is important to note that such small, strongly-coupled cliques do exist in the brain at different scales \cite{Sporns:2004aa,Heuvel:2011aa,Hilgetag:2004aa,Song:2005aa,Klinshov:2014aa,Sizemore:2018aa}, and the results of our analysis are thus pertinent to those situations. One might even speculate that it could be functionally beneficial for these dense, highly-symmetric clusters to be widespread, since they can both engender a large multiplicity of states due to their symmetries \cite{Battaglia2012:DynamicEffective,Battaglia2007:TemporalDecorrelation} and are also highly amenable to control. That said, it would be interesting to study functional state control in the presence of some realistic asymmetries in both the interareal connectivity weights and time delays. It is also well-known that large-scale, whole brain networks exhibit significantly more complex and heterogeneous wiring patterns. Another important direction would thus be to consider collective routing states and their control in larger brain networks with more varied connectivity. Actually, \citet{Kirst2016:Dynamic} recently explored this avenue in a more general context (i.e., not just for neural dynamics) and found a number of compelling results, such as the possibility for remote control of distant functional interactions via local manipulations. 

One could also study the control of collective states in interareal circuits that emerge from multifrequency oscillations \cite{Helmer:2015aa}. While particular focus has been placed on the role of gamma oscillations for information routing between brain areas \cite{Bosman2014:Functions,Pascal2009:NeuronalGamma}, there is also evidence that other frequency bands may play a role in long-distance communication as well \cite{Siegel2012:Spectral,Brovelli:2004aa,Varela2001:TheBrainweb,Canolty2010:TheFunctional}. For example, gamma band coherence is thought to regulate bottom-up information routing, whereas alpha and beta band interactions are thought to regulate top-down communication \cite{Bastos2015:Visual,Kerkoerle:2014aa}. It would therefore be of interest to build computational network models capable of multifrequency interareal coherence \cite{Mejias2016:FeedforwardAndFeedback,Helmer:2015aa}, and to again study the potential for flexible reconfiguration of these more complex communication pathways via targeted control signals. 

Finally, we note that interareal coherence and the corresponding phase relations do not in and of themselves imply the presence of interareal communication and effective functional interactions. Concretely demonstrating that particular phase-locking states enable particular patterns of information flow through a network with certain directionalities and efficiencies requires further work. For example, one way of detecting and quantifying causal interactions in interareal brain circuits is via information theoretic measures such as transfer entropy \cite{Schreiber2000:MeasuringInformation,Vicente:2011aa, Battaglia2012:DynamicEffective, Palmigiano2017:FlexibleInformation}. By applying such information theoretic analyses to the multiarea brain circuits considered here, one could attempt to show that specific modulations of the collective phase-locking states induced by local perturbations correspond to specific modulations of the directions and levels of interareal information transmission. We are currently exploring this direction in ongoing work.

\section*{Acknowledgements}

This work was supported by National Institutes of Mental Health RF1MH116920 (Theodore D. Satterthwaite, Dani S. Bassett, Desmond J. Oathes).

\clearpage
\newpage

\renewcommand*{\thesection}{Supplementary figures}
\counterwithin{figure}{section}
\renewcommand*{\thefigure}{S.\arabic{figure}}

\section{}
\label{s:parameter_variation}

Fig.~\ref{f:081620201420_pulseStim_varyDelta} examines the consequences of increasing the width parameter $\delta$ that determines how much the phase difference $\Delta \theta$ can vary around the preferred value $\Delta \theta^{*}$ within a candidate perturbation window for the stochastic 2-area networks (see Sec.~\ref{s:2area_stochastic_pulse}). More specifically, we consider the effects of $\delta$ for the working point examined in Fig.~\ref{f:081620201420_pulseStim}, where the default value of $\delta = \pi/6$ was used. As expected, increasing $\delta$ results in a widening of the distributions of $\Delta \theta$ that manifest at the beginning of the time-windows selected for the application of perturbations (Fig.~\ref{f:081620201420_pulseStim_varyDelta}A). Moreover, while the maximum switching probability $\max[P_{\mathrm{switch}}]$ across onset phases tends to decrease with increasing $\delta$ (Fig.~\ref{f:081620201420_pulseStim_varyDelta}B), the estimated probabilities remain well above the chance levels (which are less than 0.07 for all considered values of $\delta$). 

Fig.~\ref{f:4node_stochastic_pulseStim_100320201120_varyDelta} demonstrates the effect of varying the width parameter $\delta$ on the efficacy of state-switching in a stochastic 4-area network. More specifically, we consider the effects of $\delta$ for the working point examined in Fig.~\ref{f:4node_stochastic_pulseStim_100320201120}, where the default value of $\delta = \pi/6$ was used. As anticipated, increasing $\delta$ decreases the maximum phase-advance and phase-delay transition probabilities for the chosen fixed pulse amplitude and fixed pulse duration, but the transition probabilities remain above chance levels across the range of examined width parameters.

\clearpage

\begin{figure}[h!]
	\centering
	\includegraphics[width=1\textwidth]{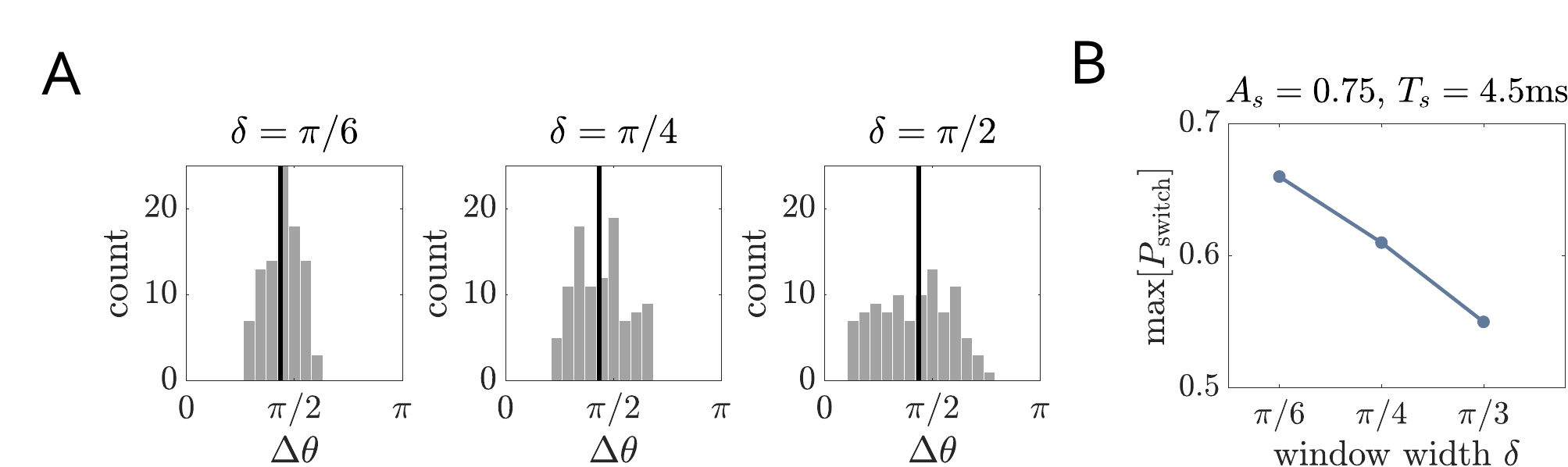}
	\caption[Effect of varying the state-selection width parameter $\delta$ on the response of a stochastic 2-area network to brief pulse perturbations.]{\textbf{Effect of varying the state-selection width parameter $\delta$ on the response of a stochastic 2-area network to brief pulse perturbations.} The key network parameters are: $\overline{P_{E}} = 1.30$, $T_{D} = 1.5$ ms, $G_{EE} = 0.2$, $\alpha = 10$, $\sigma = 0.2$. The perturbation is applied to the phase-leader. \textbf{(A)} Distributions of $\Delta \theta$ at the beginning of the oscillation cycles during which a pulse perturbation is applied. The three different panels correspond to three different thresholds $\delta$, which determine how much the phase difference between the two areas can vary around its peak baseline value across a candidate window for perturbation onset. \textbf{(B)} Across all onset phases $\theta_{\mathrm{on}}$, the blue line shows the maximum switching probability $\mathrm{max}[P_{\mathrm{switch}}]$ that a local pulse perturbation changes the phase-locking state as a function of the width parameter $\delta$. This analysis was carried out for input pulses with amplitude $A_{s} = 0.75$ and duration $T_{s} = 4.5$ ms. A state-switch was considered successful if the sign of the interareal phase relation switched within 2 cycles and if the new configuration lasted for at least 5 more cycles following the initial switch.}
	\label{f:081620201420_pulseStim_varyDelta}
\end{figure}

\begin{figure}[h!]
	\centering
	\includegraphics[width=0.7\textwidth]{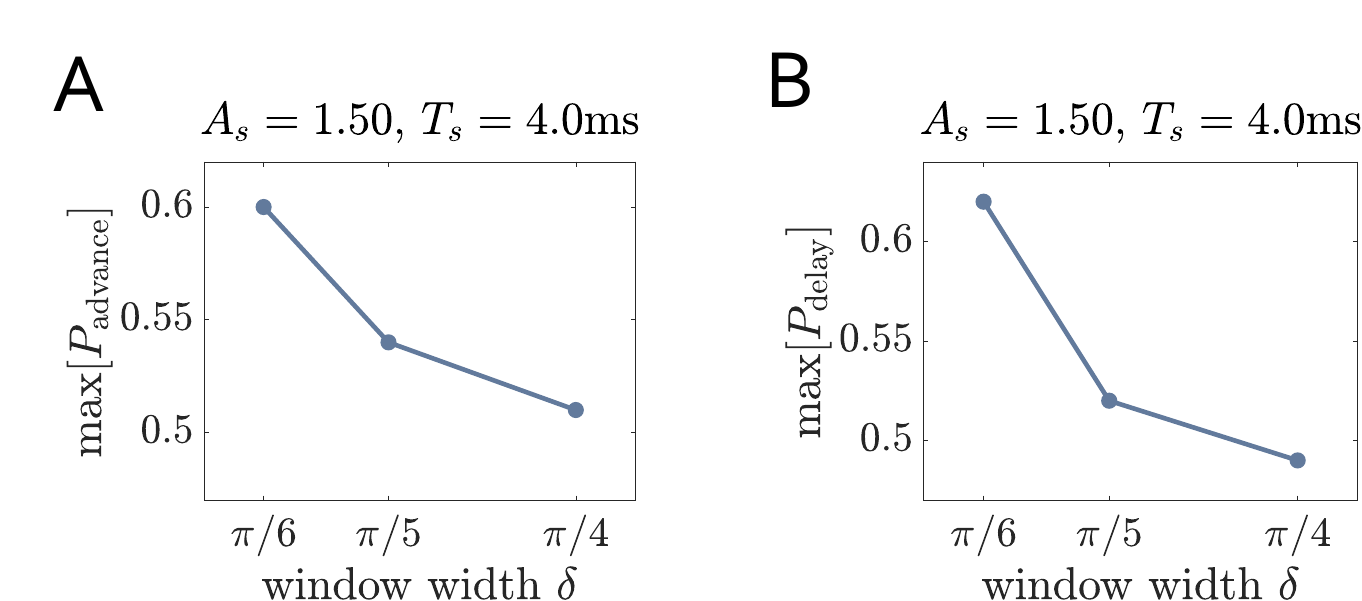}
	\caption[Effect of varying the state-selection width parameter $\delta$ on the response of a stochastic 4-area network to brief pulse perturbations.]{\textbf{Effect of varying the state-selection width parameter $\delta$ on the response of a stochastic 4-area network to brief pulse perturbations.} The key network parameters are: $\overline{P_{E}} = 1.3$, $T_{D} = 1.5$ ms, $G_{EE} = 0.2$, $\alpha = 5$, $\sigma = 0.1$. \textbf{(A)} The maximum phase-advance transition probability across all onset phases, $\max[P_{\mathrm{advance}}]$, for three different values of the state-selection width $\delta$. A state transition was considered successful if the ordering of the phases switched to the correct configuration within 2 cycles and if the new ordering lasted for at least 5 more cycles following the initial switch. This analysis was carried out for input pulses with amplitude $A_{s} = 1.5$ and duration $T_{s} = 4.0$ ms. \textbf{(B)} The maximum phase-delay transition probability across all onset phases, $\max[P_{\mathrm{delay}}]$, for three different values of the state-selection width $\delta$. A state transition was considered successful if the ordering of the phases switched to the correct configuration within 2 cycles and if the new ordering lasted for at least 5 more cycles following the initial switch. This analysis was carried out for input pulses with amplitude $A_{s} = 1.5$ and duration $T_{s} = 4.0$ ms.}
	\label{f:4node_stochastic_pulseStim_100320201120_varyDelta}
\end{figure}


\bibliographystyle{unsrtnat}
\bibliography{your_bib_file}

\end{document}